\documentclass[%
 reprint,
superscriptaddress,
showpacs,preprintnumbers,
 amsmath,amssymb,
 aps,
prb,
]{revtex4-1}

\usepackage{graphicx}
\usepackage{dcolumn}
\usepackage{bm}
\usepackage{subfig}
\usepackage{hyperref}
\usepackage{booktabs}
\hypersetup{
	colorlinks=true,
	linkcolor=blue,
	filecolor=magenta,
	urlcolor=cyan,
}
\usepackage{xcolor}

\newcommand{\vt}[1]{\boldsymbol{#1}}

\newcommand{\td}{\text{d}}
\newcommand{\ket}[1]{\left| #1 \right\rangle}
\newcommand{\bra}[1]{\left\langle #1 \right|}
\newcommand{\braket}[2]{\left\langle #1 | #2 \right\rangle}
\newcommand{\ketbra}[2]{\left| #1 \rangle \langle #2 \right|}

\newcommand{\Schrodinger}{Schr\"{o}dinger }

\begin{document}

\title{Optical nonlinearities of excitons in monolayer MoS$_2$}

\author{Daniel B. S. Soh}
\email[]{dansoh@stanford.edu}
\affiliation{Edward L. Ginzton Laboratory, Stanford University, Stanford, CA 94305, USA}
\affiliation{Sandia National Laboratories, Livermore, CA 94550, USA}
\author{Christopher Rogers}
\affiliation{Edward L. Ginzton Laboratory, Stanford University, Stanford, CA 94305, USA}
\author{Dodd J. Gray}
\affiliation{Edward L. Ginzton Laboratory, Stanford University, Stanford, CA 94305, USA}
\author{Eric Chatterjee}
\affiliation{Edward L. Ginzton Laboratory, Stanford University, Stanford, CA 94305, USA}
\author{Hideo Mabuchi}
\affiliation{Edward L. Ginzton Laboratory, Stanford University, Stanford, CA 94305, USA}

\date{\today}

\begin{abstract}
We calculate linear and nonlinear optical susceptibilities arising from the excitonic states of monolayer MoS$_2$ for in-plane light polarizations, using second-quantized bound and unbound exciton operators. Optical selection rules are critical for obtaining the susceptibilities. We derive the valley-chirality rule for the second-harmonic generation in monolayer MoS$_2$, and find that the third-harmonic process is efficient only for linearly polarized input light while the third-order two-photon process (optical Kerr effect) is efficient for circularly polarized light using a higher order exciton state. The absence of linear absorption due to the band gap and the unusually strong two-photon third-order nonlinearity make the monolayer MoS$_2$ excitonic structure a promising resource for coherent nonlinear photonics.
\end{abstract}

\maketitle


\section{Introduction}

Design bottlenecks arising from energy dissipation and heat generation in the dense on-chip interconnect of CMOS computing architectures have led to renewed interest in approaches to all-optical information processing. The preeminent challenge remains to develop materials with low loss and large optical nonlinearity, which are suitable for incorporation with integrated nanophotonic structures \cite{sipahigil2016integrated, alam2016large, sipahigil2016single, benson2011assembly,liu2010mid}. Looking to the future, the development of coherent nonlinear photonics may be regarded as preliminary work towards quantum photonic architectures that represent and process information utilizing non-classical states of light \cite{hacker2016photon, reimer2016generation, kockum2017deterministic, mabuchi2012qubit}. Beyond computation per se, nonlinear optical materials are fundamental for many other integrated photonics applications including on-chip frequency comb generation \cite{brasch2016photonic, del2007optical}, frequency conversion \cite{guo2016chip} and supercontinuum generation \cite{carlson2017photonic, hsieh2007supercontinuum}.

Atomically thin 2D materials are promising candidates for providing optical nonlinearity in integrated photonic circuits, especially as growth techniques have advanced in recent years to enable the deposition on top of lithographically fabricated devices \cite{ajayan2016two}. In particular, monolayer MoS$_2$ has attracted great interest following the discovery that it is indeed a direct band gap semiconductor \cite{mak2010atomically} with intriguing optical properties such as valley optical selectivity \cite{xiao2012coupled}. The nonlinear optical properties of monolayer MoS$_2$ are now being explored; in this article we aim to characterize important contributions from its excitonic bound states.

Large collective optical responses from excitonic states are well known \cite{haug2009quantum, wang2017excitons}. Reduced dimensionality further increases the optical response of excitons since the most significant contribution to exciton formation comes from the band edges where density of states in 2D is much larger than that in 3D. Hence, excitonic states of the transition metal (Mo, W) dichalcogenides (S$_2$, Se$_2$) (TMDs) are expected to contribute substantial optical nonlinearity even with their atomically thin layer thickness. Unlike in graphene, which has a significant linear absorption everywhere in the optical spectrum, nonlinear processes utilizing the excitonic states of monolayer MoS$_2$ may preserve a sufficient level of coherence due to the band gap. This unique set of features make monolayer MoS$_2$ an attractive material for coherent nonlinear photonics.

The direct band gap around the $\pm \vt{K}$ points in the first Brillouin zone of monolayer MoS$_2$ is analytically best modeled by a gapped Dirac cone \cite{xiao2012coupled, wang2017excitons, wang2016radiative, wang2015fast, wang2012electronics}. We answer the natural question whether the monolayer MoS$_2$ has an optical nonlinearity comparable to that of graphene. Although there are numerous results available for the optical properties of monolayer MoS$_2$ \cite{zhang2014absorption, hill2015observation, wang2012electronics, wang2016radiative, selig2016excitonic}, only a few results on the optical nonlinearities of monolayer MoS$_2$ are available \cite{kumar2013second, clark2014strong, merano2016nonlinear, pedersen2015intraband, trolle2014theory, gruning2014second}. All of these focused on the second-harmonic generation process, which, according to our study, turns out to be a weak perturbative effect stemming from the threefold rotational symmetry, while the third-order nonlinearity may be more significant considering the symmetries of the excitons.

We calculate the optical susceptibilities of monolayer MoS$_2$ when the frequency of the output light is nearly resonant with the highly optically responsive exciton energy levels. We show that, while the optical selection rule dictates the substantially contributing channels in nonlinear processes, that of MoS$_2$ excitonic states inherits the threefold rotational symmetry of the atomic structure. As a result, several unusual high order transition channels can be formed in the excitonic level transitions, which appear to violate the usual valley selection rule. Although previously an empirical nonlinear selection rule was adopted,\cite{seyler2015electrical, xiao2015nonlinear} we explain the optical selection rules through the actual calculation of dipole moments based on massive Dirac Hamiltonian with the perturbative contribution from the threefold rotational symmetry of the atomic system. The same reason leads to unusually efficient third-harmonic generation and the Kerr nonlinearity with certain polarization configurations.

We restrict our analysis to the case where the higher harmonic frequencies fall below the excition levels so that the linear absorption of this higher harmonic frequencies can be avoided. These transitions are of particular interest for all-optical information processing as well as quantum dynamical applications.

This paper consists in the following. Section 2 contains the theory of light matter interaction for the monolayer MoS$_2$, clearly presenting the assumptions made, the Hamiltonians, and the perturbative approach. Section 3 presents the calculation of the linear and the nonlinear susceptibilities for the interesting linear and nonlinear processes, resolved by the input light polarizations. Finally, a conclusion with discussions follows.

The appendix presents a clear derivation of the exciton creation operator based on our defined completeness relations in the Hilbert spaces, which necessarily clarifies the dimension of constants. We also rigorously derived the second quantized operators for the unbound exciton states in the same appendix.

\section{Interaction of monolayer M$\mathbf{o}$S$_2$ with a light field}

We assume zero temperature for simplicity. We count only the radiative transitions, ignoring the coupling to phonon excitations from the radiatively excited states. Most of the practical nonidealities are collected phenomenologically in the linewidth broadening factor. Our primary interest is the linear and the nonlinear optical processes that involve the bound exciton states of the monolayer MoS$_2$. We particularly assume a low-density exciton so that we address only the regime of a single exciton over the sample. Consequently, we ignore the exciton-exciton interaction. This makes the perturbative approach valid. We also ignore the trions, and focus solely on the exciton states.

\subsection{Unperturbed Hamiltonian}

\subsubsection{Excitons}

The band structure of MoS$_2$ is well known \cite{xiao2012coupled}. Due to the mismatch between the Mo atom and the S atom, the spatial inversion symmetry is broken, and hence, the degeneracy at $\pm \vt{K}$ points of the monolayer MoS$_2$ is lifted. The band structure is best described by the gapped Dirac Hamiltonian (see the details in Appendix \ref{sec:band-structure}). We then proceed to the exciton description below.

At zero-degree temperature, the ground state is the Fermi sea $\ket{0}$ where all the electrons are in the valence band. A photon may be absorbed to produce an electron in the conduction band and a hole in the valence band. The Coulomb attraction between the two creates an exciton state. Considering that the exciton size in the monolayer MoS$_2$ is approximately \cite{zhang2014absorption} $\sim$ 1 nm, which is larger than the unit cell, we adopt the Wannier exciton \Schrodinger equation \cite{haug2009quantum, dresselhaus1956effective}:
\begin{equation}
	\left[ - \frac{\hbar^2 \nabla^2}{2 m_r} + V(r)   \right] \psi_\nu (\vt{r}) = E_\nu \psi_\nu (\vt{r}), \label{eq:schrodinger-exciton}
\end{equation}
where $\psi_\nu (\vt{r}) = \braket{\vt{r}}{\textrm{x}_\nu}$ with an exciton state ket $\ket{\textrm{x}_\nu}$ is the wave function of an electron-hole pair with the relative position $\vt{r} = \vt{r}_e - \vt{r}_h$ with the position of the electron $\vt{r}_e$ and the hole $\vt{r}_h$, respectively, $m_r = (1/m_c + 1/|m_v|)^{-1}$ is the reduced mass where we calculate approximately $m_c = 0.55 m_e$ and $m_v = - 0.56 m_e$ from the energy dispersion equation \eqref{eq:threefold-rotational-sym}  with the electron rest mass $m_e$  are the effective masses of the conduction and the valence band electrons, respectively, $V(\vt{r})$ is the Coulomb potential between the electron and the hole, and $E_\nu$ is the energy eigenvalue with the quantum number $\nu$. We note that this \Schrodinger equation includes the Bloch state solutions of the electron and the hole through the renormalized particle mass $m_r$ that reflects the dispersions of the conduction and the valence bands.

The monolayer MoS$_2$ is a 2D sheet. The Coulomb potential, however, is not strictly 2D due to the dielectric screening effect, and the more appropriate potential for an isolated 2D sheet is the Keldysh-type screened potential \cite{cudazzo2011dielectric}. The main differences between the strictly 2D Coulomb potential and the Keldysh-type screened potential are the binding energies and the oscillation strengths \cite{wu2015exciton, robert2017optical, kylanpaa2015binding}, but the corrections are relatively small (of order unity) for the isolated MoS$_2$ 2D sheet when we fit the binding energy of the lowest exciton state to an empirical data (see for example Fig. 3 of Robert \textit{et al.}\cite{robert2017optical} When the lowest binding energies are equalized, the difference in upper level energies is not significant). In addition, the exciton wave functions using the Keldysh-type screened potential are obtained usually through sophisticated numerical methods. Our main goal is to estimate the magnitude of the nonlinear response of the exciton states, and the simple strictly 2D Coulomb potential turns out to be sufficient for our purpose with the advantage of easier calculation of the transition matrix elements among the exciton states, based on the well-known analytic 2D hydrogen-type wave functions. Due to these reasons, we rather adopt the simple 2D Coulomb potential to obtain the 2D solution to the Wannier \Schrodinger equation whose energy eigenvalues are, for the quantum number $\nu = (n,m)$ with $n = 0,1,2, \cdots$ and $m = -n, -(n-1), \cdots, n$:\cite{haug2009quantum}
\begin{equation}
	E_\nu = - E_0 \frac{1}{(n+1/2)^2},
\end{equation}
where
\begin{equation}
	E_0 = \frac{e^4 m_r}{2 ( 4 \pi \epsilon_0 \epsilon_r)^2 \hbar^2} = \left(\frac{m_r}{m_e}\right) \left(\frac{1}{\epsilon_r^2}\right) \text{Ry},
\end{equation}
with the electron charge $e = -|e| = -1.6 \times 10^{-19}$ C, and the vacuum and the relative material permittivity $\epsilon_0, \epsilon_r$, respectively. Here, Ry = $13.6$ eV is the hydrogen Rydberg energy. Indeed, later in the text, our calculated results will be shown to be surprisingly close to the experimental results even with this simplified picture.

The eigenvalues of \Schrodinger equation in equation \eqref{eq:schrodinger-exciton} are the binding energies. Therefore, the actual exciton energy levels are given through $E_c (\vt{0}) + E_\nu$ where $E_c(\vt{0})$ is the lowest energy of the conduction band (see equation \eqref{eq:uqv}). It is also noteworthy that the band structure calculated in Appendix \ref{sec:band-structure} is essential in calculating all orders of susceptibilities since it constructs the exciton creation operator (see Appendix \ref{sec:exciton-creation}). 

The wave function is \cite{haug2009quantum}
\begin{align}
	\psi_{n,m} (\vt{r}) =& \sqrt{\frac{1}{ \pi a_0^2 (n+1/2)^3} \frac{(n-|m|)!}{[(n+|m|)!]^3}} \nonumber \\
	&\times \rho^{|m|} \text{e}^{-\rho/2} L^{2|m|}_{n+|m|} (\rho) \text{e}^{i m \phi}, \label{eq:psir}
\end{align}
where $a_0 = 4 \pi \hbar^2 \epsilon_0 \epsilon_r / (e^2 m_r)$, $\rho = 2r / ((n+1/2)a_0)$, and $L^p_q (\rho)$ is the Laguerre polynomials defined by
\begin{equation}
L^p_q (\rho) = \sum_{\nu = 0}^{q-p} (-1)^{\nu + p} \frac{(q!)^2 \rho^\nu}{(q - p - \nu)! (p + \nu)! \nu!}.
\end{equation}
These wave functions satisfy the normalization $\delta_{\nu, \nu'} = \braket{\textrm{x}_{\nu}}{\textrm{x}_{\nu'}} = \int \td^2 r \braket{\textrm{x}_\nu}{\vt{r}} \braket{\vt{r}}{\textrm{x}_{\nu'}} = \int \td^2 r \psi_{\nu}^* (\vt{r}) \psi_{\nu'} (\vt{r}) $ where we used the completeness relation $\int \td^2 r \ketbra{\vt{r}}{\vt{r}} = \mathbf{1}$. Also one can consider the Fourier transform pair using an additional completeness relation $\sum_{\vt{q}} \ketbra{\vt{q}}{\vt{q}} = \vt{1}$:
\begin{align}
	\psi_\nu (\vt{r}) &= \frac{1}{\sqrt{A}} \sum_{\vt{q}} \psi_\nu (\vt{q}) \text{e}^{i \vt{q} \cdot \vt{r}}, \nonumber \\
	\psi_\nu (\vt{q}) &= \frac{1}{\sqrt{A}} \int_{A} \td^2 r \psi_\nu (\vt{r}) \text{e}^{- i \vt{q} \cdot \vt{r}},
\end{align}
where $A$ is the entire sample area of the monolayer MoS$_2$. As an example, for $\nu = (0,0)$, we have $\psi_{(0,0)} (\vt{r}) = (2 \sqrt{2/\pi}/a_0) \text{e}^{-2 r / a_0}$ and the Fourier transform is $\psi_{(0,0)} (\vt{q}) = \sqrt{2 \pi/A} (8 a_0 / (4 + a_0^2 k^2)^{3/2})$. The corresponding energy eigenvalue is $-4 E_0$.

According to this wave function, the radius of the lowest exciton state is calculated to be $\bra{\psi_{(0,0)}} r \ket{\psi_{(0,0)}} = a_0/2$. The exciton radius of the monolayer MoS$_2$ is experimentally measured as $6 \sim 10 \; \text{\AA}$ at zero temperature \cite{zhang2014absorption}. In addition, the binding energy of the lowest exciton state of the monolayer MoS$_2$ is estimated as $-0.5 \sim -0.3$ eV \cite{zhang2014absorption, klots2014probing, hill2015observation, chiu2015determination}. These two lead to the value of $\epsilon_r$, and we chose $\epsilon_r$ to be 7, which implies the exciton radius of 6.7 \AA$\;$ and the binding energy of $-0.31$ eV.

\subsubsection{Second quantization of excitons}

The total Hamiltonian is the sum of the band Hamiltonian $H$ and the Coulomb potential $V(\vt{r})$ such that $\mathcal{H}_0 = H + V(\vt{r})$. When additional light-matter interaction Hamiltonian $\mathcal{H}_I$ is present, one faces the situation where two interaction Hamiltonians, $V(r)$ and $\mathcal{H}_I$, are both present. This makes the problem complicated. One approach is to absorb the Coulomb potential into the \emph{unperturbed} Hamiltonian and deal with $\mathcal{H}_I$ as a perturbing Hamiltonian.

The Hilbert subspace of the single particle excited states is spanned by the band basis of a pair of an electron and a hole: $\ket{\vt{q}, -\vt{q}'} = \alpha^\dag_{\vt{q}} \beta^\dag_{-\vt{q}'} \ket{0}$ where $\alpha^\dag_{\vt{q}}$ and $\beta^\dag_{-\vt{q}'}$ are the creation operators for the electron Bloch state in the conduction and the hole Bloch state in the valence band, with the momentum $\hbar \vt{q}$ and $-\hbar \vt{q}'$, respectively. Because we know that the exciton states diagonalize the unperturbed Hamiltonian $\mathcal{H}_0$, we now represent it using the second quantized exciton creation and annihilation operators.

Following the procedure in Haug \emph{et al}.\cite{haug2009quantum}, we first define the creation operator of a bound exciton $B^\dag_{\nu \vt{Q}} = \ketbra{\nu \vt{Q}}{0}$ where $\nu$ is the quantum number of the exciton state, and $\hbar \vt{Q}$ is the combined momentum of the electron hole pair. Then, using the completeness $\int \td^2 r \ketbra{\vt{r}}{\vt{r}} = \vt{1}$ and $\sum_{\vt{q}} \ketbra{\vt{q}}{\vt{q}} = \vt{1}$, it is straightforward to show (see the Appendix \ref{sec:exciton-creation} and the equation \eqref{eq:A-3}) that
\begin{equation}
	B^\dag_{\nu \vt{Q}} = \sum_{\vt{q}} \psi_\nu \left( \vt{q} - \frac{\vt{Q}}{2}\right) \alpha^\dag_{\vt{q}} \beta^\dag_{\vt{Q} - \vt{q}}, \label{eq:creation-bound-exciton}
\end{equation}
At zero temperature, the exciton momentum $\hbar \vt{Q}$ must be equal to the momentum of the incoming photon since no phonon is available. Considering the negligibly small photon momentum compared to the crystal momentum $\hbar \vt{q}$, we can approximately set $\vt{Q} \approx \vt{0}$. Then, the bound exciton creation operator is
\begin{equation}
	B^\dag_{\nu} = \sum_{\vt{q}} \psi_\nu \left( \vt{q}\right) \alpha^\dag_{\vt{q}} \beta^\dag_{- \vt{q}}. \label{eq:bound-creation}
\end{equation}

Appendix \ref{sec:exciton-creation} also derives the creation operator for the unbound exciton states as $C^\dag_{\vt{q}} = \alpha^\dag_{\vt{q}} \beta^\dag_{- \vt{q}}$. Setting the energy of the ground state Fermi sea $\ket{0}$ as zero, and using the fact that the entire Hilbert subspace of the single excitation is spanned by the bound and the unbound exciton states such that the completeness relation is (Appendix \ref{sec:exciton-creation})
\begin{equation}
	\sum_{\nu} \ketbra{\mathrm{x}_\nu}{\mathrm{x}_\nu} + \sum_{\vt{q}} \ketbra{C_{\vt{q}}}{C_{\vt{q}}} = \vt{1}, \label{eq:completeness-single-excitation}
\end{equation}
we finally obtain the second quantized Hamiltonian for the exciton states:
\begin{equation}
	\mathcal{H}_0 = \hbar \sum_{\nu} e_\nu B^\dag_{\nu} B_{\nu} + \hbar \sum_{\vt{q}} \omega_{\vt{q}} C^\dag_{\vt{q}} C_{\vt{q}}, \label{eq:second-quantized}
\end{equation}
where the energy is given by $\hbar e_\nu = E_g + E_\nu$ for bound state excitons ($E_\nu < 0$), and $\hbar \omega_{\vt{q}} = E_g + \hbar^2 q^2 / (2 m_r)$ for the unbound exciton.

\subsection{Interaction Hamiltonian}

Let us now consider a monochromatic external field $\vt{E} (t) = \hat{\vt{\varepsilon}} \mathcal{E}(\vt{\kappa}) \text{e}^{-i \omega_\kappa t}$ where $\hat{\vt{\varepsilon}}$ is the unit vector of polarization and each photon has a momentum $\hbar \vt{\kappa}$. The nature of the interaction between the external field and the monolayer MoS$_2$ is the dipole interaction represented by an interaction Hamiltonian \cite{haug2009quantum}
\begin{equation}
	\mathcal{H}_I = - \sum_{\vt{q}} \left[ d_{cv} (\vt{q}) \alpha^\dag_{\vt{q}} \beta^\dag_{- \vt{q}} \mathcal{E} (\vt{\kappa}) \text{e}^{- i \omega_\kappa t} + \text{h.c.}    \right], \label{eq:2}
\end{equation}
where h.c. stands for the Hermitian conjugate. Momentum is conserved in this interaction as $\hbar \vt{\kappa} = \hbar \vt{Q}  \approx \hbar \vt{q} + (- \hbar \vt{q})$ since the crystal momentum $\vt{q}$ is much larger than $\vt{\kappa}$. Hence, in principle the incoming photon can excite an electron-hole pair with any $\vt{q}$. Here, the dipole moment for the interband transition is given by
\begin{equation}
	d_{cv} (\vt{q}) = \bra{c_{\vt{q}}} e \vt{r} \cdot \hat{\vt{\varepsilon}} \ket{v_{\vt{q}}},
\end{equation}
where $\ket{c_{\vt{q}}}$ and $\ket{v_{\vt{q}}}$ are the conduction and the valence band state with a crystal momentum $\pm \hbar \vt{q}$, respectively. Particularly for the $\sigma_+$ circularly polarized light with $\hat{\vt{\varepsilon}} = \hat{\vt{\varepsilon}}^+ = (1/\sqrt{2}) (\hat{\vt{x}} + i \hat{\vt{y}})$, the dipole moment is $d_{cv}^+ (\vt{q}) = (e/\sqrt{2}) \bra{c_{\vt{q}}} (r \cos \phi + i r \sin \phi) \ket{v_{\vt{q}}} = (e/\sqrt{2}) \bra{c_{\vt{q}}} r \text{e}^{i \phi} \ket{v_{\vt{q}}}$ where we adopted the polar position coordinate $\vt{r} = (r, \phi)$.

We wish to use the second quantized bound and unbound exciton operators in the interaction Hamiltonian since our basis kets are the exciton states. We then calculate the following:
\begin{align}
	\alpha^\dagger_{\vt{q}} \beta^\dagger_{- \vt{q}} &= \sum_{\vt{A}} \delta_{\vt{q}, \vt{A}} \alpha^\dagger_{\vt{A}} \beta^\dagger_{- \vt{A}} \nonumber \\
	&= \sum_{\vt{A}, \nu} \psi^*_\nu (\vt{q}) \psi_\nu \left(\vt{A}\right) \alpha^\dagger_{\vt{A}} \beta^\dagger_{- \vt{A}} \nonumber \\
	&~~~+ \sum_{\vt{q}'', \vt{A}} \braket{\vt{q}}{C_{\vt{q}''}} \braket{C_{\vt{q}''}}{\vt{A}} \alpha^\dagger_{\vt{A}} \beta^\dagger_{-\vt{A}} \nonumber \\
	&= \sum_{\nu} \psi^*_\nu (\vt{q}) B^\dagger_{\nu}  + \sum_{\vt{q}'', \vt{A}} \braket{\vt{q}}{C_{\vt{q}''}} \braket{C_{\vt{q}''}}{\vt{A}} \alpha^\dagger_{\vt{A}} \beta^\dagger_{-\vt{A}}, \label{eq:3}
\end{align}
where the last equation follows from equation \eqref{eq:bound-creation}, and the second equation follows from the following, using the completeness in equation \eqref{eq:completeness-single-excitation}:
\begin{align}
	\delta_{\vt{q}, \vt{q}'} &= \langle \vt{q} | \vt{q}' \rangle = \sum_{\nu} \langle \vt{q} | \textrm{x}_\nu \rangle \langle \textrm{x}_\nu | \vt{q}' \rangle + \sum_{\vt{q}''} \braket{\vt{q}}{C_{\vt{q}''}} \braket{C_{\vt{q}''}}{\vt{q}'} \nonumber \\
	&= \sum_{\nu} \psi_\nu^* (\vt{q}) \psi_\nu (\vt{q}') + \sum_{\vt{q}''} \braket{\vt{q}}{C_{\vt{q}''}} \braket{C_{\vt{q}''}}{\vt{q}'}.
\end{align}
Following the treatment of Haug \emph{et al.}\cite{haug2009quantum}, approximating the band states as the free states allows $\braket{\vt{q}}{C_{\vt{q}''}} \approx \delta_{\vt{q}, \vt{q}''}$. We then obtain
\begin{align}
	&\mathcal{H}_I = \nonumber \\
	&- \left[ \sum_{\nu} g_\nu B^\dag_{\nu} \mathcal{E}(\vt{\kappa}) \textbf{e}^{-i \omega_\kappa t} + \sum_{\vt{q}} d_{cv} (\vt{q}) C_{\vt{q}}^\dag \mathcal{E}(\vt{\kappa}) \textbf{e}^{-i \omega_\kappa t} \right] \nonumber \\
	&+ \textrm{h.c.}, \label{eq:interaction-hamiltonian}
\end{align}
where we defined
\begin{equation}
	g_\nu = \sum_{\vt{q}} d_{cv} (\vt{q}) \psi^*_\nu (\vt{q}) = e \bra{\mathrm{x}_\nu} \hat{\vt{\varepsilon}} \cdot \vt{r} \ket{0}. \label{eq:gnu}
\end{equation}

\subsection{Optical selection rules and dipole moments} \label{sec:selection-rule}

\subsubsection{Interband transition}

The well-known valley selection rule for the first-order interband transition is explained as follows: The $\sigma_+$ polarized light couples only to $\vt{K}$ valley whereas $\sigma_-$ polarized light couples only to $-\vt{K}$ valley. This chiral selection rule can be deduced from the symmetry considerations. The monolayer MoS$_2$ at $\pm \vt{K}$ points belong to $C_{3h}$ point symmetry group. Then, the Bloch wave functions of the valence bands transform like the states with angular momentum $\mp \hbar$ for $\pm \vt{K}$ valley, respectively, whereas the conduction bands transform like the states with zero angular momentum for both valleys\cite{xiao2012coupled, cao2012valley, mak2012control, sallen2012robust, sallen2014erratum}. This explains the chiral optical selection rule in the angular momentum conservation scheme. We note, however, that the $\sigma_+$ photon couples to the excitation of either a spin up or down electron at $+\vt{K}$ valley, depending on the optical frequency. Therefore, a broadband $\sigma_+$ photon will see the absorption peak at both transitions separated by the spin orbit coupling energy.

It is, however, important to recognize that the symmetry argument is only for $\pm \vt{K}$ points (valley bottoms). For other $\vt{k} \neq \pm \vt{K}$, the valley can interact with the opposite circularly polarized photon as we will confirm below. This is a critical difference between the interband transitions and the transitions involving the exciton states as the exciton is a collective superposition from various $\vt{q}$ as shown in equation \eqref{eq:creation-bound-exciton}.

In order to calculate the dipole moment, one can use the velocity operator $\vt{\mathrm{v}} = (1/\hbar) \vt{\nabla}_{\vt{k}} H$, which leads to $d_{cv} (\vt{q}) = -(i e / \hbar \omega_{\vt{q}}) \bra{u_{\vt{q},c}} \hat{\vt{\varepsilon}} \cdot \vt{\nabla}_{\vt{q}} H \ket{u_{\vt{q},v}}$, where $H$ is the band Hamiltonian for the Bloch functions. An equivalent expression is the well-known Blount formula \cite{blount1962formalisms}:
\begin{align}
	&\bra{ \psi_{\vt{k}, \lambda}} \vt{r} \ket{\psi_{\vt{k}', \lambda'}} = \nonumber \\
	&-i \vt{\nabla}_{\vt{k}'} \braket{\psi_{\vt{k},\lambda}}{\psi_{\vt{k}', \lambda'}} +  i \delta_{\vt{k}, \vt{k}'} \bra{u_{\vt{k}, \lambda}} \vt{\nabla}_{\vt{k}} \ket{u_{\vt{k}, \lambda'}}. \label{eq:blount}
\end{align}
Here, $\ket{\psi_{\vt{k}, \lambda}} = \textrm{e}^{i \vt{k} \cdot \vt{r}} \ket{u_{\vt{k}, \lambda}}$ is the wave function of a Bloch state at band $\lambda$ with a periodic Bloch function $u_{\vt{k}, \lambda} (\vt{r}) = \braket{\vt{r}}{u_{\vt{k}, \lambda}}$. For interband transition, the first term vanishes. Now we can calculate the dipole moment $d_{cv} (\vt{q})$ by diagonalizing the band Hamiltonian and finding the eigenvectors.

If we use the analytical solution for the band states (derived in equation \eqref{eq:uqv} of Appendix \ref{sec:band-structure}), we find the dipole moment for the $\sigma_+$ light to be $d_{cv}^+ (\vt{q}) = \bra{c_{\vt{q}}} e \vt{r} \cdot \hat{\vt{\varepsilon}}^+ \ket{v_{\vt{q}}} = - i (\sqrt{2} e \hbar v/\Delta) (1 - 4 \hbar^2 v^2 q^2 / \Delta^2)$ for $\tau = +1$ ($+\vt{K}$ valley), but $d_{cv}^+ (\vt{q}) = - i \sqrt{2} e \hbar^3 v^3 q^2 \mathrm{e}^{i 2 \phi_q}/\Delta^3$ for $\tau = -1 $ ($-\vt{K}$ valley). Here, $\Delta = E_g \pm \tau E_\textrm{soc}/2$ for up or down spin subspace, respectively, with the energy band gap $E_g$ and the spin-orbit coupling energy $E_\textrm{soc}$. Also, $\vt{q} = (q_x, q_y) = \vt{k} - \tau \vt{K}$, and $q \mathrm{e}^{i \phi_q} = q_x + i q_y$. On the contrary, the $\sigma_-$ light produces $d_{cv}^- (\vt{q}) = \bra{c_{\vt{q}}} e \vt{r} \cdot \hat{\vt{\varepsilon}}^- \ket{v_{\vt{q}}} =  i \sqrt{2} e \hbar^3 v^3 q^2 \mathrm{e}^{-i 2 \phi_q}/\Delta^3$ for $\tau = +1$, but $d_{cv}^- (\vt{q}) = i (\sqrt{2} e \hbar v/\Delta) (1 - 4 \hbar^2 v^2 q^2 / \Delta^2)$ for $\tau = -1$. Here, $\hat{\vt{\varepsilon}}^- = (1/\sqrt{2}) (\hat{x} - i \hat{y}) = \hat{\vt{\varepsilon}}^{+*}$. For the $\pm \vt{K}$ points where $q = 0$, this dipole moment explains the valley selection rule.

\begin{figure}[!tb]
	\centering
	\subfloat[$\sigma_+$ polarization]{\includegraphics[width=0.4\textwidth]{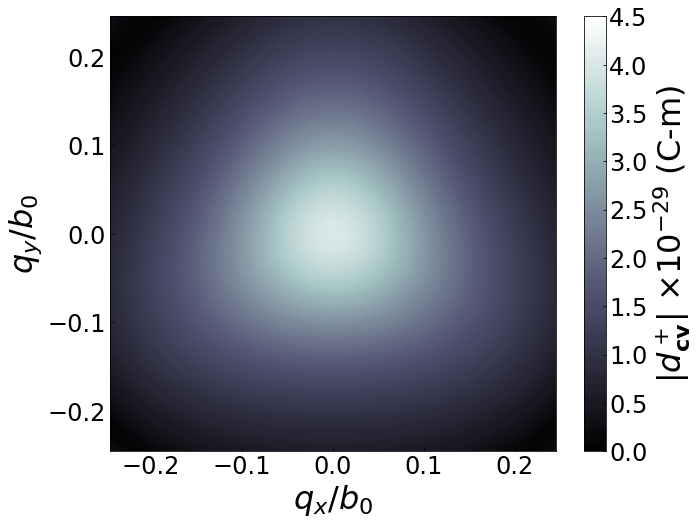}}\\
	\subfloat[$\sigma_-$ polarization]{\includegraphics[width=0.4\textwidth]{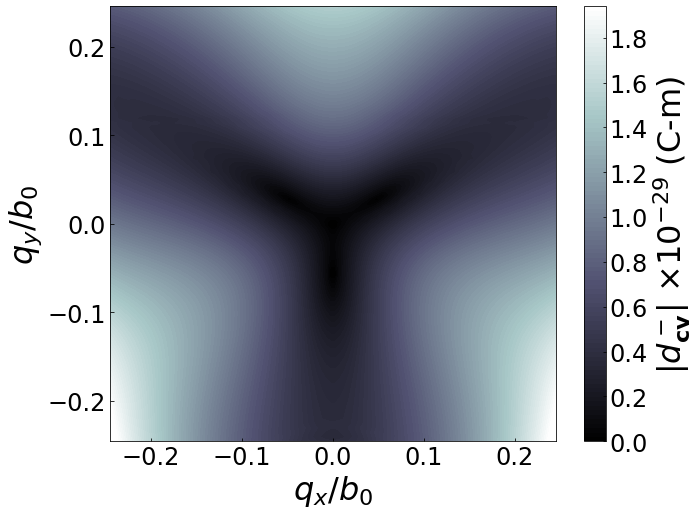}}
	\caption{\label{fig:dcv} Numerically evaluated $d_{cv}^+ (\vt{q})$ for $\sigma_+$ polarization (up) and $d_{cv}^- (\vt{q})$ for $\sigma_-$ polarization (down) around $+\vt{K}$ point, based on the higher-order corrected Dirac Hamiltonian.}
\end{figure}

For a more accurate result, we numerically obtain the eigenvectors from the higher-order corrected band Hamiltonian (see Appendix \ref{sec:band-structure}). Fig. \ref{fig:dcv} shows the numerically evaluated $d_{cv\pm} (\vt{q})$ around $+\vt{K}$ point. Qualitatively the numerical solution $d_{cv+} (\vt{q})$ has a negligibly small real value, matching the analytical result. The maximum is also similar to the analytical solution. On the other hand, the threefold rotational symmetry is clearly shown. We note that $d_{cv-} (\vt{q} = \vt{0}) = 0$ while $d_{cv-} (\vt{q} \neq \vt{0}) \neq 0$ at $+\vt{K}$ point. This confirms that the chiral valley selection rule is only for $\pm \vt{K}$ points. Indeed, the symmetry argument breaks on the points away from $\pm \vt{K}$.

\subsubsection{Transition between Fermi sea ground state $\ket{0}$ and bound exciton states $\ket{\mathrm{x}_\nu}$}

\begin{table}[!tb]
	\centering
	\caption{Numerically calculated $g_{\nu}^\pm$ for $+\vt{K}$ valley. For $-\vt{K}$ valley, $g_{\nu}^+$ and $g_{\nu}^-$ are switched. All values are relative to $g_{(0,0)}^+/\sqrt{A} = i 2.28 \times 10^{-20}$ C. The symbol ``$\approx 0$" implies that the parameter is negligibly small ($\lesssim 10^{-3}$).} \label{tab:gnu}
	\begin{ruledtabular}
		\begin{tabular}{ccc}
			$\nu = (n,m)$ & $g_{\nu}^+/g_{(0,0)}^+$ & $g_{\nu}^-/g_{(0,0)}^+$ \\[1mm]
			\colrule \\[-3mm]
			$(0,0)$ & $\mathbf{1}$ & $\approx 0$ \\
			\colrule \\[-3mm]
			$(1,1)$ & $\approx 0$ & $\mathbf{0.022}$ \\
			$(1,0)$ & $\mathbf{0.148}$ & $\approx 0$ \\
			$(1,-1)$ & $\approx 0$ & $\approx 0$ \\
			\colrule \\[-3mm]
			$(2,2)$ & $ \approx 0$ & $\approx 0$ \\
			$(2,1)$ & $\approx 0$ & $ \mathbf{0.010}$ \\
			$(2,0)$ & $\mathbf{0.068}$ & $\approx 0$ \\
			$(2,-1)$ & $\approx 0$ & $\approx 0$ \\
			$(2,-2)$ & $\approx 0$ & $\approx 0$ \\
			\colrule \\[-3mm]
			$(3,3)$ & $\approx 0$ & $\approx 0$ \\
			$(3,2)$ & $\approx 0$ & $\approx 0$ \\
			$(3,1)$ & $\approx 0$ & $\approx 0$ \\
			$(3,0)$ & $\mathbf{0.042}$ & $\approx 0$ \\
			$(3,-1)$ & $\approx 0$ & $ \approx 0$ \\
			$(3,-2)$ & $\approx 0$ & $ \approx 0$ \\
			$(3,-3)$ & $\approx 0$ & $\approx 0$ \\
			\colrule \\[-3mm]
			$(4,0)$ & $\approx 0$ & $\approx 0$ \\
			$(5,0)$ & $\mathbf{0.053}$ & $\approx 0$\\
			$(6,0)$ & $\mathbf{0.086}$ & $\approx 0$\\
			$(7,0)$ & $\mathbf{0.055}$ &$\approx 0$\\
			$(8,0)$ & $\approx 0$ & $\approx 0$\\
			$(9,0)$ & $\mathbf{-0.034}$ & $\approx 0$\\
			$(10,0)$ & $\mathbf{-0.046}$ & $\approx 0$\\
			$(11,0)$ & $\mathbf{-0.040}$ & $\approx 0$\\
			$(12,0)$ & $\mathbf{-0.026}$ & $\approx 0$\\
			$(13,0)$ & $\mathbf{-0.011}$ & $\approx 0$\\
			$(n(>13),m)$ & $\approx 0$ &$\approx 0$\\
		\end{tabular}
	\end{ruledtabular}
\end{table}

The dipole moment $g_\nu$ defined in equation \eqref{eq:gnu} is often approximated as $g_\nu \approx \sum_{\vt{q}} d_{cv} (\vt{0}) \psi^*_\nu (\vt{q}) = \sqrt{A} d_{cv} (\vt{0}) \psi_\nu^* (\vt{r} = \vt{0})$ in an understanding that $\psi^*_\nu (\vt{q})$ is significant only for $|\vt{q}| \ll 1/a_0$ (Haug \emph{et al.} \cite{haug2009quantum}). Notably for a given quantum number $\nu = (n,m)$ the wave function is $\psi_\nu (\vt{r}) \propto r^m$, and consequently, the substantial $g_\nu$ occurs for $\nu = (n,0)$. Using these, we arrive at an approximate analytical solution for the monolayer MoS$_2$:
\begin{equation}
	g_{(n,0)} = -i \sqrt{\frac{A}{(2n + 1) \pi}} \left(\frac{4 e \hbar v}{(2n + 1) a_0 \Delta}\right). \label{eq:gnu-analytic}
\end{equation}
We numerically calculated both $g_{\nu}^\pm$ for $\sigma_\pm$, respectively, based on the dipole moments shown in Fig. \ref{fig:dcv}. The result is shown in table \ref{tab:gnu}. The values $g_{\nu}^-$ are generally small compared to the substantial $g_{\nu}^+$'s. Recall that $g_{\nu}^- = \sum_{\vt{q}} \psi_\nu^* (\vt{q}) d_{cv}^- (\vt{q})$. The envelope of $\psi_\nu (\vt{q})$ decays as $q$ increases. Since $d_{cv}^- (\vt{q} = \vt{0}) = 0$, $g_{\nu}^-$ must be significantly smaller than $g_\nu^+$.

We find that only two transition dipoles $g_{(1,1)}^-$ and $g_{(2,1)}^-$ are substantial. The reason for this is as follows: we recall that the analytical solution $d_{cv}^- (\vt{q}) \propto \mathrm{e}^{-2 i \phi_q}$ at $+\vt{K}$ valley. The higher order correction, however, imposed the weak threefold rotational symmetry (see the equation \eqref{eq:threefold-rotational-sym} in Appendix \ref{sec:band-structure} and the text underneath). In perturbative treatment, the dipole moment can be expressed as \cite{simon1968second}
\begin{align}
	&d_{cv}^- (\vt{q}) = \mathrm{e}^{-2 i \phi_q} \times \nonumber \\
	& \left( \xi^{(0)} (\vt{q}) + \cos (3 \phi_q) \xi^{(1)} (\vt{q})  + \mathcal{O} (\cos^2 (3 \phi_q)) \right),
\end{align}
where the zeroth order term does not possess the threefold rotational symmetry, but the first order term has a factor $\cos (3 \phi_q) = (1/2) (\mathrm{e}^{i 3 \phi_q} - \mathrm{e}^{-i 3 \phi_q})$. The net effect is $\delta d_{cv}^- (\vt{q}) = d_{cv}^- (\vt{q}) - \mathrm{e}^{-2 i \phi_q} \xi^{(0)} (\vt{q}) \propto \mathrm{e}^{+ i \phi_q}$ while we discard the faster term $\mathrm{e}^{-i 5 \phi_q}$ that will later result in zero while integrating over $\phi_q$. The Fourier transformed wave function of exciton is $\psi_{(1,1)}^* (\vt{q}) = 288 i a_0^2 \sqrt{3 \pi} q \mathrm{e}^{-i \phi_q}/(4 + 9 a_0^2 q^2)^{5/2} \propto \mathrm{e}^{-i \phi_q}$. Hence, these two cooperate such that $\psi^*_{(1,1)} (\vt{q}) \delta d_{cv}^- (\vt{q})$ does not depend on $\phi_q$, resulting in nontrivial value after integrating over $\phi_q$. This nontrivial integral over $\phi_q$ produces a substantial value for $g_{(1,1)}^-$, and $g_{(2,1)}^-$, although the amplitude of the latter is smaller due to a faster oscillation of $\psi^*_{(2,1)}$ in the radial direction than $\psi^*_{(1,1)}$. For a large $n$, however, the envelope of $\psi_\nu (\vt{q})$ quickly oscillates in the radial direction, resulting in small values for $g_{(n,1)}^-$ for large $n$. The same reason causes decreasing $g_{(n,0)}$ as $n$ increases.

This weak opposite chiral valley response of the bound exciton states leads to some nontrivial optical nonlinearities in the monolayer MoS$_2$ as will be presented in the following sections. Unlike the usual chiral valley selection rule, the excitons respond to the opposite circularly polarized light since they are collective excitations including $\vt{k} \neq \pm \vt{K}$ (see equation \eqref{eq:gnu}). Nonetheless, we note that this opposite chiral response is rather weak as they only exists in a weak perturbative fashion.

\subsubsection{Transition between bound exciton states $\ket{\mathrm{x}_\nu}$}

The transition dipole moment between two bound exciton states follow the usual angular momentum conservation rule, which can be deduced from the spherical symmetry of excitons, since the Wannier Schr\"{o}dinger equation in equation \eqref{eq:schrodinger-exciton} is rotationally symmetric, thus the bound exciton states have well defined angular momenta such that the angular momentum of $\ket{\mathrm{x}_{(n,m)}}$ state is $\hbar m$. Then, the optical selection rule is such that the transitions $\ket{\mathrm{x}_{n,m}} \rightarrow \ket{\mathrm{x}_{n',m\pm 1}}$ are allowed and mediated by the $\sigma_\pm$ circularly polarized photons, respectively, while all others are forbidden.

Let us define the dipole moment $h^\pm_{\nu_1 \nu_2} \equiv e \bra{\mathrm{x}_{\nu_1}} \hat{\vt{\varepsilon}}^\pm \cdot \vt{r} \ket{\mathrm{x}_{\nu_2}}$ between the two bound exciton states. Then, the optical selection rule is such that
\begin{equation}
	h^\pm_{(n',m')(n,m)} \left\{  \begin{array}{ll} \neq 0, & \mathrm{if}~~ m' = m \pm 1, \\ = 0, & \mathrm{otherwise.}  \end{array}   \right. \label{eq:selection-rule-exciton-levels}
\end{equation}
Some selected dipole moment $h^+_{(n',m')(n,m)}$ are shown in the table \ref{tab:hnu1nu2-twophoton}, which we will use later for the Kerr nonlinearity calculation.

\begin{table}[!tb]
	\caption{	\label{tab:hnu1nu2-twophoton}
		Examples of the dipole moment $h^+_{(1,1) \nu_1}$ between the bound exciton states. The value is relative to $|e|a_0$.
	}
	\begin{tabular}{ c @{\qquad} c }
		\toprule
		$ \nu_1$ & $h^+_{(1,1) \nu_1} / (|e| a_0)$\\
		\colrule\\[-3mm]
		$(0,0)$ &  $0.344$\\
		$(1,0)$ & $-3.18$\\
		$(2,0)$ & $0.752$\\
		$(3,0)$ & $0.320$\\
		$(4,0)$ & $0.194$\\
		$(5,0)$ & $0.135$\\
		$(6,0)$ & $0.102$\\
		$(7,0)$ & $0.080$\\
		\botrule
	\end{tabular}
\end{table}

\subsubsection{Transition between bound exciton states $\ket{\mathbf{x}_\nu}$ and the unbound exciton states $\ket{C_{\mathbf{q}}}$}

The relevant dipole moment of this transition is defined as $f_\nu (\vt{q}) = e \bra{\mathrm{x}_\nu} \hat{\vt{\varepsilon}} \cdot \vt{r} \ket{C_{\vt{q}}}$. This dipole moment turns out to be negligibly small, which is rigorously shown in Appendix \ref{sec:fnu}.

\subsubsection{Summary of optical selection rules}

\begin{figure}[!tb]
	\centering
	\includegraphics[width=0.5\textwidth]{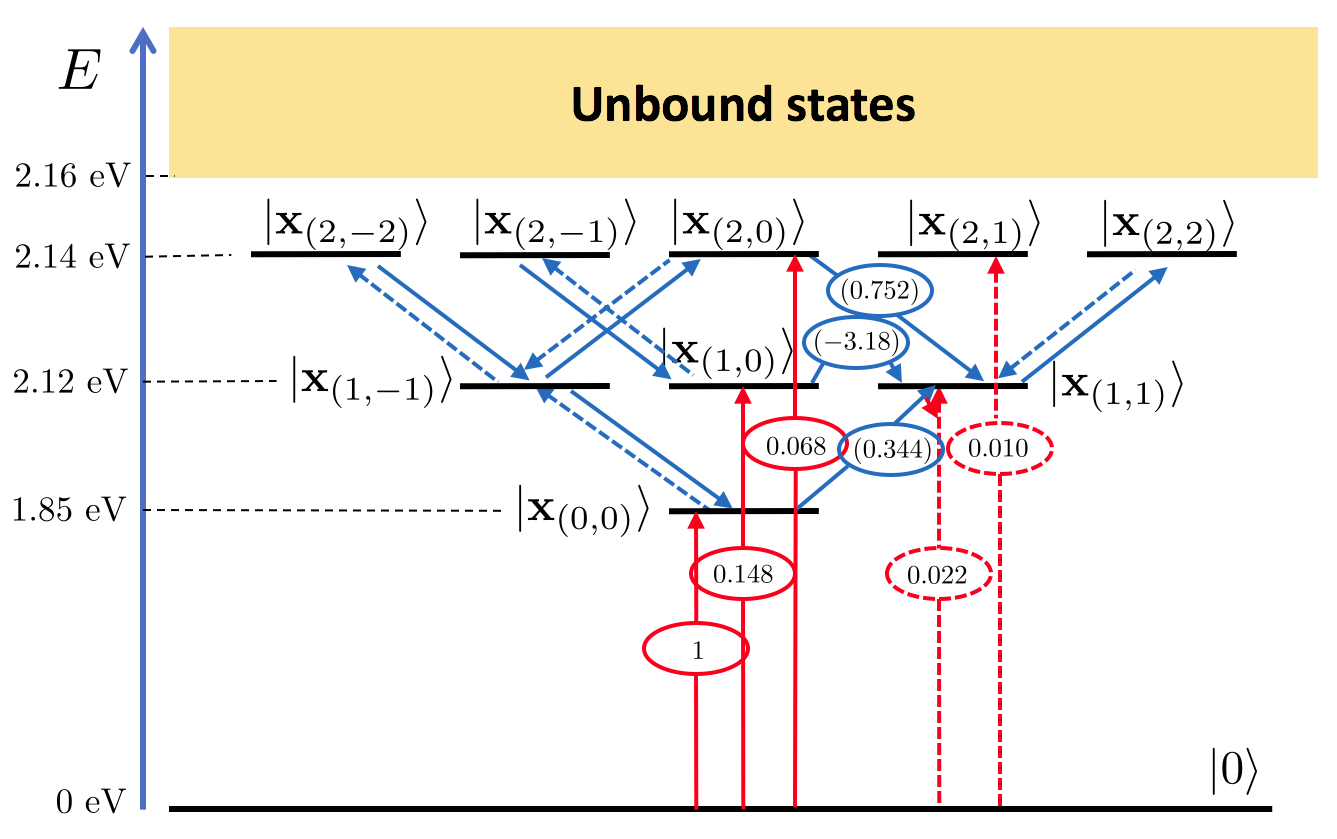}
	\caption{Summary of allowed optical transitions in the $+\vt{K}$ valley. The red and blue transitions correspond to $\ket{0} \rightarrow \ket{\mathrm{x}_\nu}$ and  $\ket{\mathrm{x}_{\nu_1}} \rightarrow \ket{\mathrm{x}_{\nu_2}}$, respectively. The solid and dotted lines are mediated by $\sigma_+$ and $\sigma_-$ photons, respectively. The values in the circles represent the dipole moments ($g_\nu$ are in unit of $g_{(0,0)}^+$ while $h_{\nu_1 \nu_2}$ are in unit of $|e|a_0$.) In the $-\vt{K}$ valley, the roles of $\sigma_\pm$ photons are switched.}  \label{fig:selection-rule}
\end{figure}

The optical selection rule, quantified through the appropriate dipole moments, plays the central role in the optical susceptibility calculations. While the transitions between the bound excitons $(h_{\nu_1 \nu_2})$ follow usual angular momentum conservation rule, the transitions from the ground state $\ket{0}$ to any bound exciton states $(g_\nu)$ are not trivial since the corresponding dipole moments $g_\nu$ weakly inherit the threefold rotational symmetry from the band states. Fig. \ref{fig:selection-rule} summarizes the optical selection rule. It also reveals the values of the dipole moments for some transitions we will use later to calculate the optical susceptibilities.

\subsection{Induced current density and susceptibility}

For clarity, we present the following procedure to calculate the susceptibilities, which is indeed well known in the literature \cite{boyd2003nonlinear}. We will extensively use the calculated dipole moments to evaluate the susceptibilities in the next section.

When an external field is present, an induced current is produced as a result of the dipole interaction. It is obtained as
\begin{equation}
	\vt{J} =  e N_e \langle \vt{\mathrm{v}} \rangle = e N_e \text{tr}[\vt{\mathrm{v}} \rho], \label{eq:J}
\end{equation}
where $N_e$ is the free carrier density, $\vt{\mathrm{v}}$ is the velocity operator, and $\rho$ is the quantum mechanical density operator. The density operator follows the von Neumann equation $i \hbar \dot{\rho} = [\mathcal{H}_{0} + \mathcal{H}_{I}, \rho]$. The solution is recursively obtained:
\begin{align}
	&\rho (t) = - \frac{i}{\hbar} \int_{-\infty}^t \td t' [\mathcal{H}_0 + \mathcal{H}_I, \rho(t')] \nonumber \\
	&= - \frac{i}{\hbar} \int_{-\infty}^t \td t' \left[ \mathcal{H}_0 + \mathcal{H}_I, \left( - \frac{i}{\hbar} \int_{-\infty}^{t'} \td t'' [\mathcal{H}_0 + \mathcal{H}_I, \rho(t'')] \right)  \right] \nonumber \\
	&\hspace{5cm} \vdots
\end{align}
Since $\mathcal{H}_I \propto \mathcal{E}(\vt{\kappa})$, one can expand the perturbative order of $\rho$ such that $\rho(t) = \sum_{n=0}^{\infty} \rho^{(n)} (t)$ where $\rho^{(n)} (t)$ involves only $\mathcal{O} (\mathcal{E}^n(q))$ terms. We then use $\vt{J} = \sigma \vt{E} = \left( \sum_{n=0}^{\infty} \sigma^{(n)} \right) \vt{E}$ to resolve $\sigma^{(n)}$ order by order.  Combining the relations $\vt{J} = \partial \vt{P}/ \partial t $ and $\vt{P} = \epsilon_0 \chi \vt{E}$, one obtains
\begin{equation}
\frac{\partial}{\partial t}\left( \epsilon_0 (\chi^{(1)} + \chi^{(2)} + \cdots) \vt{E} (t) \right) = (\sigma^{(1)} + \sigma^{(2)} + \cdots) \vt{E} (t). \label{eq:chi-sigma}
\end{equation}
Equating term by term, the relation between the susceptibility and the conductivity for each order is obtained, which finally resolves the optical susceptibilities for various orders.

\subsection{Perturbative solution}

The advantage of using the second quantized exciton Hamiltonian in equation \eqref{eq:second-quantized} is that the exciton states already diagonalize the unperturbed Hamiltonian $\mathcal{H}_0$. Then, solving the \Schrodinger equation perturbatively becomes straightforward. To obtain the physical quantities such as the induced current, however, one must represent the operators in the exciton basis. It is our task to calculate the velocity operator $\vt{\mathrm{v}}$ in this exciton basis. For example, in the linear response theory where the incoming light photon energy is close to the energy of a bound exciton state $\ket{\textbf{x}_\nu}$, our Hilbert space is essentially two dimensional, with the basis $\{ \ket{\text{x}_\nu}, \ket{0} \}$. Consequently, the velocity operator and the density operator are now $2 \times 2$ matrices:
\begin{equation}
	\vt{\mathrm{v}} = \left( \begin{array}{cc} \vt{\mathrm{v}}_{\text{xx}} & \vt{\mathrm{v}}_{\text{x0}} \\ \vt{\mathrm{v}}_{\text{0x}} & \vt{\mathrm{v}}_{\text{00}} \end{array}\right), \quad \rho = \left( \begin{array}{cc} \rho_{\text{xx}} & \rho_{\text{x0}} \\ \rho_{\text{0x}} & \rho_{\text{00}} \end{array}\right),
\end{equation}
where each element is such that, for example, $\vt{\mathrm{v}}_{\text{x0}} = \bra{\text{x}_\nu} \vt{\mathrm{v}} \ket{0}$. To obtain the matrix elements of the velocity operator, we move to the Heisenberg picture and connect to the dipole moment as follows:
\begin{align}
	\vt{\mathrm{v}}_{\text{0x}} &= \bra{0} \dot{\vt{r}} \ket{\text{x}_\nu} = - \frac{i}{\hbar} \bra{0} [\vt{r}, \mathcal{H}_0 + \mathcal{H}_I] \ket{\text{x}_\nu} \nonumber \\
	&= - \frac{i}{\hbar} \bra{0} [\vt{r}, \mathcal{H}_0] \ket{\text{x}_\nu} = -i e_{\nu} \bra{0} \vt{r} \ket{\text{x}_\nu}. \label{eq:vfe}
\end{align}
Here, we used the fact that $[\vt{r}, \mathcal{H}_I] = 0$ since $\mathcal{H}_I \propto \vt{r}$ as it involves the dipole moment element. It is also noteworthy that the diagonal terms of the velocity operator $\vt{\mathbf{v}}$ are all zero according to the above derivation since the same energies of the same state cancel each other. We thus need only the off-diagonal terms of the density matrix to calculate the induced current:
\begin{equation}
	\vt{J} =  e N_e (\vt{\mathrm{v}}_{\text{x0}} \rho_{\text{0x}} + \vt{\mathrm{v}}_{\text{0x}} \rho_{\text{x0}} ). \label{eq:Jnew}
\end{equation}

Next, since the normalization of the polarization vectors is $\hat{\vt{\varepsilon}}^- \cdot \hat{\vt{\varepsilon}}^+ = 1,$ the velocity matrix component in $\hat{\vt{\varepsilon}}^+$ is $\vt{\mathrm{v}}_{0\mathrm{x}} = - i e_\nu \bra{0} \hat{\vt{\varepsilon}}^- \cdot \vt{r}  \ket{\mathrm{x}_\nu} \hat{\vt{\varepsilon}}^+$. We calculate
\begin{align}
	&\bra{0} \hat{\vt{\varepsilon}}^- \cdot \vt{r} \ket{\text{x}_\nu} = \sum_{\vt{q}} \psi_{\nu} (\vt{q})  \bra{0} \hat{\vt{\varepsilon}}^- \cdot \vt{r} \alpha^\dagger_{\vt{q}} \beta^\dagger_{- \vt{q}} \ket{0} \nonumber \\
	&= \sum_{\vt{q}} \psi_\nu (\vt{q}) \bra{v(\vt{q})} \hat{\vt{\varepsilon}}^- \cdot \vt{r} \ket{c(\vt{q})} = \frac{g_\nu^{+*}}{e}.
\end{align}
This leads to $\vt{\mathrm{v}}_{\text{0x}} =  \hat{\vt{\varepsilon}}^+ (- i e_\nu g_\nu^{+*} /e) $.

All we have left is to solve the \Schrodinger equation for $\rho$. We first note that $\bra{\text{x}_\nu}[\mathcal{H}_0, \rho] \ket{0} = \hbar e_{\nu} \rho_{\text{x0}}$. We then establish a differential equation for $\rho_{\text{x0}}$ in the \Schrodinger picture:
\begin{equation}
	\dot{\rho_{\text{x0}}} (t) = - i e_\nu \rho_{\text{x0}}(t) - \frac{i}{\hbar} \bra{\text{x}_\nu}[\mathcal{H}_I, \rho(t)] \ket{0}.
\end{equation}
From this, we carry out bookkeeping for each order on the differential equations for $n = 0,1,2,\cdots$:
\begin{align}
	\dot{\rho}^{(0)}_\text{x0} (t) &= - i e_\nu \rho^{(0)}_\text{x0} (t), \nonumber \\
	\dot{\rho}^{(n)}_\text{x0} (t) &= - i e_\nu \rho^{(n)}_{\text{x0}} (t) - \frac{i}{\hbar} \bra{\text{x}_\nu}[\mathcal{H}_I, \rho^{(n-1)}] \ket{0}. \label{eq:perturbation-DE}
\end{align}
Other matrix elements for $\rho^{(n)}$ can be obtained in a similar manner.

\section{Linear and nonlinear optical susceptibilities}

In this section, we calculate the optical susceptibilities of the excitonic states from monolayer MoS$_2$ . We will first resolve the linear susceptibility and the resulting linear absorption and refractive index. Then, we proceed to the higher order nonlinear susceptibilities.

\subsection{Linear susceptibility} \label{sec:linear-susceptibility}

\begin{figure}[!tb]
	\centering
	\includegraphics[width=0.23\textwidth]{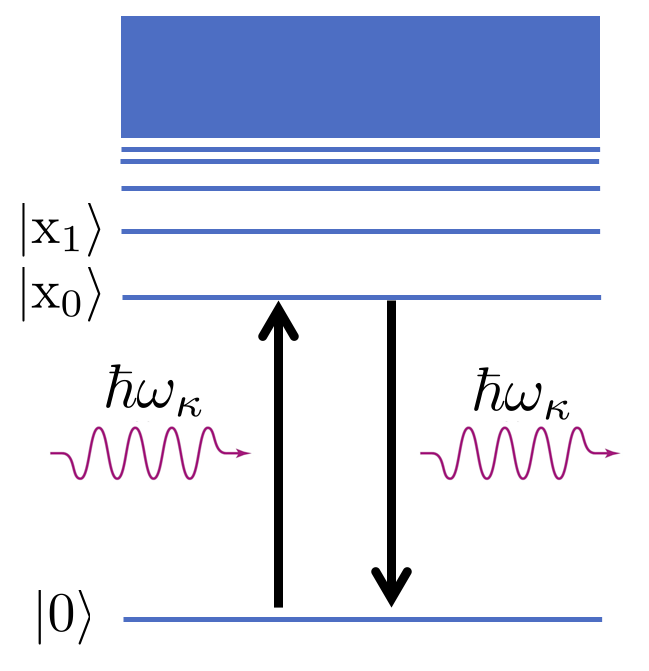}
	\caption{\label{fig:first-trans} Schematic of excitonic energy levels and the first order radiative transition of excitonic states. The continuum is the unbound electron-hole pair states.}
\end{figure}

We are interested in a case where an incoming photon has an energy closely resonant with an exciton state $\ket{\textrm{x}_\nu}$ energy level (see Fig. \ref{fig:first-trans}). The first equation in \eqref{eq:perturbation-DE} describes the dynamics of $\rho_{\text{x0}}^{(0)}$ in the absence of any external perturbation. It is a free rotation. We then need to solve $\rho^{(1)}_{\textrm{x0}}$ to resolve $\chi^{(1)}$. For this, we first calculate for the case of $\sigma_+$ photon at $+\vt{K}$ valley:
\begin{align}
	&\bra{\text{x}_\nu}[\mathcal{H}_I, \rho^{(0)}] \ket{0} \nonumber \\
	&= - \bra{\text{x}_\nu } \left( \sum_{\nu'} g^+_{\nu'} \left( B^\dagger_{\nu'} \rho^{(0)} - \rho^{(0)} B^\dagger_{\nu'} \right) \right)  \ket{0} \mathcal{E}(\vt{\kappa}) \text{e}^{-i \omega_\kappa t}  \nonumber \\
	&~~~~ \quad - \bra{\text{x}_\nu } \left( \sum_{\nu'} g^{+*}_{\nu'} \left( B_{\nu'} \rho^{(0)} - \rho^{(0)} B_{\nu'} \right) \right)  \ket{0} \mathcal{E}^*(\vt{\kappa}) \text{e}^{i \omega_\kappa t} \nonumber \\
	&= -g^+_\nu (\rho_{\text{00}}^{(0)} - \rho_{\text{xx}}^{(0)}) \mathcal{E}(\vt{\kappa}) \text{e}^{- i \omega_\kappa t} \nonumber \\
	&= - g^+_\nu \mathcal{E}(\vt{\kappa}) \text{e}^{-i \omega_\kappa t} , \label{eq:comm101}
\end{align}
where we used the fact that $B_{\nu'} \ket{0} = \ketbra{\mathrm{x}_{\nu'}}{0}$ with $\rho_{\text{00}}^{(0)} = 1$ and $\rho_{\text{xx}}^{(0)} = 0$ since the state without the external field at zero temperature is the Fermi sea. From this, the first order differential equation is now
\begin{equation}
	\dot{\rho}_\text{x0}^{(1)} (t') = - i e_\nu \rho^{(1)}_\text{x0} (t') + \frac{i}{\hbar} g^+_\nu \mathcal{E} (\vt{\kappa}) \text{e}^{-i \omega_\kappa t'},
\end{equation}
Integrating over $-\infty < t' < t$ yields the following first order solution:
\begin{equation}
	\rho^{(1)}_\text{x0} (t) = \frac{g^+_\nu}{\hbar} \frac{1}{(e_\nu - \omega_\kappa) - i \epsilon} \mathcal{E}(\vt{\kappa}) \text{e}^{- i \omega_\kappa t}, \label{eq:first-order}
\end{equation}
where $\epsilon$ is a positive infinitesimal parameter regulating the integral at $t' \rightarrow -\infty$. 

From $\rho^{(1)}_{\mathrm{x0}} (t) = \rho^{(1)}_{\mathrm{x0}} (\omega_\kappa) \mathrm{e}^{- i \omega_\kappa t}$, we easily obtain
\begin{equation}
	\rho^{(1)}_\mathrm{0x} (\omega_\kappa) = \rho^{(1)*}_\mathrm{x0} (-\omega_\kappa) = \frac{g^+_\nu}{\hbar} \frac{1}{ (e_\nu + \omega_\kappa) + i \epsilon} \mathcal{E} (\omega_\kappa),
\end{equation}
where we used $\mathcal{E}^* (- \omega_\kappa) = \mathcal{E} (\omega_\kappa)$. This is a nonresonant term, which must be much less than the resonant term $\rho^{(1)}_{\mathrm{x0}}$. Then, using equations \eqref{eq:Jnew} and \eqref{eq:vfe}, we obtain
\begin{equation}
	\vt{J}^{(1)} = e N_e \frac{-i e_\nu g^{+*}_\nu}{e} \frac{g^+_\nu}{\hbar} \sum_{p_1 = \pm 1}\frac{1}{e_\nu +p_1( \omega_\kappa + i \epsilon)} \hat{\vt{\varepsilon}}\mathcal{E}(\vt{\kappa}) \text{e}^{- i \omega_\kappa t}.
\end{equation}
From this, we obtain the linear conductivity $\sigma^{(1)}$, and then, using the relation in equation \eqref{eq:chi-sigma}, we obtain the linear susceptibility of the exciton state:
\begin{equation}
	\chi^{(1)} (\omega_\kappa) = \frac{e_\nu |g^+_\nu|^2 N_e}{\hbar \epsilon_0 \omega_\kappa} \sum_{p_1 = \pm 1} \frac{1}{ e_\nu +p_1( \omega_\kappa + i \epsilon)}.
\end{equation}

We now explain how to handle the free carrier density $N_e$ in the following. The value of $g^\pm_\nu$ is generally numerically evaluated. If, however, we adopt the previous approximation $g^\pm_\nu \approx \sqrt{A} d^\pm_{cv} (\vt{q} = \vt{0}) \psi_\nu^* (\vt{r} = 0)$, we obtain
\begin{align}
	\chi^{(1)} (\omega_\kappa) =& \left( \frac{\hbar e_\nu}{\hbar \omega_\kappa}\right) \frac{A N_e}{\epsilon_0} |d^+_{cv} (\vt{0})|^2 | \psi_\nu (\vt{r} = \vt{0})|^2 \nonumber \\
	& \times \sum_{p_1 = \pm 1}\frac{1}{\hbar e_\nu + p_1 ( \hbar \omega_\kappa + i \hbar \epsilon)}.
\end{align}
The induced current density $\vt{J} = \text{tr}[e (N_e \rho) \mathbf{v}]$ in equation \eqref{eq:J} captures the density of charge carriers and their movements. Particularly $N_e \rho$ with the quantum mechanical density $\rho$ (with unity maximum value) captures the density of the excited exciton. Since each exciton carries one excitation and thus one charge carrier, it is correct to replace $N_e \rightarrow 1/A d_\text{eff}$.  Here, $d_{\text{eff}} \approx 6.5 ~\text{\AA}$ \cite{wang2012electronics, radisavljevic2011single} is the effective thickness of the monolayer MoS$_2$. The resulting formula exactly matches the single spin electron results in Elliott's seminal paper \cite{elliott1957intensity} as well as the formula appearing in Haug, et al. \cite{haug2009quantum} (see equation 10.103) and also the formula appearing in Klingshirn \cite{klingshirn2012semiconductor} (see equation 27.52). The agreement confirms that our replacement $N_e \rightarrow 1/A d_\textrm{eff}$ is reasonable.

One must add the responses from the different exciton levels, resulting in the contribution from the bound exciton levels as
\begin{align}
	\chi^{(1)}_B (\omega_\kappa) = \sum_\nu \sum_{p_1 = \pm 1} \frac{e_\nu |\overline{g^+_\nu}|^2}{\hbar \epsilon_0 \omega_\kappa d_\mathrm{eff}} \left(   \frac{1}{ e_\nu + p_1 (\omega_\kappa + i \gamma_B/2)} \right), \label{eq:chi1_bound}
\end{align}
where we used $\overline{g^+_\nu} = g^+_\nu/\sqrt{A}$, which does not depend on the sample size since $g_\nu \propto \sqrt{A}$. We also introduced the phenomenological replacement $\epsilon \rightarrow \gamma_B/2$ where $\gamma_B$ is the decay rate of the bound exciton $\ket{\mathrm{x}_\nu}$. Wang \emph{et al.} \cite{wang2016radiative} and Selig \emph{et al.} \cite{selig2016excitonic} calculated the radiative lifetime of the exciton at a temperature of 5 K to be $\sim$ 200 fs. From the radiative decay perspective, it is expected that the line broadening will depend on $\nu$. However, other broadening mechanisms including the phonon-exciton scattering and the disorder further broadens the spectrum \cite{wang2016radiative, moody2015intrinsic} in real samples, and the difference among various $\nu$ from the radiative decay alone is washed out. To account for the phenomenological linewidth, various values were used ranging from 1 meV to 50 meV \cite{wang2017excitons,mak2010atomically,molina2013effect}. We particularly choose 10 meV that matches our own experimentally measured data at 4 K temperature \cite{rogers2017absorption}, as well as the qualitative curves of the absorption spectra found in low-temperature experimental results \cite{zhang2014absorption,he2013experimental, qiu2013optical, moody2015intrinsic}.

The contribution from the unbound excitons is easily deduced as
\begin{align}
	&\chi^{(1)}_U (\omega_\kappa) = \nonumber \\
	&\int \td^2 q \frac{\omega_{\vt{q}} |d^+_{cv} (\vt{q})|^2}{4 \pi^2 \hbar \epsilon_0 \omega_\kappa d_\mathrm{eff}} \sum_{p_1 = \pm 1 }\frac{1}{\omega_{\vt{q}} +p_1( \omega_\kappa + i \gamma_U/2)}, \label{eq:chi1_unbound}
\end{align}
where we used the replacement $\sum_{\vt{q}} \rightarrow (A/(2 \pi^2)) \int \td^2 k$. Here, $\gamma_U$ is the radiative decay rate (inverse of the radiative lifetime) of the conduction bands. Using Fermi's golden rule, we obtain $\gamma_U = \omega^3_{\vt{q}}|d^+_{cv} (\vt{q})|^2/(2 \pi \epsilon_0 \hbar c^3)$. With the monolayer MoS$_2$ parameters, we obtain the radiative lifetime of the conduction band to be approximately 4 ns.

Finally, we obtain the linear susceptibility: $\chi^{(1)A} (\omega_\kappa) = \chi^{(1)}_B (\omega_\kappa) + \chi^{(1)}_U (\omega_\kappa)$. For a single optical frequency $\omega_\kappa$, the contribution comes from all the bound and the unbound exciton states. Note, however, that $\chi^{(1)A} (\omega_\kappa)$ is the contribution only from the spin up electrons. The exciton states from the up spin in valley $+\vt{K}$ are called the \emph{A excitons}. One must add the contribution from the \emph{B excitons}, which comes from the spin down electrons. The major difference between the A and the B excitons is the energy eigenvalues. The B excitons have higher energy by $E_\mathrm{soc}$. Consequently, all the exciton level energies are offset by the similar amount. Finally, we obtain the true physical linear susceptibilities as
\begin{align}
	\chi^{(1)} (\omega_\kappa) &= \chi^{(1)A} (\omega_\kappa) + \chi^{(1)B} (\omega_\kappa)\nonumber \\
	&\approx \chi^{(1)A} (\omega_\kappa) + \chi^{(1)A} (\omega_\kappa - E_\mathrm{soc}/\hbar).
\end{align}
This response is only for the $\sigma_+$ polarized light, coming from $+\vt{K}$ valley. Indeed, $\sigma_-$ polarized light sees the linear response from $+\vt{K}$, too. The relative strength of $g^-_{(1,1)}$ and $g^-_{(2,1)}$ are, however, only 2\% and 1\% of $g^+_{(0,0)}$, respectively. Therefore, the relative strength of the response will be only $\sim 10^{-4}$, compared to the strong $g_{(0,0)}$. The same applies to the case for $\sigma_+$ polarized light and the $-\vt{K}$ valley. Hence, the linear response of $\sigma_+$ light is mostly from $+\vt{K}$ valley.  On the other hand, the contribution $\chi^{(1)}_U (\omega_\kappa)$ from $\sigma_-$ will increase as $\omega_\kappa$ increases well beyond $\Delta/\hbar$ since $d^-_{cv} (\vt{q}) \propto q^2$.

\begin{figure}[!tb]
	\centering
	\subfloat[Near exciton resonances]{\includegraphics[width=0.45\textwidth]{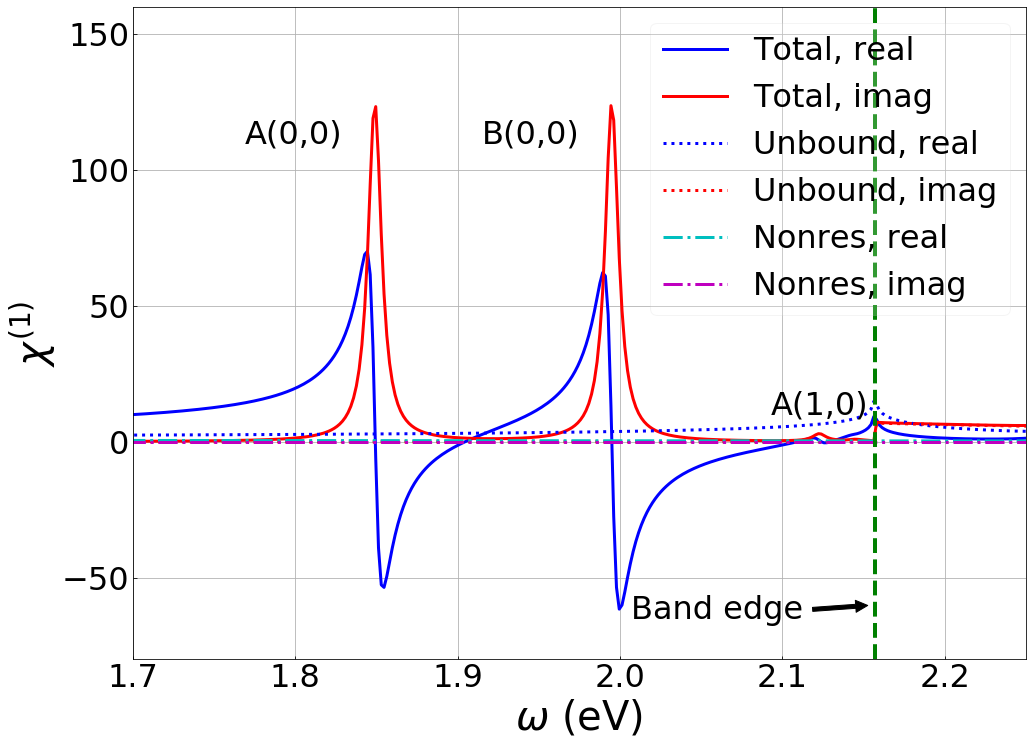}}\\
	\subfloat[Below exciton resonances]{\includegraphics[width=0.45\textwidth]{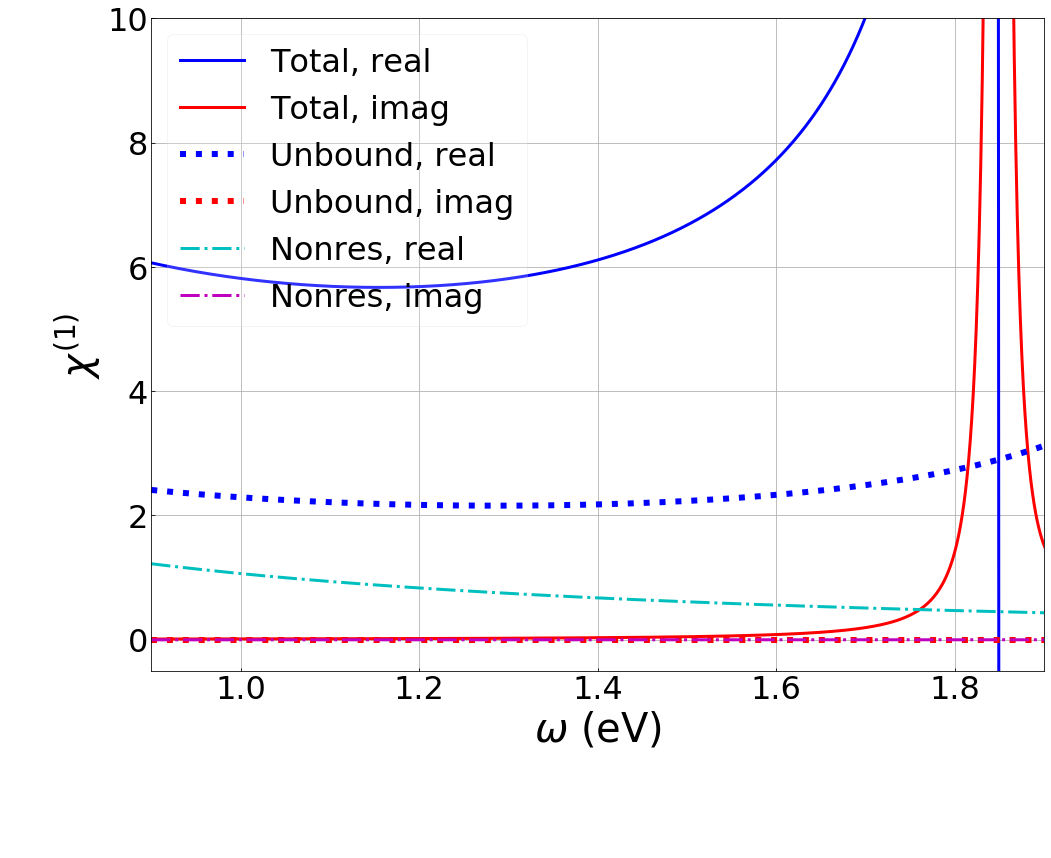}}
	\caption{\label{fig:chi1} Calculated $\chi^{(1)}$ near (up) and below (down) exciton resonances. The real and the imaginary parts of the nonresonant (dash-dot, sum of $p_1 = +1$ terms in equations \eqref{eq:chi1_bound} and \eqref{eq:chi1_unbound} ), unbound states (dot, equation \eqref{eq:chi1_unbound}), and the total sum (solid) are shown separately. The resonance labels indicate either A or B exciton with the quantum number $(n,m)$.}
\end{figure}

\begin{figure}[!tb]
	\centering
	\includegraphics[width=0.45\textwidth]{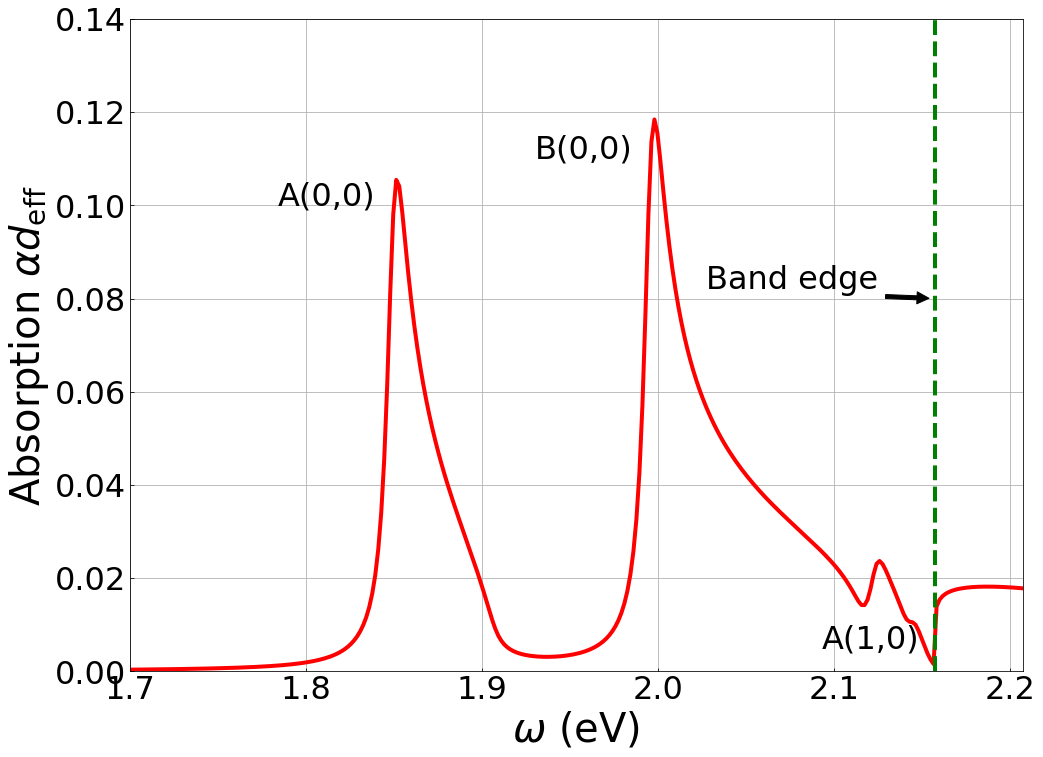}
	\caption{\label{fig:absorption} Inferred absorption from the calculated $\chi^{(1)}$. The resonance labels indicate either A or B exciton with the quantum number $(n,m)$.}
\end{figure}

We calculated the $\chi^{(1)}$ as shown in Fig. \ref{fig:chi1}. The plot shows that the contribution only from the nonresonant terms (the sum of $p_1 = +1$ terms in equations \eqref{eq:chi1_bound} and \eqref{eq:chi1_unbound},  dash-dot curves) is negligibly small. That from the unbound states (dot curves, \eqref{eq:chi1_unbound}) leaves a long tail in the real part only. Far below the exciton resonances, the contribution from the nonresonant term starts gaining. On the other hand, the absorption decays fast below the exciton resonances. The contribution from the bound excitons dominates in the spectral range below the band edge. Near the band edge, the higher order excitons contribute significantly. The band edge for the A excitons (spin up electron) occurs at 2.16 eV, while that of the B excitons at 2.31 eV. The contribution from the unbound states reaches the spectrum below the band edge. Our model does not include higher conduction levels, which diminishes the influence of this unbound state contribution in the bound state resonances.

We also calculate the linear absorption and the reflectance from the excitonic states (Fig. \ref{fig:absorption}). The complex refractive index is given as $n = \sqrt{1 + \chi^{(1)}}$. The imaginary part produces the absorption coefficient $\alpha = 2 \text{Im} [n] \omega_\kappa / c$. The linear absorption from the 2D sheet is given by $\alpha d_\text{eff} = 2 d_\text{eff} \text{Im}[\sqrt{1 + \chi^{(1)}}] \omega_\kappa / c$. The single pass absorption does not depend on $d_\mathrm{eff}$ on the bound exciton resonances due to the large value of $|\chi^{(1)}|$. Fig. \ref{fig:chi1} (b) shows the calculated absorption spectrum. The calculated absorption peaks for the lowest A and B exciton resonance match reasonably well the measured absorptions of 10\% $\sim$ 15\%, having the similar broadening \cite{he2013experimental, qiu2013optical, zhang2014absorption, moody2015intrinsic}. We note that the distortion of the curves are due to the excessive negative real value of $\chi^{(1)}$, caused by underestimated contribution from the unbound exciton states as we mentioned above. As a result, the blue side of the resonance curves are much more exaggerated than the real situation. Nevertheless, both the absorption and the reflection curves match qualitative features of the published results.

\subsection{Second order susceptibility}

\begin{figure}[!tb]
	\centering
	\includegraphics[width=0.23\textwidth]{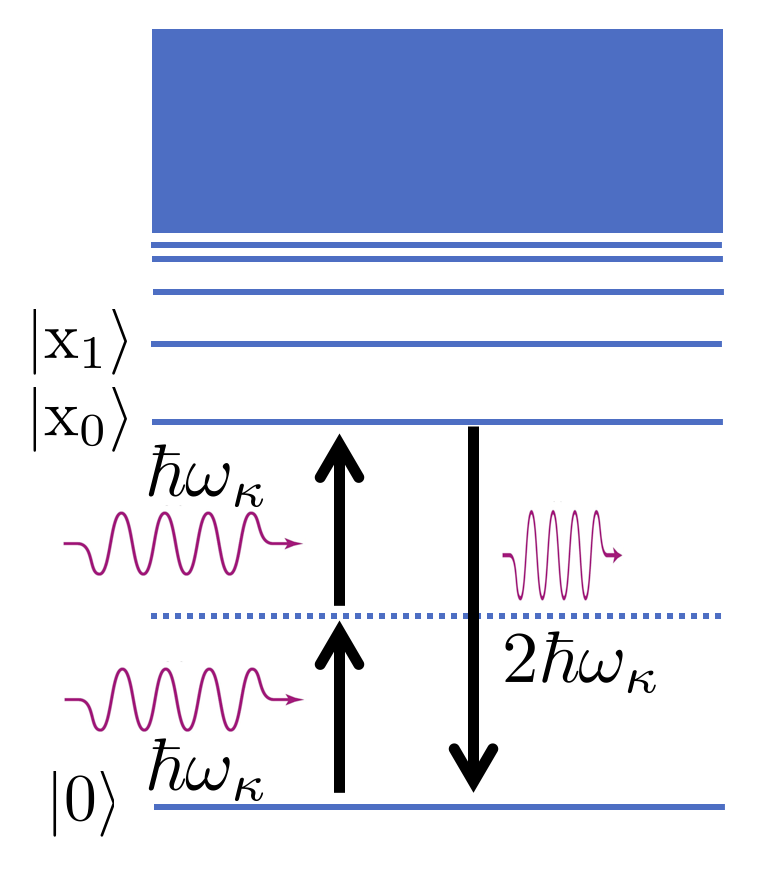}
	\caption{\label{fig:second} Schematic of the second-harmonic process where the second harmonic is near resonant with an exciton level.}
\end{figure}

Let us consider the second-harmonic generation for which the output second-harmonic frequency is nearly resonant with the exciton energy levels (see the Fig. \ref{fig:second}). Due to the energy gap, one can avoid the direct linear absorption for the fundamental pump light. If one also avoids the direct linear absorption for the second harmonic by slightly detuning from the resonance, one can accomplish a coherent and efficient second-harmonic process. The same applies to the degenerate optical parametric amplifier pumped at the exciton resonance, amplifying the signal at the half frequency.

This second-harmonic transition involves the virtual levels, which sum all possible intermediate levels linking the initial Fermi sea ground state $\ket{0}$ to the final exciton state $\ket{\mathrm{x}_\nu}$. We are particularly interested in the resonant second-harmonic frequency $2 \omega_\kappa \sim e_{0} (= e_{(0,0)})$ (the frequency of the state $\ket{\mathrm{x}_{(0,0)}})$ since it involves the largest dipole moment $g_{(0,0)}$. The virtual level can be either the bound or the unbound exciton states.

\subsubsection{Bound exciton virtual states}

Let us first consider the bound exciton virtual levels. The composite transition must obey the optical selection rule explained in section \ref{sec:selection-rule}. Let us consider the case where the highest level is $\ket{\mathrm{x}_{(0,0)}}$. For $+\vt{K}$ valley, where the second-harmonic light is in $\sigma_+$, the second order transition involving two $\sigma_+$ fundamental photons is not allowed since $h^+_{(0,0)(n,0)} =0$ due to the angular momentum conservation rule. This implies the tensor element $\chi^{(2)}_{+;++} = 0$. Instead, the transition $\ket{0} \rightarrow \ket{\mathrm{x}_{(1(2),1)}} \rightarrow \ket{\mathrm{x}_{(0,0)}}$ is allowed by absorbing two $\sigma_-$ photons because the first transition relies on the dipole moment $g^-_{1(2),1} (\neq 0)$, and the second transition relies on the dipole moment $h^-_{(0,0)(1(2),1)}$, which is nonzero. The transition $\ket{\mathrm{x}_{(0,0)}} \rightarrow \ket{0}$ emits a $\sigma_+$ photon as explained in previous section. This corresponds to the susceptibility tensor $\chi^{(2)}_{+;--}$. We note that $\chi^{(2)}_{+;-+} = 0$ since the dipole element $h^+_{(0,0)(1(2),1)} = 0$. Also, $\chi^{(2)}_{+;+-} = 0$ since $h^+_{(0,0)(n,0)} = 0$.

For $-\vt{K}$ valley, the opposite circularly polarized photons are used in the same transitions. Since the second-harmonic output from $-\vt{K}$ valley is always $\sigma_-$ photon as we explained in the previous section, we conclude that $\chi^{(2)}_{-;--} = \chi^{(2)}_{-;+-} = \chi^{(2)}_{-;-+}= 0$ and $\chi^{(2)}_{-;++} \neq 0$.

In summary, we have only two nonzero second-order susceptibility tensor elements, $\chi^{(2)}_{-;++}$  and $ \chi^{(2)}_{+;--}$. This result is consistent with the well known experimental results for the second-harmonic generation in TMDs, where the output second-harmonic polarization has the opposite chirality relative to the input circular polarization \cite{seyler2015electrical, xiao2015nonlinear}.

Let us quantify the tensor element $\chi^{(2)}_{+;--}$ from $+\vt{K}$ valley. For this, we solve the second order differential equation for the density matrix elements. First, the basis for the Hilbert space is $\{ \ket{0}, \ket{\mathrm{x}_{(s,1)}}, \ket{\mathrm{x}_{(0,0)}} \}$ where $s =$ 1 or 2. Since we now involve the exciton-exciton transition, we have an additional interaction Hamiltonian:
\begin{equation}
	\mathcal{H}'_I = - \sum_{s = 1,2} \left[ h^-_{(0,0)(s,1)} B^\dag_{0} B_{(s,1)} \mathcal{E} (\vt{\kappa}) \mathrm{e}^{- i \omega_\kappa t} + \mathrm{h.c.}  \right]. \label{eq:interact3}
\end{equation}
We need to calculate the matrix elements such as $\rho^{(2)}_{(1,2) 0} (t) = \bra{ \mathrm{x}_{(s,1)}} \rho^{(2)} (t) \ket{0}$, $\rho^{(2)}_{(0,0)(s,1)} (t) = \bra{ \mathrm{x}_{(0,0)}} \rho^{(2)} (t) \ket{\mathrm{x}_{(s,1)}}$, and $\rho^{(2)}_{(0,0)0} (t) = \bra{ \mathrm{x}_{(0,0)} } \rho^{(2)} (t) \ket{0}$. Using the operator properties and their action on the states, we obtain that the only substantial term among three is $\rho^{(2)}_{\mathrm{x}0} (t)$, given as (see Appendix \ref{sec:matrix-rho})
\begin{align}
    & \rho^{(2)}_{\mathrm{x}0} (t) = \nonumber \\
    & \frac{\mathcal{E}^2(\vt{\kappa}) \text{e}^{- i 2 \omega_\kappa t}}{\hbar^2} \frac{g^-_\nu h^-_{(0,0)\nu}}{\left( e_\nu -( \omega_\kappa + i \epsilon)  \right) \left(e_0 -( 2 \omega_\kappa + i \epsilon') \right)} . \label{eq:rho2}
\end{align}
We already calculated the velocity element $\vt{\mathrm{v}}_{\mathrm{0x}} =  \hat{\vt{\varepsilon}}^+ (-i e_0 g^{+*}_{(0,0)}/e)$. Using $\vt{J}^{(2)} = \sum_{\nu} e N_e (\vt{\mathrm{v}}_{\mathrm{0x}} \rho^{(2)}_{\mathrm{x0}} + \rho^{(2)}_{\mathrm{0x}} \vt{\mathrm{v}}_{\mathrm{x0}} )$ and  $\vt{J}^{(2)} = \sigma^{(2)} \hat{\vt{\varepsilon}}^+ \mathcal{E}^2 (\vt{q}) \text{e}^{-i 2 \omega_\kappa t}$, we obtain
\begin{align}
	&\sigma^{(2)} = \nonumber \\
	& \sum_{\nu} \sum_{p_1 = \pm 1} \left( \begin{array}{l} \cfrac{-i N_e g^-_\nu  h^-_{(0,0)\nu} g^{+*}_{(0,0)}}{\hbar^2} \\
	\times \cfrac{1}{\left( e_\nu + p_1 ( \omega_\kappa + i \epsilon)  \right) \left(e_{0} +p_1( 2 \omega_\kappa + i \epsilon')\right)} \end{array} \right).
\end{align}

From the equation \eqref{eq:chi-sigma}, the second order susceptibility for the second-harmonic generation is obtained through
\begin{equation}
	\chi^{(2)} (\omega_\kappa \sim e_\nu) = \frac{\sigma^{(2)}}{- i 2 \epsilon_0 \omega_\kappa}. \label{eq:sigma2}
\end{equation}
Then, we finally obtain the contribution of the bound virtual exciton states:
\begin{align}
	&\chi^{(2)}_{B,+;--} (\omega_\kappa \sim e_0/2)  = \sum_{\nu} \sum_{p_1 = \pm 1} \cfrac{e_0 g^-_\nu  h^-_{(0,0)\nu} g^{+*}_{(0,0)}}{2 \omega_\kappa \hbar^2 \epsilon_0 d_\mathrm{eff}} \nonumber \\
	&\times  \cfrac{1}{\left( e_\nu + p_1 ( \omega_\kappa + i \gamma_B/2)  \right) \left(e_{0} + p_1( 2 \omega_\kappa + i \gamma_B/2) \right)}  . \label{eq:chi2B}
\end{align}
Recall that that $g^-_\nu$ is substantial only for $\nu = (1,1)$ and $\nu = (2,1)$. Note that this contains the resonant ($p_1 = -1$) and the nonresonant ($p_1 = +1$) term. 

Calculating $\chi^{(2)}_{B,-;++}$ from $-\vt{K}$ valley produces the same result since the only difference between the two valleys is the switched role between $\pm\sigma$.

\subsubsection{Unbound exciton virtual states}

We now calculate the contribution from the unbound exciton virtual states. Let us first consider the case of $\sigma_+$ polarized light. The cascaded second-order transition is $\ket{0} \rightarrow \ket{C(\vt{q})} \rightarrow \ket{ \mathrm{x}_{(0,0)} } \rightarrow \ket{0}$. In order to address the second transition, we need the following interaction Hamiltonian:
\begin{equation}
	\mathcal{H}_I'' = - \sum_{\vt{q}} \left[ f_{(0,0)} (\vt{q}) B^\dag_0 C_{\vt{q}} \mathcal{E}(\vt{\kappa}) \text{e}^{-i \omega_\kappa t} + \text{h.c.}  \right], \label{eq:second-interaction-Hamiltonian}
\end{equation}
where the new dipole transition element $f_\nu (\vt{q})$ is given as
\begin{align}
	f_\nu (\vt{q}) &= e \bra{\mathrm{x}_\nu} \hat{\vt{\varepsilon}} \cdot \vt{r} \ket{C_{\vt{q}}} \nonumber \\
	&= e \sum_{\vt{q}'} \psi^*_{\nu} (\vt{q}') \bra{C_{\vt{q}'}}  \hat{\vt{\varepsilon}} \cdot \vt{r} \alpha^\dag_{\vt{q}} \beta^\dagger_{-\vt{q}} \ket{0} \nonumber \\
	&= e \sum_{\vt{q}'} \psi^*_\nu (\vt{q}') \bra{C_{\vt{q}'}} \hat{\vt{\varepsilon}} \cdot \vt{r} \ket{C_{\vt{q}}}. \label{eq:fnu}
\end{align}
The physical intuition is that this dipole moment is a superposition of all intraband dipole moment weighted by the (Fourier-transformed) exciton wave function.

We can easily deduce $\chi^{(2)}$ from this channel based on equation \eqref{eq:chi2B}:
\begin{align}
	&\chi^{(2)}_U (\omega_\kappa \sim e_0/2) = \int \td^2 q \cfrac{e_\nu d_{cv} (\vt{q}) f_{\nu} (\vt{q}) g_\nu^*}{8 \pi^2 \omega_\kappa \epsilon_0 \hbar^2 d_\textrm{eff}} \times \nonumber \\
	& \sum_{p_1 = \pm 1} \cfrac{1}{(\omega_{\vt{q}} +p_1( \omega_\kappa  + i \gamma_U/2))(e_\nu + p_1( 2 \omega_\kappa  + i \gamma_B/2))} .
\end{align}

Appendix \ref{sec:fnu} derives and conclude that $f^\pm_\nu (\vt{q})$ vanishes due to the symmetry. Hence, the virtual transition through the unbound exciton to land on a bound exciton state is negligible. This allows us to ignore in the future any virtual channel involving the unbound exciton states.

\begin{figure}[!tb]
	\centering
	\includegraphics[width=0.45\textwidth]{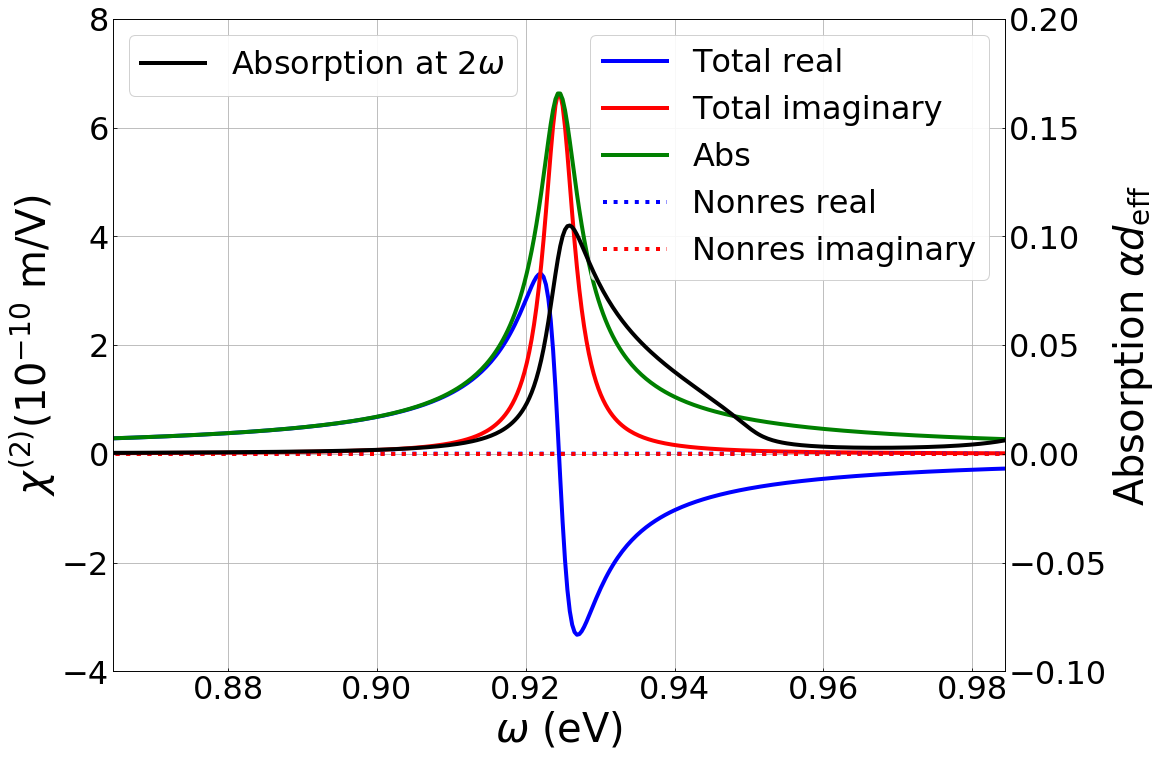}
	\caption{Numerically evaluated $\chi^{(2)}_{+;--} (\omega_\kappa \sim e_0/2)$ based on the higher-order corrected gapped Dirac Hamiltonian. Also shown is the second-harmonic absorption at $2 \omega$ in black.}
	\label{fig:chi2}
\end{figure}

\subsubsection{Overall second-order susceptibility}

We showed the opposite chirality rule between the fundamental light and the second-harmonic light for the second-harmonic generation. Since the virtual channels from the unbound excitons can be ignored, the second-order susceptibility is $\chi^{(2)} (\omega_\kappa \sim e_0/2) = \chi^{(2)}_B (\omega_\kappa \sim e_0/2)$. Fig. \ref{fig:chi2} shows the calculated $\chi^{(2)}$ for a single polarized second-harmonic output from a linearly polarized pump light. The intensity of the second-harmonic light depends on the absolute value $|\chi^{(2)}|$ whereas the phase of $\chi^{(2)}$ explains the phase delay of the second-harmonic light \cite{boyd2003nonlinear}. The maximum value of the calculated $|\chi^{(2)}|$ at frequency $e_0/2$ is $6.6 \times 10^{-10}$ m/V. Fig. \ref{fig:chi2} also shows the linear absorption at the second harmonic $2 \omega$. In order to avoid it, one may want to operate at slight red detuning from the resonance. The figure also shows the contribution from the nonresonant term ($p_1 = +1$ in equation \eqref{eq:chi2B}), which is negligibly small in both real and imaginary parts. This is expected since the second-order susceptibility is concentrated near resonance and both the two factors in the denominator of the resonant term in equation \eqref{eq:chi2B} diverge around the resonance.

A few more experimental results on the monolayer MoS$_2$ second-harmonic generation that quantified the second-order susceptibility were reported: Malard \emph{et al.}\cite{malard2013observation} reported a sheet susceptibility of $8 \times 10^{-20}$ m$^2$/V, equivalent to a bulk $\chi^{(2)}$ of $1.2 \times 10^{-10} $ m/V, and Clark \emph{et al.}\cite{clark2014strong} experimentally obtained $2 \times 10^{-9}$ m/V while Woodward \emph{et al.}\cite{woodward2016characterization} reported $3 \times 10^{-11}$ m/V, all with the second harmonic at the A exciton resonance of 1.9 eV. These match our result within an order of magnitude. Trolle \emph{et al.} theoretically calculated $\chi^{(2)}$ through the tight binding band structures and obtained $4 \times 10^{-9}$ m/V \cite{trolle2014theory}, which also agrees with our result approximately within an order of magnitude, although the approach was different.

Compared to the typical $\chi^{(2)}$ value $2 \times 10^{-11}$ m/V of lithium niobate, which is the common material for the second-harmonic generation, the single pass second-order effect in the monolayer MoS$_2$ is equivalent to approximately only nanometer thick lithium niobate material. Hence, the monolayer MoS$_2$ does not appear to be a strong second-harmonic nonlinear material.

\subsection{Third order susceptibility}

\begin{figure}[tb]
	\centering
	\subfloat[Third-harmonic generation]{\includegraphics[width=0.23\textwidth]{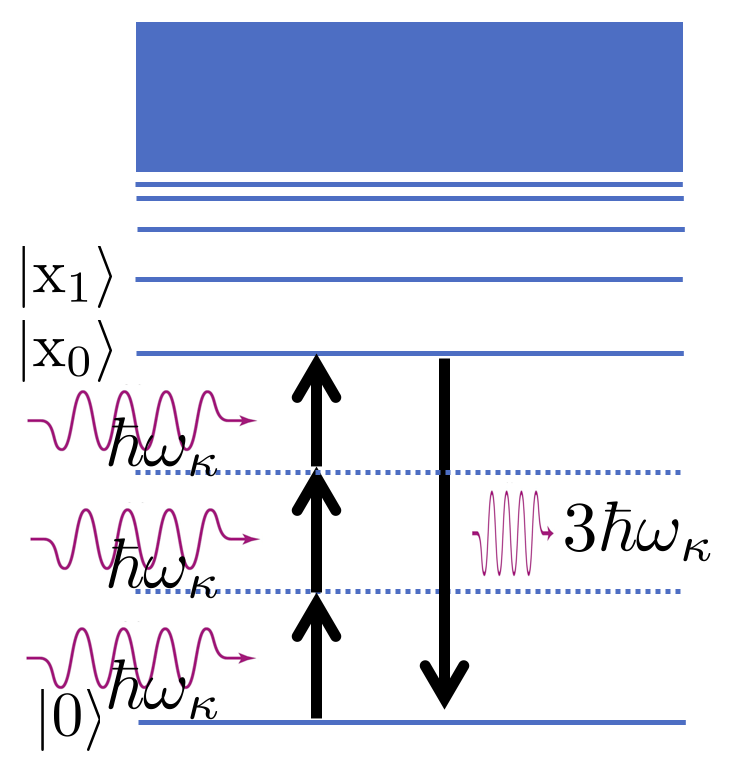}} \hspace{2mm}
	\subfloat[Two-photon process]{\includegraphics[width=0.23\textwidth]{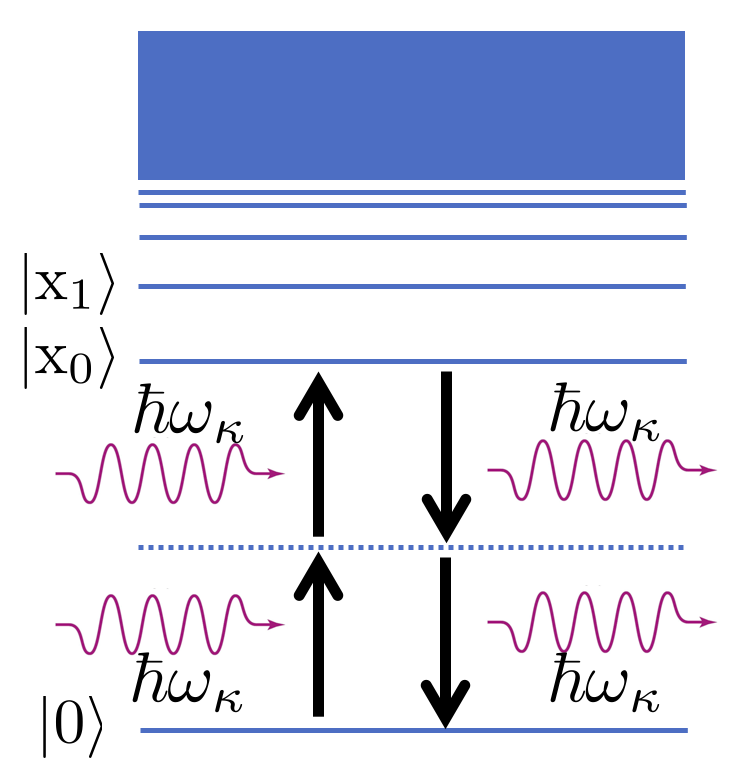}}
	\caption{Third order processes with low frequency input light. (a) Third-harmonic generation where $3 \omega_\kappa \sim e_0$. (b) Two-photon process where $2 \omega_\kappa \sim e_0$. }
	\label{fig:low-third}
\end{figure}

The third order processes that can avoid the direct linear absorption are the third-harmonic generation and the two-photon process (i.e., Kerr effect and two-photon absorption) as shown in  Fig. \ref{fig:low-third}.

\subsubsection{Third-harmonic generation}

We first consider the third-harmonic generation process where $\omega_\kappa \sim e_0/3$ (see Fig. \ref{fig:low-third} (a)). This process involves two virtual levels between $\ket{0}$ and $\ket{\mathrm{x}_{(0,0)}}$. As we have seen from the previous calculation for $\chi^{(2)}$, the virtual contribution from the unbound excitons is negligible. We then count only the virtual levels from the bound exciton states. This requires a modification of the second interaction Hamiltonian in equation \eqref{eq:interact3} as
\begin{equation}
	\mathcal{H}'_I = - \sum_{\nu_1, \nu_2} \left[ h_{\nu_1 \nu_2} B^\dag_{\nu_1} B_{\nu_2} \mathcal{E} (\vt{\kappa}) \mathrm{e}^{- i \omega_\kappa t} + \mathrm{h.c.}  \right].
\end{equation}
This third-harmonic generation process involves the four  states $\ket{0}, \ket{\mathrm{x}_{\nu_1}}, \ket{\mathrm{x}_{\nu_2}}, \ket{\mathrm{x}_{(0,0)}}$ with the successive transition $\ket{0} \rightarrow \ket{\mathrm{x}_{\nu_1}} \rightarrow \ket{\mathrm{x}_{\nu_2}} \rightarrow \ket{\mathrm{x}_{(0,0)}} \rightarrow \ket{0}$.

The optical selection rule where only $\ket{\mathrm{x}_{(n,m)}} \rightarrow \ket{\mathrm{x}}_{(n,m\pm1)}$ are allowed from the polarization $\sigma_\pm$, respectively, applies here as well for efficient virtual transitions. For $\sigma_+$ input light alone, there are no cascaded transitions to arrive at $\ket{\mathrm{x}_{(0,0)}}$ through the two virtual bound exciton states. The same applies to $\sigma_-$. This forces the tensor elements $\chi^{(3)}_{TH, \pm; +++} = \chi^{(3)}_{TH, \pm;---} = 0$. On the other hand, if both $\sigma_\pm$ photons are present, they can cooperate and incur the following transition: $\ket{0} \rightarrow \ket{\mathrm{x}_{(s,0)}} \rightarrow \ket{\mathrm{x}_{s',-1}} \rightarrow \ket{\mathrm{x}_{(0,0)}} \rightarrow \ket{0}$ with $s = 0,1,2, \cdots$ and $s' = 1,2, \cdots$. The sequential transitions are mediated by $\sigma_+$, $\sigma_-$, $\sigma_+$, $\sigma_+$ for $+\vt{K}$ valley involving the dipole moments $g^+_{(s,0)}, h^-_{(s',-1)(s,0)}, h^+_{(0,0)(s,-1)}, g^{+*}_{(0,0)}$, respectively, leaving the output polarization in $\sigma_+$ light of the third harmonic. The opposite polarization sequence applies to the $-\vt{K}$ valley, leaving the output third-harmonic light in $\sigma_-$.

Let us consider the tensor element $\chi^{(3)}_{TH, +;+-+} (= \chi^{(3)}_{TH, +;++-} = \chi^{(3)}_{TH, +;-++})$ from the $+\vt{K}$ valley. The detailed calculations reveal that the only nonzero matrix element in the density matrix $\rho^{(3)}$ are $\rho^{(3)}_{(0,0)0} = \bra{\mathrm{x}_{(0,0)}} \rho^{(3)} \ket{0}$ and $\rho^{(3)}_{(s',-1)(0,0)} = \bra{\mathrm{x}_{(s',-1)}} \rho^{(3)} \ket{\mathrm{x}_{(0,0)}}$ (see Appendix \ref{sec:matrix-rho}):
\begin{align}
	&\rho^{(3)+-+ }_{(0,0)0} = \sum_{s', s} \frac{h^+_{(0,0)(s',-1)}h^-_{(s',-1)(s,0)}g^+_{(s,0)} }{\hbar^3} \times \nonumber \\
	& \frac{\varepsilon^3(\kappa) \mathrm{e}^{- 3 i \omega_\kappa t}}{(e_s - \omega_\kappa - i \epsilon)(e_{s'} - 2 \omega_\kappa - i \epsilon')(e_0 - 3 \omega_\kappa - i \epsilon'')},
\end{align}
and
\begin{align}
	&\rho^{(3)+-+ }_{(s',-1)(0,0)} = -\sum_{s', s} \frac{g^{+*}_{(0,0)}h^-_{(s',-1)(s,0)}g^+_{(s,0)} }{\hbar^3} \times \nonumber \\
	& \frac{\varepsilon^3(\kappa) \mathrm{e}^{- 3 i \omega_\kappa t}}{(e_s - \omega_\kappa - i \epsilon)(e_{s'} - 2 \omega_\kappa - i \epsilon')(e_{s'} - e_0 + \omega_\kappa + i \epsilon'')}.
\end{align} 
We then calculate the induced current for the third-harmonic generation: $	\vt{J}^{(3)} =  \sum_{\nu_1} e N_e ( \mathbf{v}_{(0,0)0} \rho^{(3)}_{0(0,0)} + \rho^{(3)}_{(0,0)0} \mathbf{v}_{0(0,0)} + \mathbf{v}_{(s',-1)(0,0)} \rho^{(3)}_{(0,0)(s',-1)} + \rho^{(3)}_{(s',-1)(0,0)} \mathbf{v}_{(0,0)(s',-1)}  )$.  After resolving the velocity matrix elements in a similar way to equations \eqref{eq:vfe} and \eqref{eq:velement}, we use $\vt{J}^{(3)} = \sigma^{(3)} \hat{\vt{\varepsilon}}^+ \mathcal{E}^3 (\vt{\kappa}) \text{e}^{- i 3 \omega_\kappa t}$ with the following relation:
\begin{equation}
	\frac{\partial}{\partial t} \epsilon_0 \chi^{(3)}_{TH} (\omega_\kappa \sim e_\nu) \mathcal{E}^3 (\vt{\kappa}) \text{e}^{- i 3 \omega_\kappa t} = \sigma^{(3)} \mathcal{E}^3 (\vt{\kappa}) \text{e}^{- i 3 \omega_\kappa t},
\end{equation}
which leads to $\chi^{(3)}_{TH} (\omega_\kappa \sim e_\nu) = \sigma^{(3)}/(-i 3 \epsilon_0 \omega_\kappa)$, we finally obtain the third-order susceptibility for the third-harmonic generation as
\begin{widetext}
\begin{align}
	\chi^{(3)}_{TH, B, +;+-+} & (\omega_\kappa \sim e_0/3) = \sum_{s,s'} \cfrac{g^{+*}_{(0,0)} h^+_{(0,0)(s',-1)} h^-_{(s',-1)(s,0)} g^+_{(s,0)}}{3 \omega_\kappa \epsilon_0 \hbar^3 d_\mathrm{eff}} \times \nonumber \\
	&   \left( \begin{array}{l}  \sum_{p_1 = \pm 1}  \cfrac{e_0}{\left( e_s + p_1( \omega_\kappa + i \gamma_{B}/2) \right)\left( e_{s'} + p_1( 2 \omega_\kappa + i \gamma_{B}/2)   \right)\left( e_0 + p_1 ( 3\omega_\kappa + i \gamma_{B}/2)  \right)} \\
	- \sum_{p_2 = \pm 1}  \cfrac{e_{s'}-e_0}{\left( e_s + p_2( \omega_\kappa + i \gamma_{B}/2) \right)\left( e_{s'} + p_2( 2 \omega_\kappa + i \gamma_{B}/2)   \right)\left( e_{s'} - e_0 - p_2 ( \omega_\kappa + i \gamma_{B}/2)  \right)} \end{array} \right). \label{eq:chi3TH}
\end{align}
\end{widetext}
Here, $s = 0,1, \cdots$ and $s' = 1,2, \cdots$. There are four terms in the above for a given $s,s'$ pair. The first term with $p_1 = - 1$ is the resonant term with all frequency difference denominator factors, while the other three terms are nonresonant terms with at least one frequency sum in the denominator. This is the response from $+\vt{K}$ valley only. Since we ignore the virtual channel through the unbound exciton states, we obtain $\chi^{(3)}_{TH, +;++-} (\omega_\kappa \sim e_0/3) = \chi^{(3)}_{TH, B, +;++-} (\omega_\kappa \sim e_0/3)$. The response from the other valley is identical since $\sigma_\pm$ polarizations switch roles. Hence, we obtain the tensor elements
\begin{align}
	&\chi^{(3)}_{TH, \pm;\pm \pm \mp} (\omega_\kappa \sim e_0/3) = \chi^{(3)}_{TH, \pm;\mp \pm \pm} (\omega_\kappa \sim e_0/3) \nonumber \\
	&= \chi^{(3)}_{TH, \pm;\pm \mp \pm} (\omega_\kappa \sim e_0/3),
\end{align}
all having the same result as in equation \eqref{eq:chi3TH}. All the other tensor elements are negligible.

\begin{figure}[!tb]
	\centering
	\includegraphics[width=0.45\textwidth]{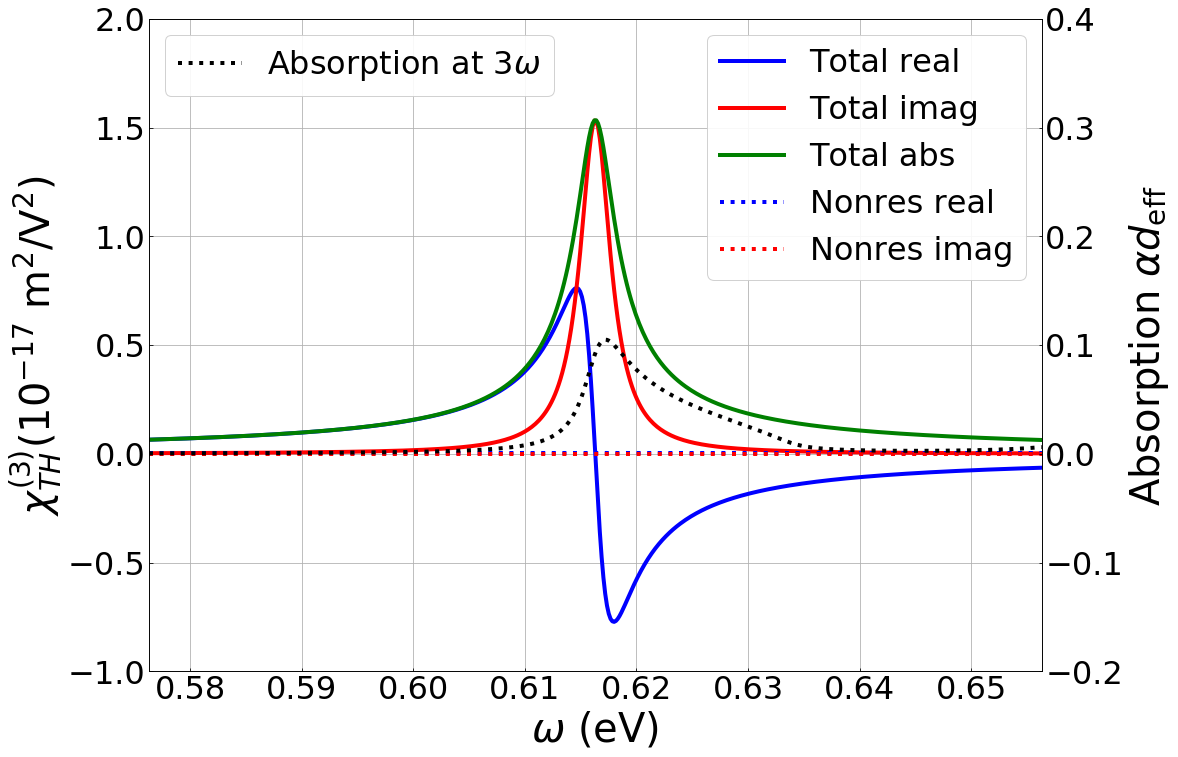}
	\caption{Numerically evaluated $\chi^{(3)}_{TH,+;+-+} (\omega_\kappa \sim e_0/3)$ based on the higher-order corrected gapped Dirac Hamiltonian. The real value (blue solid), the imaginary value (red solid), and the absolute value (green solid) of the total $\chi^{(3)}$ (sum of resonant and nonresonant terms) are shown. Also the separate contributions from the nonresonant terms (blue and red dotted) are shown to be negligibly small.}
	\label{fig:chi3TH}
\end{figure}

We evaluated this susceptibility tensor element numerically (see Fig. \ref{fig:chi3TH}). Just as the second-harmonic generation, what matters in the third-harmonic generation efficiency is the amplitude $|\chi^{(3)}_{TH}|$, while the phase of $\chi^{(3)}_{TH}$ determines the phase of the third-harmonic output light. The maximum $|\chi^{(3)}_{TH}|$ of the monolayer MoS$_2$ is $1.5 \times 10^{-17}$ m$^2$/V$^2$, which can be favorably compared to the typical nonlinear bulk crystal third order susceptibility\cite{boyd2003nonlinear} $\sim 10^{-24}$  m$^2$/V$^2$. The linear absorption at the third harmonic (dotted black) shows a significant absorption at near resonance. Hence, for an efficient third-harmonic generation, one would operate at slight red detuning.

The figure also shows the contribution from the nonresonant terms (dotted in red and blue). Both the real and the imaginary values from the nonresonant terms are negligible. The reason is as follows: the biggest contribution in the nonresonant term is from the second term in equation \eqref{eq:chi3TH} with $p_2 = - 1$. However, the magnitude of the resonant denominator's real values $|e_s - \omega_\kappa|$ and $|e_{s'} - 2 \omega_\kappa|$ are still quite large since $\omega_\kappa \sim e_0/3$. In addition, the third-harmonic generation susceptibility is concentrated near resonance.

\subsubsection{Two-photon process}

Next, let us turn to the two-photon transition shown in Fig. \ref{fig:low-third} (b). We consider the case where the input light frequency is such that $\omega_\kappa \sim e_0/2$. This process involves two virtual levels, one mediating the upward transition and the other the downward transition, corresponding to $\ket{0} \rightarrow \ket{\mathrm{x}_{\nu_1}} \rightarrow \ket{\mathrm{x}_{(0,0)}} \rightarrow \ket{\mathrm{x}_{\nu_2}} \rightarrow \ket{0}$. For $+\vt{K}$ valley, the circularly polarized input light $\sigma_-$ alone can make a second order transition since the virtual levels can be $\nu_1, \nu_2 = (1(2),1)$. Then, sequential transitions involve the corresponding dipole moment $g^-_{(1(2),1)}$, $h^-_{(0,0)(1(2),1)}$, $h^{-*}_{(0,0)(1(2),1)}$, $g^{-*}_{(1(2),1)}$, respectively, leaving the output photon in $\sigma_-$ polarization from $+\vt{K}$ valley. For $-\vt{K}$ valley, $\sigma_\pm$ polarizations switch roles, accepting $\sigma_+$ photons and leaving the output in $\sigma_+$.

This sequence of transition, however, is not the most efficient two-photon transition: the transition dipole moment for $\ket{0} \leftrightarrow \ket{\mathrm{x}_{(1(2),1)}}$ is indeed small (see the Table \ref{tab:gnu}). When we numerically evaluated, the maximum value of $| \chi^{(3)}_{TP} (\omega_\kappa = e_0/2)|$ was only $1.6 \times 10^{-21}$ m$^2$/V$^2$. Rather, involving an intermediate level whose dipole moment to and from the ground state is large must be much more efficient. This is accomplished if the upper state is $\ket{\mathrm{x}_{(1,1)}}$, through the the circularly polarized input light $\sigma_+$ in $+\vt{K}$ valley. As before, we ignore the virtual channels involving the unbound exciton states. The following two-photon transition is plausible: $\ket{0} \rightarrow \ket{\mathrm{x}_{(s,0)}} \rightarrow \ket{\mathrm{x}_{(1,1)}} \rightarrow \ket{\mathrm{x}_{(s',0)}} \rightarrow \ket{0}$ where $s,s' = 0,1,2, \cdots$. These transitions involve the dipole moments $g^+_{(s,0)}, h^+_{(1,1)(s,0)}, h^{+*}_{(1,1)(s',0)}, g^{+*}_{(s',0)}$, respectively, where all the dipole moments are indeed substantial. We also listed the value of $h_{(1,1)(n,0)}$ in Table \ref{tab:hnu1nu2-twophoton}. 

We then need to calculate the tensor elements of $\rho^{(3)}$ from $+\vt{K}$ valley in the basis $\{ \ket{0}, \ket{\mathrm{x}_{(s,0)}}, \ket{\mathrm{x}_{(1,1)}}  \}$.  The only nonzero elements of $\rho^{(3)}$ are (see the derivation in Appendix \ref{sec:matrix-rho}):
\begin{align}
	&\rho^{(3)+++ }_{(s'',0)0} = \sum_{ s} \frac{h^{+*}_{(1,1)(s'',0)} h^+_{(1,1)(s,0)} g^+_{(s,0)} }{\hbar^3} \times \nonumber \\
	& \frac{|\varepsilon(\kappa)|^2 \varepsilon(\kappa) \mathrm{e}^{- i \omega_\kappa t}}{(e_{s} - \omega_\kappa - i \epsilon)(e_{1} - 2 \omega_\kappa - i \epsilon')(e_{s''} + \omega_\kappa + i \epsilon'')}.
\end{align}
and
\begin{align}
	&\rho^{(3)+++ }_{(1,1)(s'',0)} = -\sum_{ s} \frac{g^{+*}_{(s'',0)} h^+_{(1,1)(s,0)} g^+_{(s,0)} }{\hbar^3} \times \nonumber \\
	& \frac{|\varepsilon(\kappa)|^2 \varepsilon(\kappa) \mathrm{e}^{- i \omega_\kappa t}}{(e_{s} - \omega_\kappa - i \epsilon)(e_{1} - 2 \omega_\kappa - i \epsilon')(e_{1} - e_{s''} - \omega_\kappa - i \epsilon'')}.
\end{align}

The two-photon induced current is $	\vt{J}^{(3)} =  \sum_{\nu_1} e N_e ( \mathbf{v}_{(s'',0)(1,1)} \rho^{(3)}_{(s'',0)(1,1)} + \rho^{(3)}_{(1,1)(s'',0)} \mathbf{v}_{(s'',0)(1,1)} + \mathbf{v}_{(s'',0)0} \rho^{(3)}_{(s'',0)0} + \rho^{(3)}_{0(s'',0)} \mathbf{v}_{(s'',0)0}  )$. We then need to resolve the following velocity matrix element:
\begin{align}
	\mathbf{v}_{\nu_1 (1,1) } &= \bra{\mathrm{x}_{\nu_1}} \dot{\vt{r}} \ket{\mathrm{x}_{(1,1)}} = - \frac{i}{\hbar} \bra{\mathrm{x}_{\nu_1}} [\vt{r}, \mathcal{H}_0] \ket{\mathrm{x}_{(1,1)}} \nonumber \\
	&= - i (e_{1} - e_{\nu_1}) \bra{\mathrm{x}_{\nu_1}} \vt{r} \ket{\mathrm{x}_{(1,1)}}. \label{eq:velement}
\end{align}
Hence, we obtain the component parallel to $\hat{\vt{\varepsilon}}^+$ as $\vt{\mathrm{v}}_{(1,1)\nu_1} = -i (e_{1} - e_{\nu_1}) \bra{\mathrm{x}_{\nu_1}} \hat{\vt{\varepsilon}}^- \cdot \vt{r} \ket{\mathrm{x}_{(1,1)}} \hat{\vt{\varepsilon}}^+ =  -i (e_{1} - e_{\nu_1}) (h^{+*}_{(1,1)\nu_1}/e) \hat{\vt{\varepsilon}}^+$. We then use $\vt{J}^{(3)}_{TP} = \sigma^{(3)}_{TP} |\mathcal{E} (\vt{\kappa})|^2 \hat{\vt{\varepsilon}}^+   \mathcal{E} (\vt{\kappa}) \text{e}^{- i \omega_\kappa t}$. We also use the fact that the two-photon susceptibility is obtained through
\begin{align}
	\frac{\partial}{\partial t} \epsilon_0 \chi_{TP}^{(3)} (\omega_\kappa \sim e_\nu) &| \mathcal{E} (\vt{\kappa})|^2 \mathcal{E}(\vt{\kappa}) \text{e}^{- i \omega_\kappa t} \nonumber \\
	&= \sigma^{(3)}_{TP} | \mathcal{E}(\vt{\kappa})|^2 \mathcal{E} (\vt{\kappa}) \text{e}^{- i \omega_\kappa t},
\end{align}
which leads to $\chi^{(3)}_{TP} (\omega_\kappa) = \sigma^{(3)}_{TP}/(- i \epsilon_0 \omega_\kappa)$. From all these we finally obtain the two-photon susceptibility tensor element
\begin{widetext}
	\begin{align}
	\chi^{(3)}_{TP, B, +;+++} & (\omega_\kappa \sim e_1/2) = \sum_{s,s'} \cfrac{\overline{g^{+*}_{(s',0)}} h^{+*}_{(s'0)(1,1)} h^{+}_{(1,1)(s,0)} \overline{g^+_{(s,0)}}}{\omega_\kappa \epsilon_0 \hbar^3 d_\mathrm{eff}} \times \nonumber \\
	&   \left( \begin{array}{l}  -\sum_{p_1 = \pm 1}  \cfrac{(e_1 - e_{s'})}{\left( e_s + p_1( \omega_\kappa + i \gamma_{B}/2) \right)\left( e_{1} + p_1( 2 \omega_\kappa + i \gamma_{B}/2)   \right)\left( e_1 - e_{s'} + p_1 ( \omega_\kappa + i \gamma_{B}/2)  \right)} \\
	+ \sum_{p_2 = \pm 1}  \cfrac{e_{s'}}{\left( e_s + p_2( \omega_\kappa + i \gamma_{B}/2) \right)\left( e_1 + p_2( 2 \omega_\kappa + i \gamma_{B}/2)   \right)\left( e_{s'} - p_2 ( \omega_\kappa + i \gamma_{B}/2)  \right)} \end{array} \right). \label{eq:TP-low1}
	\end{align}
\end{widetext}
The above contains four terms: one resonant term with $p_1 = - 1$ from the first sum, and the other three nonresonant terms $(p_1 = +1, p_2 = \pm 1)$. Here, $\overline{g^+_{\nu}} = g^+_{\nu}/\sqrt{A}$, which does not depend on the sample area $A$ (see the Table \ref{tab:gnu}). This is the response from $+\vt{K}$ valley with both the input and the output lights in $\sigma_+$ polarization. As was before, we ignore the virtual channels through the unbound excitons, and hence, we obtain the two-photon response $\chi^{(3)}_{TP} = \chi^{(3)}_{TP,B}$. The response from $-\vt{K}$ valley is identical to this since $\sigma_\pm$ polarizations switch roles, and both the input and the output from $-\vt{K}$ valley are in $\sigma_-$ polarization. All the tensor elements other than $\chi^{(3)}_{TP, \pm ; \pm \pm \pm}$ are negligible.

\begin{figure}[!tb]
	\includegraphics[width=0.50\textwidth]{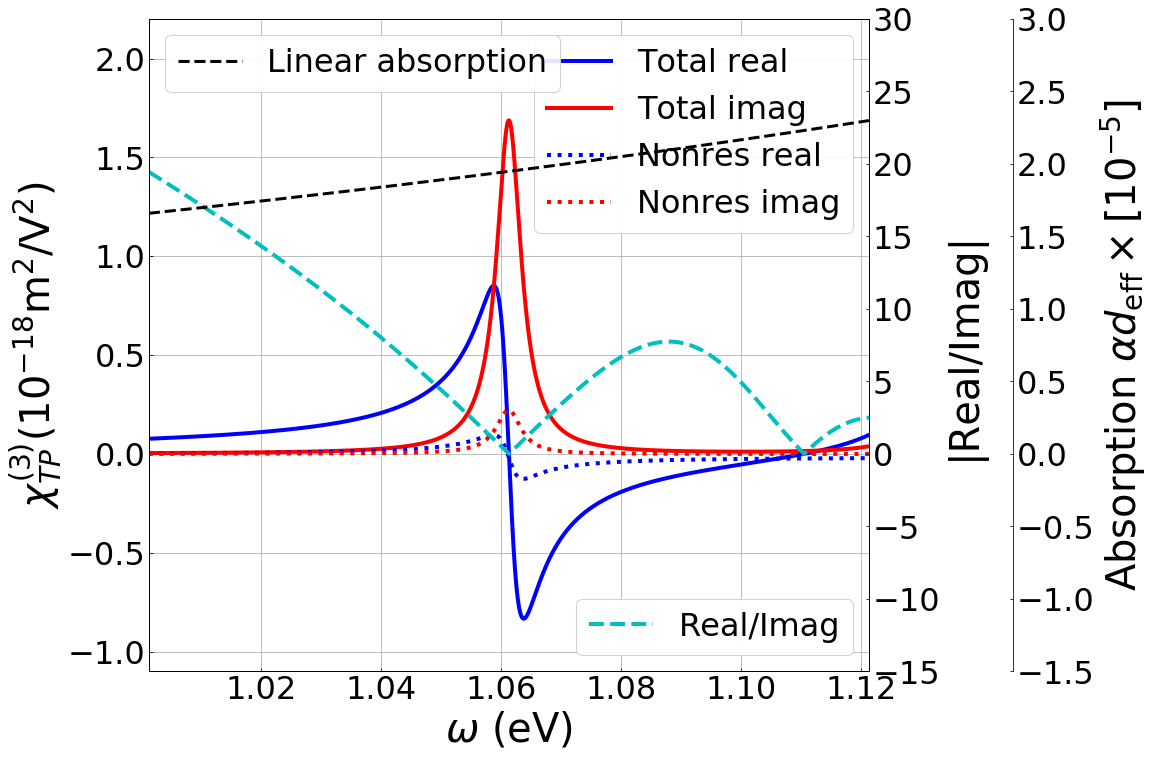}
	\caption{\label{fig:chi3TPcirc} Calculated $\chi^{(3)}_{TP,++++} (\omega_\kappa \sim e_1/2)$ for the $\sigma_+$ input light polarization. We plot the ratio $\mathrm{Re}[\chi^{(3)}_{TP}]/\mathrm{Im}[\chi^{(3)}_{TP}]$ (cyan dotted), as well as the linear absorption (black dotted). Also shown are the contribution only from the nonresonant term (red and blue dots).}
\end{figure}

Fig. \ref{fig:chi3TPcirc} shows the calculated values of $\chi^{(3)}_{TP} (\omega_\kappa \sim e_1 / 2)$. The imaginary value of the two-photon third order susceptibility is related to the actual two-photon absorption, implying the loss of the incoming light in pairs. The real value of the two-photon third order susceptibility is related to the Kerr nonlinearity where the refractive index varies proportionally to the incoming light intensity. This is best seen by the relation \cite{boyd2003nonlinear}:
\begin{equation}
	\chi_\textrm{eff} = \chi^{(1)} + 3 \chi^{(3)}_{TP} | \mathcal{E} (\omega_\kappa)|^2.
\end{equation}
The negative sign in equation \eqref{eq:TP-low1} is physically substantial since it produces a positive imaginary value for $\chi^{(3)}_{TP}$ implying real two-photon absorption. The maximum of the real value $\chi^{(3)}_{TP} (\omega_\kappa \sim e_{1}/2)$ is $8.5 \times 10^{-19}$ m$^2$/V$^2$ around $e_1/2$ of A excitons. This value is approximately six orders of magnitude larger than the typical bulk material. The figure shows the influence of the same transition for the B excitons on the blue side (spin down electrons). Additionally, it also shows the linear absorption, which comes from the off-resonant contribution from the nearest exciton states absorption $\ket{\mathrm{x}_{(0,0)}}$. The linear absorption is only order of $\sim 10^{-5}$, which is sufficiently small.

The optical Kerr effect is a valuable resource for coherent optical switching. Hence, avoiding the incoherent two-photon absorption is important. We plotted the figure of merit $|\mathrm{Re}[\chi^{(3)}_{TP}]/\mathrm{Im}[\chi^{(3)}_{TP}|]$ in Fig. \ref{fig:chi3TPcirc}. Let us compare the two-photon process results of the monolayer MoS$_2$ with those of graphene \cite{soh2016comprehensive}. The graphene exhibits $\chi^{(3)}_{TP} (e_1/2) \sim 4.8 \times 10^{-17}$ m$^2$/V$^2$, which is larger than the monolayer MoS$_2$. The ratio $|\mathrm{Re}[\chi^{(3)}_{TP}]/\mathrm{Im}[\chi^{(3)}_{TP}|]$ of graphene at the same frequency, however, is only 0.06, whereas the monolayer MoS$_2$ has quite a favorable ratio, much larger than unity over broadband at certain frequency regions. This is because MoS$_2$ exciton responses are narrow band resonances whereas that of the graphene is broadband interband transitions. In addition, the graphene also suffers from the broadband linear absorption of 2.3\% for the pumping photon\cite{soh2016comprehensive} while such linear absorption is completely absent from the monolayer MoS$_2$ thanks to the band gap. This makes the monolayer MoS$_2$ a superior material for the coherent Kerr optical nonlinearity.

It is noteworthy that the contribution from the nonresonant terms for the two-photon third-order susceptibility is much larger than others (dotted lines in Fig. \ref{fig:chi3TPcirc}). The reason is as follows: the biggest contribution comes from the term with $p_2 = - 1$ from the second term in equation \eqref{eq:TP-low1}. The magnitude of the resonant denominators' real values $|e_s - \omega_\kappa|$ and $|e_1 - 2 \omega_\kappa|$ is relatively small since $\omega_\kappa \sim e_1/2$. Hence, the contribution from the nonresonant terms in the two-photon susceptibility is significantly larger than other cases. Nevertheless, it is fair to say that the major contribution still comes from the resonant term.

\section{Conclusion and discussions}

We calculated the linear and nonlinear optical susceptibilities of excitonic states in monolayer MoS$_2$, based on the second-order corrected Dirac Hamiltonian around $\pm \vt{K}$ points in the first Brillouin zone. We derived and utilized the second quantized bound and unbound exciton operators and efficiently calculated the perturbative solutions of the density matrix. This connected to the induced current, the optical conductivity, and eventually the optical susceptibilities in a perturbative order. We showed that the simple higher-order corrected Dirac gapped Hamiltonian produced linear and second-order susceptibilities that reasonably match experimental results. An alternative route would be the detailed computationally heavy DFT-based calculation.

The reasonable agreement of our theoretical results with experimental data may be somewhat surprising considering that we have approximated the physical system as completely two dimensional, whereas the detailed atomic positions are indeed in three dimensions, and hence, the detailed electron density distribution might have played an important role. However, the exciton is a collective excitation spanning the entire sample area and atomic details may be blurred over the large exciton size (several times the unit cell). It is thus plausible to consider our physical system as being approximately circularly symmetric, and the angular momentum based optical selection rules of our bound exciton solution played a vital role. We emphasize that such an averaging effect is indeed a nature of the Wannier excitons with a large size.

The second-harmonic process of the exciton states from the monolayer MoS$_2$, on the other hand, is well expected to be small since the exciton states are approximately centro-symmetric where only a very minor centro-symmetry breaking feature is provided through the weak threefold rotational symmetry, connecting the Fermi sea and a couple of the high order excitons. We also note that we resolved quantitatively the previously known opposite chirality rule for the second-harmonic generation in the monolayer TMDs materials through directly calculating the dipole moments and the susceptibilities.

The obtained third-order nonlinear optical susceptibility of monolayer MoS$_2$ merits further investigation for potential photonics applications. The excitonic states of this material are promising for device designs utilizing coherent nonlinear optical processes, such as the coherent Kerr-type optical operation in an extremely small strong cavity (\cite{mabuchi2012qubit}), since one can avoid incoherent linear loss while strong optical response is provided via collecting the broadband responses of the bands into a narrow band exciton resonance.

It is worth mentioning that, while the center frequency of the lowest exciton state of our result is based on empirically measured binding energy, those of higher exciton states may need to be adjusted slightly according to either the more accurate Keldysh-type binding energies of exciton states or the actual experimental results, albeit the difference is small as we mentioned above.

\begin{acknowledgments}
Sandia National Laboratories is a multi-program laboratory managed and operated by Sandia Corporation, a wholly owned subsidiary of Lockheed Martin Corporation, for the U.S. Department of Energy\textsc{\char13}s National Nuclear Security Administration under contract DE-AC04-94AL85000. This work was supported by the National Science Foundation under award PHY-1648807 and by a seed grant from the Precourt Institute for Energy at Stanford University.
\end{acknowledgments}

\appendix

\section{Electronic band structure of MoS$_2$} \label{sec:band-structure}

For the band structure of the monolayer MoS$_2$, we assume a gapped Dirac cone model that was adopted in many of the theoretical works of the TMD material calculations \cite{xiao2012coupled, kormanyos2013monolayer, ridolfi2015tight, fang2015ab, rasmussen2015computational, kormanyos2015k, selig2016excitonic, wang2015fast, wang2016radiative}. This approach assumes the tight binding approximation, where the Bloch wave is $\psi_{\vt{k}, \lambda} (\vt{r}) = \textrm{e}^{i \vt{k} \cdot \vt{r}} u_{\vt{k}, \lambda} (\vt{r})$ with the band index $\lambda = c,v$ for the conduction and the valence bands, respectively. Here, $\vt{k}$ is a Bloch wave vector, and the Bloch function is represented as $u_{\vt{k}, \lambda} (\vt{r}) = (1/\sqrt{N}) \sum_{m} \text{e}^{i \vt{k} \cdot (\vt{R}_m - \vt{r})} \phi_{\vt{k}, \lambda} (\vt{r} - \vt{R}_m)$, where $N$ is the total number of atoms in the sample, $\vt{R}_m$ is the lattice site position, and $\phi_\lambda(\vt{r})$ is the atomic orbital. At the $\pm \vt{K}$ points, it is conventional \cite{xiao2012coupled, kormanyos2013monolayer, ridolfi2015tight, fang2015ab, rasmussen2015computational, kormanyos2015k, selig2016excitonic, wang2015fast, wang2016radiative} to approximate $\phi_{\tau \vt{K}, c} (\vt{r}) = \braket{\vt{r}}{d_{z^2}}$ and $\phi_{\tau \vt{K}, v} (\vt{r}) = (1/\sqrt{2}) \left( \braket{\vt{r}}{d_{x^2 - y^2}} + i \tau \braket{\vt{r}}{d_{xy}}  \right)$ where $\ket{d_{z^2}}$, $\ket{d_{x^2 - y^2}}$, $\ket{d_{xy}}$ are the 4d shell atomic orbitals of the Mo atom. Here, $\tau = \pm 1$ is the valley index corresponding to $\pm \vt{K}$ points, respectively. In fact, the conduction and the valence bands at $\pm \vt{K}$ points consist of both the $d$ orbitals of Mo atoms and the $p$ orbitals of S atoms. The relative contributions of Mo atom $d$ orbitals are 92\% in the conduction band and 84\% in the valence band \cite{ridolfi2015tight}.

Let us consider the Bloch waves around either of $\pm \vt{K}$ points. Adopting the basis $\{ \ket{u_{\vt{0}, c}}, \ket{u_{\vt{0}, v}} \}$, and considering only the subspace of either up or down electron spin, the Hamiltonian is given as \cite{xiao2012coupled}
\begin{align}
H_0 = \left( \begin{array}{cc}  \Delta/2 & \hbar v (\tau q_x - i q_y) \\ \hbar v (\tau q_x + i q_y) & - \Delta/2  \end{array}  \right),
\end{align}
where $\Delta = E_g \pm \tau E_\textrm{soc}/2$ for up or down spin subspace, respectively, with the energy band gap $E_g$ and the spin-orbit coupling energy $E_\textrm{soc}$. Here $\vt{q} = (q_x, q_y) = \vt{k} - \tau \vt{K}$. This is a Hamiltonian for a gapped Dirac cone. The values we use are the results of the detailed DFT calculations \cite{kormanyos2013monolayer, ridolfi2015tight}, namely, $\hbar v = 3.82$ eV \AA$~$ ($v = 5.8 \times 10^5$ m/s), $E_{g} = 2.23$ eV (DFT-HSE06) (and experimentally measured\cite{zhang2014direct}  as 2.15 eV), and $E_{\text{soc}} = 146$ meV. If we expand the solution up to the second order with respect to $\vt{q}$, we obtain an analytical formula for the \emph{uncorrected} band Hamiltonian $H_0$:
\begin{align}
E_c (q) &= \frac{\Delta}{2} + \frac{\hbar^2 v^2 q^2}{\Delta}, E_v (q) = -\left(\frac{\Delta}{2} + \frac{\hbar^2 v^2 q^2}{\Delta} \right), \nonumber \\
\ket{u_{\vt{q}, c}} &= \left(  1 - \frac{\hbar^2 v^2 q^2}{\Delta^2} \right) \ket{u_{\vt{0},c}}  + \frac{\hbar v q \tau}{\Delta} \text{e}^{i \tau \phi_q}   \ket{u_{\vt{0},v}}, \nonumber \\
\ket{u_{\vt{q}, v}} &= - \frac{\hbar v q \tau}{\Delta} \text{e}^{- i \tau \phi_q} \ket{u_{\vt{0}, c}} + \left( 1 - \frac{\hbar^2 v^2 q^2}{\Delta^2} \right) \ket{u_{\vt{0}, v}}, \label{eq:uqv}
\end{align}
where  $\phi_q = \arccos(q_x/q)$. The Dirac cone approximation inevitably produces the same effective mass for the conduction band electron and the valence band hole. For a more accurate calculation, one may adopt the higher order correction\cite{wang2015fast, wang2016radiative} such that $H = H_0 + H_\text{C}$ with
\begin{align}
&H_{C} = \nonumber \\
&\left( \begin{array}{cc} \alpha q^2 & \kappa q^2 \mathrm{e}^{2 i \tau \phi_q} - \frac{\eta}{2} q^3 \mathrm{e}^{-i \tau\phi_q} \\*[3mm] \kappa q^2 \mathrm{e}^{-2i \tau \phi_q} - \frac{\eta}{2} q^3 \mathrm{e}^{i \tau \phi_q} & \beta q^2    \end{array}  \right), \label{eq:HOC}
\end{align}
where the numerical values of the parameter based on the DFT calculations are $\alpha = 1.72$ eV \AA$^2$, $\beta = - 0.13$ eV \AA$^2$, $\kappa = - 1.02$ eV \AA$^2$, and $\eta = 8.52$ eV \AA$^3$.  The energy eigenvalues of the band Hamiltonian $H = H_0 + H_C$ are analytically solved as follows:
\begin{align}
&E_\lambda = \nonumber \\
&\frac{1}{2} (\alpha + \beta) q^2 +\lambda \frac{1}{2} \sqrt{ \begin{array}{l} 4 \hbar^2 v^2 q^2 + 2 q^2 (\alpha - \beta) \Delta \\
	+ \Delta^2 - 4 \hbar v q^4 \eta + q^6 \eta^2 \\
	+ q^4 ((\alpha-\beta)^2 + 4 \kappa^2) \\
	+ 4 q^3 (2 \hbar v - q^2 \eta) \kappa \cos(3 \phi_q) \end{array}}, \label{eq:threefold-rotational-sym}
\end{align}
where $\lambda = \pm 1$ for the conduction and the valence band, respectively. The higher order correction does not only produce different effective masses for the conduction and the valence bands, but also gives rise to the well-known threefold rotational symmetry through the dependence on $\cos(3 \phi_q)$. This threefold rotational symmetry of the energy dispersion is common in hexagonal 2D materials.

\begin{figure}[!tb]
	\centering
	\includegraphics[width=0.45\textwidth]{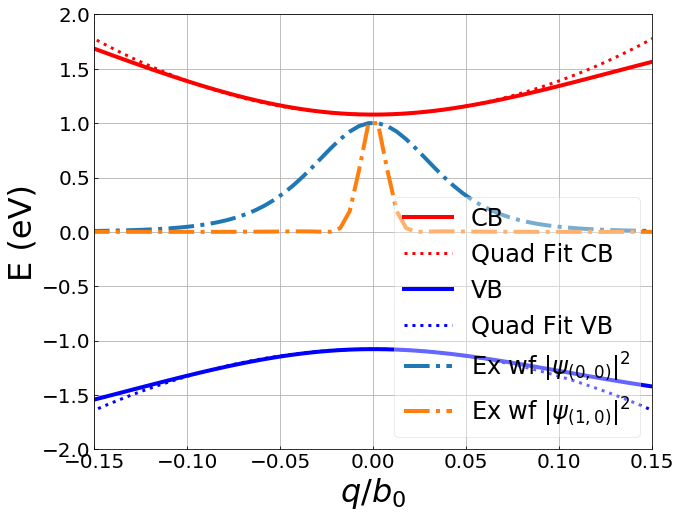}
	\caption{\label{fig:effective_mass} Band structure of monolayer MoS$_2$ near $\pm \vt{K}$ points (i.e., $\vt{q} = \vt{0})$. Only the lowest conduction and highest valence bands are shown. The conduction band (red slid) and the valence band (blue solid) are fitted with quadratic curves (dotted lines) for extracting the effective masses at each band. Also shown are the exciton wavefunctions for states $\ket{\mathrm{x}_{(0,0)}}$ (cyan) and $\ket{\mathrm{x}_{(1,0)}}$ (orange).}
\end{figure}

It is noteworthy that one extracts the effective masses for the conduction and the valence band through the energy dispersion (equation \eqref{eq:uqv}), and use them to solve the Wannier exciton equation in equation \eqref{eq:schrodinger-exciton}. Although the actual dispersion is not completely parabolic, one often approximates the band dispersion quadratically. This approximation is particularly valid for the exciton where the superposing weight $\psi_{\nu} (\vt{q})$ in equation \eqref{eq:bound-creation} is heavily concentrated in the valley bottoms. Fig. \ref{fig:effective_mass} shows the quadratic fittings of the conduction and valence bands. Also shown are the exciton wavefunctions which become weights to construct an exciton state. We note that the upper states have more concentrated wavefunctions to the valley bottom. The quadratic fitting is reasonably good even for the lowest exciton level within the exciton wavefunctions. This concretely shows that the effective mass approach is valid for the monolayer MoS$_2$ excitons, which is also consistent with literature \cite{selig2016excitonic, cheiwchanchamnangij2012quasiparticle, berkelbach2013theory, ramasubramaniam2012large, qiu2013optical}.

\section{Exciton creation operator} \label{sec:exciton-creation}

We derive the creation operators for both the bound and the unbound exciton states in terms of the band state basis. We first consider the bound exciton states, starting with the definition $B^\dag_{\nu, \vt{Q}} = \ket{\nu \vt{Q}} \bra{0}$. The exciton state $\ket{\nu \vt{Q}}$ is a dual-particle state where there is an electron-hole pair. Let us recall that the band pair state is given by $\ket{\vt{q}, -\vt{q}'} = \alpha^\dag_{\vt{q}} \beta^\dag_{-\vt{q}'} \ket{0}$. This is a composite state of an electron Bloch state in the conduction and a hole Bloch state in the valence band, having the momentum $\hbar \vt{q}$ and $-\hbar \vt{q}'$, respectively. Any single particle state lives in a Hilbert space that is spanned by basis $\{ \ket{\vt{q}, -\vt{q}'} \}$. In this subspace, the completeness relation is
\begin{equation}
	\sum_{\vt{q}, \vt{q}'} \ket{\vt{q}, -\vt{q}'} \bra{\vt{q}, -\vt{q}'} = \mathbf{1}.
\end{equation}
Then, we obtain
\begin{align}
	B^\dag_{\nu\vt{Q}} &= \sum_{\vt{q}, \vt{q}'} \ket{\vt{q}, -\vt{q}'} \braket{\vt{q}, -\vt{q}'}{\nu \vt{Q}} \bra{0} \nonumber \\
	&= \sum_{\vt{q}, \vt{q}'} \braket{\vt{q}, -\vt{q}'}{\nu \vt{Q}} \alpha^\dag_{\vt{q}} \beta^\dag_{-\vt{q}'}. 
\end{align}
Following the treatment of Haug \emph{et al.} \cite{haug2009quantum}, we then approximate the band Bloch states by free particle states such that $\braket{\vt{r}, \vt{r}'}{\vt{q}, -\vt{q}'} \approx (1/A) \textbf{e}^{i \vt{q} \cdot \vt{r} + i \vt{q}' \cdot \vt{r}'}$. Then, we calculate the following using the completeness $\int \td^2 r \td^2 r' \ketbra{\vt{r}, \vt{r}'}{\vt{r}, \vt{r}'} = \mathbf{1}$:
\begin{align}
	&\braket{\vt{q}, -\vt{q}'}{\nu \vt{Q}} = \int_A \td^2 r \td^2 r' \braket{\vt{q}, -\vt{q}'}{\vt{r}, \vt{r}'} \braket{\vt{r}, \vt{r}'}{\nu \vt{Q}} \nonumber \\
	&= \int \td^2 r \td^2 r' \frac{1}{A} \text{e}^{-i \vt{q} \cdot \vt{r}} \text{e}^{i \vt{q}' \cdot \vt{r}'} \psi_{\nu} ( \vt{r} - \vt{r}') \frac{1}{\sqrt{A}}\text{e}^{i \vt{Q} \cdot \frac{\vt{r} + \vt{r}'}{2}},
\end{align}
where $\psi_{\nu} (\vt{r}'')$ is the solution to the exciton Schr\"{o}dinger equation in equation \eqref{eq:schrodinger-exciton}. Then, we Fourier-transform $\psi_\nu (\vt{r}'')$ to obtain
\begin{widetext}
\begin{align}	
	&\braket{\vt{q}, -\vt{q}'}{\nu \vt{Q}} \nonumber \\
	&= \frac{1}{A^2}\sum_{\vt{q}''} \int \td^2 r \td^2 r' \exp \left[ i \left( \vt{Q} \cdot \frac{\vt{r} + \vt{r}'}{2} - \vt{q} \cdot \vt{r} + \vt{q}' \cdot \vt{r}' + \vt{q}'' \cdot (\vt{r} - \vt{r}')\right) \right] \psi_{ \nu } (\vt{q}'') \nonumber \\
	&= \frac{1}{A^2}\sum_{\vt{q}''}\int \td^2 r \td^2 r' \exp \left[ i \left( \vt{r} \cdot \left( \frac{\vt{Q}}{2} - \vt{q} + \vt{q}''\right) + \vt{r}' \cdot \left( \frac{\vt{Q}}{2} + \vt{q}' - \vt{q}''\right)  \right) \right] \psi_{ \nu } (\vt{q}'') \nonumber \\
	&= \sum_{\vt{q}''} \psi_\nu (\vt{q}'') \delta_{\frac{\vt{Q}}{2}, \vt{q} - \vt{q}''} \delta_{\frac{\vt{Q}}{2}, \vt{q}'' - \vt{q}'},
\end{align}
\end{widetext}
where we used $(1/A)\int \td^2 r \text{e}^{i (\vt{q} - \vt{q}') \cdot \vt{r}} = \delta_{\vt{q}, \vt{q}'}$. This leads to
\begin{align}
	\braket{\vt{q}, -\vt{q}'}{\nu \vt{Q}}= \delta_{\vt{Q},  (\vt{q} - \vt{q}')} \psi_{\nu} \left( \frac{\vt{q} + \vt{q}'}{2}\right).
\end{align}
Hence, we finally obtain
\begin{align}
	B^\dag_{\nu \vt{Q}} &= \sum_{\vt{q}, \vt{q}'}\delta_{\vt{Q}, (\vt{q} - \vt{q}')} \psi_{\nu} \left( \frac{\vt{q} + \vt{q}'}{2}\right) \alpha^\dag_{\vt{q}} \beta^\dag_{-\vt{q}'} \nonumber \\
	&= \sum_{\vt{q}} \psi_{\nu} \left( \vt{q} - \frac{\vt{Q}}{2}\right) \alpha^\dag_{\vt{q}} \beta^\dag_{\vt{Q} - \vt{q}}. \label{eq:A-3}
\end{align}
We also mention that this result matches other references \cite{wang2016radiative, wang2015fast}. Approximating $\vt{Q} \approx \vt{0}$, the exciton creation operator is $B^\dag_{\nu} \equiv B^\dag_{\nu \vt{0}}$.

Next, we consider the unbound exciton states. In the same line of thought as the bound exciton, we seek the creation operator for the unbound exciton to be a linear combination from the band basis:
\begin{equation}
	C^\dag_{\tilde{\vt{q}}} = \sum_{\vt{q}} \phi_{\tilde{\vt{q}}} (\vt{q}) \alpha^\dag_{\vt{q}} \beta^\dag_{-\vt{q}},
\end{equation}
where $\tilde{\vt{q}}$ is the canonical conjugate momentum to the relative coordinate $\vt{r} = \vt{r}_e - \vt{r}_h$. Here, $\phi_{\tilde{\vt{q}}}(\vt{q})$ is the weight to be determined. We require the two condition: orthogonality with the bound states $\braket{C_{\tilde{\vt{q}}}}{\mathrm{x}_\nu} = 0$ for all $\tilde{\vt{q}}$ and $\nu$ where $\ket{C_{\tilde{\vt{q}}}} = C^\dag_{\tilde{\vt{q}}} \ket{0}$ and $\ket{\mathrm{x}_\nu} = B^\dag_{\nu} \ket{0}$, and normalization $\braket{C_{\tilde{\vt{q}}}}{C_{\vt{\tilde{q}}'}} = \delta_{\tilde{\vt{q}} - \vt{\tilde{q}}'}$ where $\tilde{\vt{q}},\vt{\tilde{q}}'$ are continuous variables because $\ket{C_{\tilde{\vt{q}}}}$ is an unbound state. The energy eigenvalue of this unbound exciton state must be
\begin{equation}
	E_{\tilde{\vt{q}}} = E_g + \frac{\hbar^2 \tilde{q}^2}{2 m_r}.
\end{equation}
We note that this is quite similar to the energy of a band pair state of an electron and a hole: $E_{\vt{q}} = E_g + \frac{\hbar^2 q^2}{2 m_r}$. Although $\tilde{\vt{q}}$ is not directly related to the crystal momentum $\vt{q}$, we suggest the replacement $\tilde{\vt{q}} \rightarrow \vt{q}$ and $\phi_{\tilde{q}} (\vt{q}) = \delta_{\tilde{\vt{q}}, \vt{q}}$ such that
\begin{equation}
	C^\dag_{\vt{q}} = \alpha^\dag_{\vt{q}} \beta^\dag_{-\vt{q}}. \label{eq:unbound-creation}
\end{equation}
We then propose to approximate the unbound exciton state $\ket{C(\vt{q})}$ with the band pair state $\ket{\vt{q}, -\vt{q}}$ such that $\ket{C(\vt{q})} \approx \ket{\vt{q}, -\vt{q}}$.

The orthogonality from the bound state is then
\begin{equation}
	\braket{C_{\vt{q}}}{\mathrm{x}_\nu} = \sum_{\vt{q}'} \psi_\nu (\vt{q}') \bra{0} \alpha_{\vt{q}} \beta_{\vt{q}} \alpha^\dag_{\vt{q}'} \beta^\dag_{\vt{q}'} \ket{0} = \psi_\nu (\vt{q}),
\end{equation}
where we used the usual anticommutation rule for $\alpha_{\vt{q}}$ and $\beta_{\vt{q}}$, and the notation $\ket{C_{\vt{q}}} = C^\dag_{\vt{q}} \ket{0}$, $\ket{\mathrm{x}_\nu} = B^\dag_{\nu} \ket{0}$. We note that $\psi_\nu (\vt{q}) \sim a_0/\sqrt{A} \sim 1/\sqrt{N}$ where $N$ is the number of the unit cells in the sample. Hence, for a sufficiently large sample, we obtain the approximate orthogonality $\braket{C_{\vt{q}}}{\mathrm{x}_\nu} \sim 1 / \sqrt{N} \rightarrow 0$. The normalization is also easily obtained as
\begin{equation}
	\braket{C_{\vt{q}}}{C_{\vt{q}'}} = \bra{0} \alpha_{\vt{q}} \beta_{\vt{q}} \alpha^\dag_{\vt{q}'} \beta^\dag_{\vt{q}'} \ket{0} = \delta_{\vt{q}, \vt{q}'}.
\end{equation}
In addition, the energy is the same with the replacement $\tilde{\vt{q}} \rightarrow \vt{q}$. Hence, we conclude that, for a sufficiently large sample size $A$, the creation operator in equation \eqref{eq:unbound-creation} is approximately correct.

The operator $C^\dag_{\vt{q}}$ excites the electron in the valence bands to the conduction band. Hence, we can interpret as $C^\dag_{\vt{q}} = \left(\otimes_{\vt{q}' \neq \vt{q}} I_{\vt{q}'} \right) \otimes \ketbra{c_{\vt{q}}}{v_{\vt{q}}}$, where $\ket{c_{\vt{q}}}, \ket{v_{\vt{q}}}$ are the single electron Bloch states at the conduction and the valence band, respectively, with a momentum $\hbar \vt{q}$, and $I_{\vt{q}'} = \ketbra{c_{\vt{q}'}}{c_{\vt{q}'}} + \ketbra{v_{\vt{q}'}}{v_{\vt{q}'}}$. Using this representation, the anti-commutation rule for the bound and the unbound exciton creation and annihilation operators are easily obtained: $\{C_{\vt{q}}, C_{\vt{q}'}^\dag \} = \delta_{\vt{q}, \vt{q}'}$, $\{B_{\nu}, B_{\nu'}^\dag \} = \delta_{\nu, \nu'}$, $\{ C^\dag_{\vt{q}}, B_\nu \} \sim 1/ \sqrt{N} \rightarrow 0$ while all other anti-commutators are zero. It is also noteworthy that the Hilbert subspace for the single excitation is spanned by the bound and the unbound exciton states $\{ \ket{\mathrm{x}_\nu}, \ket{C_{\vt{q}}} \}$ with all possible $\nu$ and $\vt{q}$, and thus, the completeness in this single excitation subspace is
\begin{equation}
	\sum_{\nu} \ketbra{\mathrm{x}_\nu}{\mathrm{x}_\nu} + \sum_{\vt{q}} \ketbra{C_{\vt{q}}}{C_{\vt{q}}} = \vt{1}.
\end{equation}

\section{Density operator matrix elements} \label{sec:matrix-rho}
In this section, we present derivations of density matrix elements that are used in the main text. 

\subsection{Second-harmonic generation}

We resolve the matrix elements $\rho^{(2)}_{(s,1)0}, \rho^{(2)}_{(0,0)(s,1)},$ and $\rho^{(2)}_{(0,0)0}$ with $s=$ 1 or 2, with the polarization configuration of $\sigma^+;\sigma^-\sigma^-$, corresponding to the frequencies $2 \omega_\kappa$, $\omega_\kappa$, $\omega_\kappa,$ respectively. Recall that the interaction Hamiltonian is given as $\mathcal{H}_I + \mathcal{H}'_I$, where $\mathcal{H}_I$ is  given in equation \eqref{eq:interaction-hamiltonian} with an interaction coefficient $g^-_{(s,1)}$ and $\mathcal{H}'_I$ is given in equation \eqref{eq:interact3}, respectively. 

We first resolve $\rho^{(2)}_{(s,1)0}$ by considering
\begin{align}
	&\dot{\rho}^{(2)}_{(s,1) 0} (t) = \nonumber \\
	&- i e_{(s,1)} \rho^{(2)}_{(s,1) 0} (t) - \frac{i}{\hbar} \bra{\text{x}_{(s,1)}}[\mathcal{H}_I + \mathcal{H}_I', \rho^{(1)}] \ket{0}.
\end{align}
with the useful fact that the only nonzero matrix elements in $\rho^{(1)}$ in this situation are
\begin{align}
	\rho^{(1)}_{(s,1)0} &= \bra{\mathrm{x}_{(s,1)}} \rho^{(1)} \ket{0} =\frac{g^-_{(s,1)}}{\hbar} \frac{\varepsilon (\kappa) \mathrm{e}^{-i \omega_\kappa t}}{(e_{s} - \omega_\kappa - i \epsilon)} \label{eq:rho1s1}
\end{align}
and its complex conjugate, where $s = 1,2$. Using this, we insert the completeness relation and obtain
\begin{align}
	\bra{\text{x}_{(s,1)}} [\mathcal{H}_I &+ \mathcal{H}'_I,  \rho^{(1)}] \ket{0} = \nonumber \\
	& \bra{\mathrm{x}_{(s,1)}} \mathcal{H}_I + \mathcal{H}'_I \ketbra{\mathrm{x}_{(s,1)}}{\mathrm{x}_{(s,1)}} \rho^{(1)} \ket{0} \nonumber \\
	&- \bra{\mathrm{x}_{(s,1)}} \rho^{(1)} \ketbra{0}{0}  \mathcal{H}_I + \mathcal{H}'_I\ket{0}.
\end{align}
Both terms are zero since the diagonal matrix elements for $\mathcal{H}_I$ and $\mathcal{H}'_I$ are zero. This implies that $\rho^{(2)}_{(s,1)0} (t) = 0$ since $\rho^{(2)}_{(s,1)0} (-\infty) = 0$. 

Next, let us consider the matrix element $\rho^{(2)}_{(0,0)(s,1)}$. To calculate this, we evaluate the following commutator, using again that the only nonzero matrix elements of $\rho^{(1)}$ are $\rho^{(1)}_{(s,1)0}$ and its conjugate:
\begin{align}
	\bra{\text{x}_{(0,0)}} [\mathcal{H}_I &+ \mathcal{H}'_I,  \rho^{(1)}] \ket{\mathrm{x}_{(s,1)}} = \nonumber \\
	& \bra{\mathrm{x}_{(0,0)}} \mathcal{H}_I + \mathcal{H}'_I \ketbra{0}{0} \rho^{(1)} \ket{\mathrm{x}_{(s,1)}} \nonumber \\
	&- \bra{\mathrm{x}_{(0,0)}} \rho^{(1)}(\mathcal{H}_I + \mathcal{H}'_I) \ket{\mathrm{x}_{(s,1)}}.
\end{align}
The first term is zero since the only nonzero matrix elements for $\mathcal{H}_I$ are $\bra{\mathrm{x}_{(s,1)}} \mathcal{H}_I \ket{0}$ and its complex conjugate, and the only nonzero matrix elements for $\mathcal{H}'_I$ are $\bra{\mathrm{x}_{(s,-1)}} \mathcal{H}'_I \ket{\mathrm{x}_{(0,0)}}$ where $s = 1, 2, \cdots$ and its complex conjugate. The second term is zero since the bra $\bra{\mathrm{x}_{(0,0)}}$ eliminates $\rho^{(1)}$. This implies that the matrix element $\rho^{(2)}_{(0,0)(s,1)}$ is zero. 

Finally, we resolve the matrix element $\rho^{(2)}_{(0,0)0}$ by solving
\begin{align}
	&\dot{\rho}^{(2)}_{(0,0) 0} (t) = \nonumber \\
	&- i e_{(0,0)} \rho^{(2)}_{(0,0) 0} (t) - \frac{i}{\hbar} \bra{\text{x}_{(0,0)}}[\mathcal{H}_I + \mathcal{H}_I', \rho^{(1)}] \ket{0}. \label{eq:rhox01}
\end{align}
We calculate the following
\begin{align}
	\bra{\text{x}_{(0,0)}} [\mathcal{H}_I &+ \mathcal{H}'_I,  \rho^{(1)}] \ket{0} = \nonumber \\
	& \bra{\mathrm{x}_{(0,0)}} \mathcal{H}_I + \mathcal{H}'_I \ketbra{\mathrm{x}_{(s,1)}}{\mathrm{x}_{(s,1)}} \rho^{(1)} \ket{0} \nonumber \\
	&- \bra{\mathrm{x}_{(0,0)}} \rho^{(1)}(\mathcal{H}_I + \mathcal{H}'_I) \ket{0}.
\end{align}	
The second term is zero since the bra $\bra{\mathrm{x}_{(0,0)}}$ eliminates $\rho^{(1)}$. We calculate the following:
\begin{align}
	&\bra{\mathrm{x}_{(0,0)}} \mathcal{H}_I + \mathcal{H}'_I \ket{\mathrm{x}_{(s,1)}} = \bra{\mathrm{x}_{(0,0)}} \mathcal{H}'_I \ket{\mathrm{x}_{(s,1)}} \nonumber \\
	&= -\sum_{s'=1,2} h^-_{(0,0)(s,1)} \varepsilon(\kappa) \mathrm{e}^{- i \omega_\kappa t} \rho^{(1)}_{(s,1)0}. 
\end{align}
Integrating the equation \eqref{eq:rhox01} using above and equation \eqref{eq:rho1s1}, we obtain the final result in equation \eqref{eq:rho2}. 
	
\subsection{Third-harmonic generation}

We resolve the matrix elements of $\rho^{(3)}$ in the third-harmonic generation with degenerate fundamental frequencies, but with polarization configuration of $\sigma^+;\sigma^+\sigma^-\sigma^+$, corresponding to the frequencies $3 \omega_\kappa, \omega_\kappa, \omega_\kappa, \omega_\kappa$, respectively. The relevant basis for the matrix elements is $\{ \ket{0}, \ket{(s, \pm 1)}, \ket{(s', 0)} \}$ where $s = 1,2, \cdots$ and $s' = 0,1, \cdots$. We will intensively use the selection rules (dipole moments) given in Table \ref{tab:gnu} and equation \eqref{eq:selection-rule-exciton-levels}. 

We first resolve the matrix elements for $\rho^{(1)}$. For $\sigma^+$ polarization, we find the only nonzero matrix element for $\rho^{(1)}$ to be
\begin{equation}
	\rho^{(1)+}_{(s,0)0} = \bra{\mathrm{x}_{(s,0)}} \rho^{(1)} \ket{0} =\frac{g^+_{(s,0)}}{\hbar} \frac{\varepsilon (\kappa) \mathrm{e}^{-i \omega_\kappa t}}{(e_{s} - \omega_\kappa - i \epsilon)}, \label{eq:rho1plus}
\end{equation}
where $s = 0,1,2, \cdots$. 
	
Next, we will resolve the matrix elements of $\rho^{(2)}$. For $\sigma^-\sigma^+$ polarization sequence, the first order process landed on $\ket{\mathrm{x}_{(s,0)}}$. Then, the second driving from $\sigma^-$ light will bring the state to $\ket{\mathrm{x}_{(s',-1)}}$ with $s'=1,2, \cdots$. Hence, the only nonzero matrix element is $\rho^{(2)-+}_{(s',-1)0}$, which is obtained by calculating the following commutator:
\begin{align}
	&\bra{\mathrm{x}_{(s',-1)}} [\mathcal{H}_I^- + \mathcal{H}^{'-}_I, \rho^{(1)+}] \ket{0} \nonumber \\
	&= 	\bra{\mathrm{x}_{(s',-1)}} (\mathcal{H}_I^- + \mathcal{H}^{'-}_I) \ketbra{\mathrm{x}_{(s,0)}}{\mathrm{x}_{(s,0)}} \rho^{(1)+} \ket{0} \nonumber \\
	&~~~~~- 	\bra{\mathrm{x}_{(s',-1)}} \rho^{(1)+}(\mathcal{H}_I^- + \mathcal{H}^{'+}_I)\ket{0} \nonumber \\
	&= \bra{\mathrm{x}_{(s',-1)}}  \mathcal{H}^{'-}_I \ket{\mathrm{x}_{(s,0)}} \rho^{(1)+}_{(s,0)0} \nonumber \\
	&= - h^-_{(s',-1)(s,0)} \rho^{(1)+}_{(s,0)0} \mathcal{E} (\kappa) \mathrm{e}^{- i \omega_\kappa t},
\end{align}
where the second term in the first equation is zero since $\bra{\mathrm{x}_{(s',-1)}}$ eliminates $\rho^{(1)+}$. Here, we clarified that the interaction Hamiltonian is due to $\sigma^-$ light. In the second equation, we used the fact that $\mathcal{H}^{'+}_I$ connects $\ket{\mathrm{x}_{(s,0)}}$ and $\bra{\mathrm{x}_{(s',-1)}}$. Then, after integration we obtain the matrix element
\begin{align}
	&\rho^{(2)-+ }_{(s',-1)0} = \nonumber \\
	&\frac{h^-_{(s',-1)(s,0)}g^+_{(s,0)} }{\hbar^2} \frac{\varepsilon^2(\kappa) \mathrm{e}^{- 2 i \omega_\kappa t}}{(e_{s} - \omega_\kappa - i \epsilon)(e_{s'} - 2 \omega_\kappa - i \epsilon')}.
\end{align}

Next, we resolve the matrix elements for $\rho^{(3)+-+}$. Because we are solving for the third-harmonic generation, we look for the matrix elements proportional to $\mathrm{e}^{-i 3 \omega_\kappa t}$. 

It is obvious to see that one nonzero matrix element for $\rho^{(3)+-+}$ is $\rho^{(3)+-+}_{(0,0)0}$ which is given by
\begin{align}
	&\rho^{(3)+-+ }_{(0,0)0} = \sum_{s', s} \frac{h^+_{(0,0)(s',-1)}h^-_{(s',-1)(s,0)}g^+_{(s,0)} }{\hbar^3} \times \nonumber \\
	& \frac{\varepsilon^3(\kappa) \mathrm{e}^{- 3 i \omega_\kappa t}}{(e_{s} - \omega_\kappa - i \epsilon)(e_{s'} - 2 \omega_\kappa - i \epsilon')(e_{0} - 3 \omega_\kappa - i \epsilon'')}.
\end{align}
To calculate another nonzero matrix element, we consider the light with frequency $\omega'_\kappa$, which we will set later $\omega'_\kappa = - \omega_\kappa$. Consider the commutator for the matrix element $\rho^{(3)+-+}_{(s',-1)(0,0)}$:
\begin{align}
	&\bra{\mathrm{x}_{(s,-1)}} [\mathcal{H}_I^+ + \mathcal{H}^{'+}_I, \rho^{(2)-+}] \ket{\mathrm{x}_{(0,0)}} \nonumber \\
	&= 	\bra{\mathrm{x}_{(s,-1)}} (\mathcal{H}_I^+ + \mathcal{H}^{'+}_I) \rho^{(2)-+} \ket{\mathrm{x}_{(0,0)}} \nonumber \\
	&~~~ - 	\bra{\mathrm{x}_{(s,-1)}} \rho^{(2)-+} \ketbra{0}{0}(\mathcal{H}_I^+ + \mathcal{H}^{'+}_I)\ket{\mathrm{x}_{(0,0)}} \nonumber \\
	&= - \rho^{(2)-+}_{(s,-1)0}  \bra{0}  \mathcal{H}^{+}_I \ket{\mathrm{x}_{(0,0)}}  \nonumber \\
	&= g^{+*}_{(0,0)} \rho^{(2)-+}_{(s,-1)0} \mathcal{E}^* (\kappa') \mathrm{e}^{i \omega'_\kappa t}.
\end{align}
We first integrate the differential equation. Then, setting $\omega'_\kappa = -\omega_\kappa$, we obtain a result proportional to $\mathrm{e}^{-i 3 \omega_\kappa t}$:
\begin{align}
	&\rho^{(3)+-+ }_{(s',-1)(0,0)} = -\sum_{s', s} \frac{g^*_{(0,0)}h^-_{(s',-1)(s,0)}g^+_{(s,0)} }{\hbar^3} \times \nonumber \\
	& \frac{\varepsilon^3(\kappa) \mathrm{e}^{- 3 i \omega_\kappa t}}{(e_{s} - \omega_\kappa - i \epsilon)(e_{s'} - 2 \omega_\kappa - i \epsilon')(e_{s'} - e_0 + \omega_\kappa + i \epsilon'')}.
\end{align}
We note that this is a nonresonant contribution due to the factor $e_0 + \omega_\kappa + i \epsilon''$ in the denominator. 

Finally, we calculate the matrix element $\rho^{(3)+-+}_{(s',-1)0}$. For this, we calculate the following commutator:
\begin{align}
	&\bra{\mathrm{x}_{(s,-1)}} [\mathcal{H}_I^+ + \mathcal{H}^{'+}_I, \rho^{(2)-+}] \ket{0} \nonumber \\
	&= 	\bra{\mathrm{x}_{(s,-1)}} (\mathcal{H}_I^+ + \mathcal{H}^{'+}_I) \ketbra{\mathrm{x}_{(s,-1)}}{\mathrm{x}_{(s,-1)}} \rho^{(2)-+} \ket{0} \nonumber \\
	&~~~ - 	\bra{\mathrm{x}_{(s,-1)}} \rho^{(2)-+} \ketbra{0}{0}(\mathcal{H}_I^+ + \mathcal{H}^{'+}_I)\ket{0}.
\end{align}
Both terms are zero since $\mathcal{H}_I^+$ and $\mathcal{H}_I^{'+}$ have nonzero elements only on off-diagonal. This implies that $\rho^{(3)+-+}_{(s',-1)0}(t) = 0$. 

\subsection{Two-photon transition}

We resolve the matrix elements of $\rho^{(3)}$ for the two-photon transition with degenerate fundamental frequencies, with polarization configuration of $\sigma^+;\sigma^+\sigma^+\sigma^+$, corresponding to $2\omega_\kappa, \omega_\kappa, \omega_\kappa, - \omega_\kappa$, respectively. We present the result for the case where $2 \omega_\kappa \sim e_{(1,1)}$. The relevant basis for the matrix elements is $\{ \ket{0}, \ket{\mathrm{x}_{(s,0)}},  \ket{ \mathrm{x}_{(1,1)}} \}$, where $s = 0,1,2, \cdots$. 

The only nonzero matrix element for $\rho^{(1)}$ is given in equation \eqref{eq:rho1plus}. We now resolve the matrix elements of $\rho^{(2)}$. The first-order process landed on the state $\ket{\mathrm{x}_{(s,0)}}$ with $s = 0,1,2, \cdots$. According to the selection rule, the second-order process with the polarization sequence $\sigma^+\sigma^+$ needs to land on $\ket{ \mathrm{x}_{(s',1)}}$ with $s' = 1,2, \cdots$ via $\mathrm{e}^{-i \omega_\kappa t}$ term in $\mathcal{H}'_I$, or on $\ket{0}$ via $\mathrm{e}^{i \omega_\kappa t}$ term in $\mathcal{H}_I$. Let us consider $\rho^{(2) ++}_{(s',1)0}$ first. For this, let us calculate the commutator:
\begin{align}
	&\bra{\mathrm{x}_{(s',1)}} [\mathcal{H}_I^+ + \mathcal{H}^{'+}_I, \rho^{(1)+}] \ket{0} \nonumber \\
	&= 	\bra{\mathrm{x}_{(s',1)}} (\mathcal{H}_I^+ + \mathcal{H}^{'+}_I) \ketbra{\mathrm{x}_{(s,0)}}{\mathrm{x}_{(s,0)}} \rho^{(1)+} \ket{0} \nonumber \\
	&~~~~~- 	\bra{\mathrm{x}_{(s',1)}} \rho^{(1)+}(\mathcal{H}_I^+ + \mathcal{H}^{'+}_I)\ket{0} \nonumber \\
	&= \bra{\mathrm{x}_{(s',1)}}  \mathcal{H}^{'+}_I \ket{\mathrm{x}_{(s,0)}} \rho^{(1)+}_{(s,0)0} \nonumber \\
	&= - h^+_{(s',1)(s,0)} \rho^{(1)+}_{(s,0)0} \mathcal{E} (\kappa) \mathrm{e}^{- i \omega_\kappa t}.
\end{align}	
From this, it easily follows that
\begin{align}
	&\rho^{(2)++ }_{(1,1)0} = \nonumber \\
	&\frac{h^+_{(1,1)(s,0)}g^+_{(s,0)} }{\hbar^2} \frac{\varepsilon^2(\kappa) \mathrm{e}^{- 2 i \omega_\kappa t}}{(e_s - \omega_\kappa - i \epsilon)(e_1 - 2 \omega_\kappa - i \epsilon')}.
\end{align}	
We then consider $\rho^{(2)}_{00}$. For this, let us calculate the commutator:
\begin{align}
	&\bra{0} [\mathcal{H}_I^+ + \mathcal{H}^{'+}_I, \rho^{(1)+}] \ket{0} \nonumber \\
	&= 	\bra{0} (\mathcal{H}_I^+ + \mathcal{H}^{'+}_I) \ketbra{\mathrm{x}_{(s,0)}}{\mathrm{x}_{(s,0)}} \rho^{(1)+} \ket{0} \nonumber \\
	&~~~~~- 	\bra{0} \rho^{(1)+} \ketbra{\mathrm{x}_{(s,0)}}{\mathrm{x}_{(s,0)}} (\mathcal{H}_I^+ + \mathcal{H}^{'+}_I)\ket{0} \nonumber \\
	&= \bra{0}  \mathcal{H}^{'+}_I \ket{\mathrm{x}_{(s,0)}} \rho^{(1)+}_{(s,0)0} - \mathrm{h.c.} \nonumber \\
	&= - g^{+*}_{(s,0)} \rho^{(1)+}_{(s,0)0} \mathcal{E}^* (\kappa) \mathrm{e}^{i \omega_\kappa t} - \mathrm{h.c.}
\end{align}
These terms are DC drives, which is proportional to an infinitesimal constant $\epsilon$, and hence, is negligible mathematically. Hence, the only significant nonzero elements of $\rho^{(2)}$ are $\rho^{(2)++}_{(1,1)0}$ and its complex conjugate. 

Next, we resolve the matrix elements for $\rho^{(3)+++}$. Our task is to find the matrix elements proportional to $\mathrm{e}^{- i \omega_\kappa t}$. The last frequency is negative: $-\omega_\kappa$, i.e., moving downward in energy. Since the second-order process landed on the state $\ket{\mathrm{x}_{(1,1)}}$, the third-order process must land on a state $\ket{\mathrm{x}_{(s'',0)}}$. One nonzero matrix element is thus given by
\begin{align}
	&\rho^{(3)+++ }_{(s'',0)0} = \sum_{ s} \frac{h^{+*}_{(1,1)(s'',0)} h^+_{(1,1)(s,0)} g^+_{(s,0)} }{\hbar^3} \times \nonumber \\
	& \frac{|\varepsilon(\kappa)|^2 \varepsilon(\kappa) \mathrm{e}^{- i \omega_\kappa t}}{(e_{s} - \omega_\kappa - i \epsilon)(e_{1} - 2 \omega_\kappa - i \epsilon')(e_{s''} + \omega_\kappa + i \epsilon'')}.
\end{align}
This is a nonresonant term, due to the last factor in the denominator. 

We can find another nonzero matrix element $\rho^{(3)+++}_{(1,1)(s'',0)}$ as follows. Let us consider the commutator:
\begin{align}
	&\bra{\mathrm{x}_{(1,1)}} [\mathcal{H}_I^+ + \mathcal{H}^{'+}_I, \rho^{(2)++}] \ket{\mathrm{x}_{(s'',0)}} \nonumber \\
	&= 	\bra{\mathrm{x}_{(1,1)}} (\mathcal{H}_I^+ + \mathcal{H}^{'+}_I) \rho^{(2)-+} \ket{\mathrm{x}_{(s'',0)}} \nonumber \\
	&~~~ - 	\bra{\mathrm{x}_{(1,1)}} \rho^{(2)++} \ketbra{0}{0}(\mathcal{H}_I^+ + \mathcal{H}^{'+}_I)\ket{\mathrm{x}_{(s'',0)}} \nonumber \\
	&= -\rho^{(2)++}_{(1,1)0}  \bra{0}  \mathcal{H}^{+}_I \ket{\mathrm{x}_{(s'',0)}}  \nonumber \\
	&= g^{+*}_{(s'',0)} \rho^{(2)++}_{(1,1)0} \mathcal{E}^* (\kappa) \mathrm{e}^{i \omega_\kappa t}.
\end{align}
After integrating the differential equation, we obtain
\begin{align}
	&\rho^{(3)+++ }_{(1,1)(s'',0)} = -\sum_{ s} \frac{g^{+*}_{(s'',0)} h^+_{(1,1)(s,0)} g^+_{(s,0)} }{\hbar^3} \times \nonumber \\
	& \frac{|\varepsilon(\kappa)|^2 \varepsilon(\kappa) \mathrm{e}^{- i \omega_\kappa t}}{(e_{s} - \omega_\kappa - i \epsilon)(e_{1} - 2 \omega_\kappa - i \epsilon')(e_{1} - e_{s''} - \omega_\kappa - i \epsilon'')}.
\end{align}
This is a resonant term. 

Finally, let us consider the matrix element $\rho^{(3)+++}_{(1,1)0}$. Let us consider the commutator
\begin{align}
	&\bra{\mathrm{x}_{(1,1)}} [\mathcal{H}_I^+ + \mathcal{H}^{'+}_I, \rho^{(2)++}] \ket{0} \nonumber \\
	&= 	\bra{\mathrm{x}_{(1,1)}} (\mathcal{H}_I^+ + \mathcal{H}^{'+}_I) \ketbra{\mathrm{x}_{(1,1)}}{\mathrm{x}_{(1,1)}} \rho^{(2)-+} \ket{0} \nonumber \\
	&~~~ - 	\bra{\mathrm{x}_{(1,1)}} \rho^{(2)++} \ketbra{0}{0}(\mathcal{H}_I^+ + \mathcal{H}^{'+}_I)\ket{0}.
\end{align}
Both terms are zero since $\mathcal{H}_I^+$ and $\mathcal{H}_I^{'+}$ have nonzero elements only on off-diagonal. This implies that $\rho^{(3)+++}_{(1,1)0} (t) = 0$.

\section{Calculation of the dipole moment $f^\pm_{\nu} (\mathbf{q}) = e \bra{ \mathrm{x}_\nu } \mathbf{r} \cdot \hat{\mathbf{\varepsilon}}^\pm \ket{ C_{\mathbf{q}} }$} \label{sec:fnu}

Let us calculate the dipole moment $f^+_\nu (\vt{q})$ using the Blount formula in equation \eqref{eq:blount} and the analytical solution in equation \eqref{eq:uqv}:
\begin{align}
f^+_\nu(\vt{q}) =& - \hat{\vt{\varepsilon}}^+ \cdot i e \sum_{\vt{q}'} \psi^*_\nu (\vt{q}') \vt{\nabla}_{\vt{q}}\braket{c_{\vt{q}'}}{c_{\vt{q}}}\nonumber \\
& - i e \sum_{\vt{q}}\psi^*_\nu (\vt{q}) (1 + \tau) \frac{\hbar^2 v^2}{\sqrt{2}\Delta^2} q \mathrm{e}^{i \tau \phi_q}. \label{eq:fnu1}
\end{align}
The contribution coming from the second term on the right hand side is negligible due to the angular integral, if $\nu = (n,0)$. We then calculate the contribution from the first term. To evaluate this, let us use the following integration by parts.
\begin{align}
&\int_A \td^2 q  \sum_{\vt{q}'} \left( r(\vt{q}')\vt{\nabla}_{\vt{q}} \braket{\psi_{\vt{q}', \lambda}}{\psi_{\vt{q}, \lambda}} \right) s(\vt{q}) \nonumber \\
&= \int_{\partial A} \td \vt{n} r(\vt{q}) s(\vt{q}) - \int_A \td^2 q r(\vt{q})\vt{\nabla}_{\vt{q}} s(\vt{q}).
\end{align}
The first term is the boundary line integral. The contribution from the second term above vanishes due to the angular integral over $\phi_q$. This leads to
\begin{align}
	&\chi^{(2)}_U (\omega_\kappa \sim e_0/2) = \nonumber \\
	&-\hat{\vt{\varepsilon}}^+ \cdot \int_{\partial \textrm{FBZ}} \td \vt{n} \left( \begin{array}{l} \psi^{*}_\nu (\vt{q})ie \frac{e_\nu d^+_{cv} (\vt{q}) g_\nu^*}{8 \pi^2 \omega_\kappa \epsilon_0 \hbar^2 d_\textrm{eff}} \\
\times \frac{1}{(\omega_{\vt{q}}- \omega_\kappa - i (\gamma_U/2))(e_\nu - 2 \omega_\kappa - i (\gamma_B/2))}  \end{array} \right)
\end{align}

Performing the boundary line integral involves multiplying the factor $\mathrm{e}^{i \phi_q}$ since $\hat{\vt{\varepsilon}}^+ \cdot \hat{\vt{n}} = \mathrm{e}^{i \phi_q}$. Recall that the threefold rotational symmetry is perturbatively treated. The zeroth order that does not have the threefold rotational symmetry integrates to zero over $\phi_q$. The higher-order perturbative terms involving the threefold rotational symmetry also vanishes as follows. Recall that the energy $\hbar \omega_{\vt{q}}$ also has the threefold rotational symmetry. Hence, the higher order terms in the integrand have a threefold rotational symmetry. We note that
\begin{equation}
	\int_0^{2\pi} \td \phi \text{e}^{\pm i \phi} f(\cos(3 \phi)) = 0,
\end{equation}
where $f$ is any analytical function. It is easily seen by considering $\cos(3 \phi) = (1/2) (\mathrm{e}^{i 3\phi} + \mathrm{e}^{-i 3 \phi})$, and the Taylor series term $\cos^n (3 \phi)$ involves $\mathrm{e}^{\pm i 3m \phi}$ with an integer $m$ that the integral of $f(\cos(3 \phi))$ over $\phi$ after multiplying with $\mathrm{e}^{\pm i \phi}$ vanishes. This leads to a conclusion that, regardless of the polarization, this boundary integral term must be zero. Therefore, $\chi^{(2)}_U (\omega_\kappa \sim e_0/2)$ vanishes.

For $\sigma_-$ input polarization, the second term in equation \eqref{eq:fnu1} vanishes, and the boundary line integral is the same result, hence, the contribution from the unbound exciton also vanishes for $\sigma_-$ light. Overall, we conclude that the unbound exciton does not efficiently couple back to the bound exciton states. This allows us to ignore any channel of the unbound exciton virtual states to land on a bound exciton state.

\nocite{*}

\bibliography{MoS2}

\begin{thebibliography}{74}%
\makeatletter
\providecommand \@ifxundefined [1]{%
 \@ifx{#1\undefined}
}%
\providecommand \@ifnum [1]{%
 \ifnum #1\expandafter \@firstoftwo
 \else \expandafter \@secondoftwo
 \fi
}%
\providecommand \@ifx [1]{%
 \ifx #1\expandafter \@firstoftwo
 \else \expandafter \@secondoftwo
 \fi
}%
\providecommand \natexlab [1]{#1}%
\providecommand \enquote  [1]{``#1''}%
\providecommand \bibnamefont  [1]{#1}%
\providecommand \bibfnamefont [1]{#1}%
\providecommand \citenamefont [1]{#1}%
\providecommand \href@noop [0]{\@secondoftwo}%
\providecommand \href [0]{\begingroup \@sanitize@url \@href}%
\providecommand \@href[1]{\@@startlink{#1}\@@href}%
\providecommand \@@href[1]{\endgroup#1\@@endlink}%
\providecommand \@sanitize@url [0]{\catcode `\\12\catcode `\$12\catcode
  `\&12\catcode `\#12\catcode `\^12\catcode `\_12\catcode `\%12\relax}%
\providecommand \@@startlink[1]{}%
\providecommand \@@endlink[0]{}%
\providecommand \url  [0]{\begingroup\@sanitize@url \@url }%
\providecommand \@url [1]{\endgroup\@href {#1}{\urlprefix }}%
\providecommand \urlprefix  [0]{URL }%
\providecommand \Eprint [0]{\href }%
\providecommand \doibase [0]{http://dx.doi.org/}%
\providecommand \selectlanguage [0]{\@gobble}%
\providecommand \bibinfo  [0]{\@secondoftwo}%
\providecommand \bibfield  [0]{\@secondoftwo}%
\providecommand \translation [1]{[#1]}%
\providecommand \BibitemOpen [0]{}%
\providecommand \bibitemStop [0]{}%
\providecommand \bibitemNoStop [0]{.\EOS\space}%
\providecommand \EOS [0]{\spacefactor3000\relax}%
\providecommand \BibitemShut  [1]{\csname bibitem#1\endcsname}%
\let\auto@bib@innerbib\@empty
\bibitem [{\citenamefont {Sipahigil}\ \emph
  {et~al.}(2016{\natexlab{a}})\citenamefont {Sipahigil}, \citenamefont {Evans},
  \citenamefont {Sukachev}, \citenamefont {Burek}, \citenamefont {Borregaard},
  \citenamefont {Bhaskar}, \citenamefont {Nguyen}, \citenamefont {Pacheco},
  \citenamefont {Atikian}, \citenamefont {Meuwly} \emph
  {et~al.}}]{sipahigil2016integrated}%
  \BibitemOpen
  \bibfield  {author} {\bibinfo {author} {\bibfnamefont {A.}~\bibnamefont
  {Sipahigil}}, \bibinfo {author} {\bibfnamefont {R.}~\bibnamefont {Evans}},
  \bibinfo {author} {\bibfnamefont {D.}~\bibnamefont {Sukachev}}, \bibinfo
  {author} {\bibfnamefont {M.}~\bibnamefont {Burek}}, \bibinfo {author}
  {\bibfnamefont {J.}~\bibnamefont {Borregaard}}, \bibinfo {author}
  {\bibfnamefont {M.}~\bibnamefont {Bhaskar}}, \bibinfo {author} {\bibfnamefont
  {C.}~\bibnamefont {Nguyen}}, \bibinfo {author} {\bibfnamefont
  {J.}~\bibnamefont {Pacheco}}, \bibinfo {author} {\bibfnamefont
  {H.}~\bibnamefont {Atikian}}, \bibinfo {author} {\bibfnamefont
  {C.}~\bibnamefont {Meuwly}},  \emph {et~al.},\ }\href@noop {} {\bibfield
  {journal} {\bibinfo  {journal} {Science}\ }\textbf {\bibinfo {volume}
  {354}},\ \bibinfo {pages} {847} (\bibinfo {year}
  {2016}{\natexlab{a}})}\BibitemShut {NoStop}%
\bibitem [{\citenamefont {Alam}\ \emph {et~al.}(2016)\citenamefont {Alam},
  \citenamefont {De~Leon},\ and\ \citenamefont {Boyd}}]{alam2016large}%
  \BibitemOpen
  \bibfield  {author} {\bibinfo {author} {\bibfnamefont {M.~Z.}\ \bibnamefont
  {Alam}}, \bibinfo {author} {\bibfnamefont {I.}~\bibnamefont {De~Leon}}, \
  and\ \bibinfo {author} {\bibfnamefont {R.~W.}\ \bibnamefont {Boyd}},\
  }\href@noop {} {\bibfield  {journal} {\bibinfo  {journal} {Science}\ }\textbf
  {\bibinfo {volume} {352}},\ \bibinfo {pages} {795} (\bibinfo {year}
  {2016})}\BibitemShut {NoStop}%
\bibitem [{\citenamefont {Sipahigil}\ \emph
  {et~al.}(2016{\natexlab{b}})\citenamefont {Sipahigil}, \citenamefont {Evans},
  \citenamefont {Sukachev}, \citenamefont {Burek}, \citenamefont {Borregaard},
  \citenamefont {Bhaskar}, \citenamefont {Nguyen}, \citenamefont {Pacheco},
  \citenamefont {Atikian}, \citenamefont {Meuwly} \emph
  {et~al.}}]{sipahigil2016single}%
  \BibitemOpen
  \bibfield  {author} {\bibinfo {author} {\bibfnamefont {A.}~\bibnamefont
  {Sipahigil}}, \bibinfo {author} {\bibfnamefont {R.~E.}\ \bibnamefont
  {Evans}}, \bibinfo {author} {\bibfnamefont {D.~D.}\ \bibnamefont {Sukachev}},
  \bibinfo {author} {\bibfnamefont {M.~J.}\ \bibnamefont {Burek}}, \bibinfo
  {author} {\bibfnamefont {J.}~\bibnamefont {Borregaard}}, \bibinfo {author}
  {\bibfnamefont {M.~K.}\ \bibnamefont {Bhaskar}}, \bibinfo {author}
  {\bibfnamefont {C.~T.}\ \bibnamefont {Nguyen}}, \bibinfo {author}
  {\bibfnamefont {J.~L.}\ \bibnamefont {Pacheco}}, \bibinfo {author}
  {\bibfnamefont {H.~A.}\ \bibnamefont {Atikian}}, \bibinfo {author}
  {\bibfnamefont {C.}~\bibnamefont {Meuwly}},  \emph {et~al.},\ }\href@noop {}
  {\bibfield  {journal} {\bibinfo  {journal} {arXiv preprint arXiv:1608.05147}\
  } (\bibinfo {year} {2016}{\natexlab{b}})}\BibitemShut {NoStop}%
\bibitem [{\citenamefont {Benson}(2011)}]{benson2011assembly}%
  \BibitemOpen
  \bibfield  {author} {\bibinfo {author} {\bibfnamefont {O.}~\bibnamefont
  {Benson}},\ }\href@noop {} {\bibfield  {journal} {\bibinfo  {journal}
  {Nature}\ }\textbf {\bibinfo {volume} {480}},\ \bibinfo {pages} {193}
  (\bibinfo {year} {2011})}\BibitemShut {NoStop}%
\bibitem [{\citenamefont {Liu}\ \emph {et~al.}(2010)\citenamefont {Liu},
  \citenamefont {Osgood}, \citenamefont {Vlasov},\ and\ \citenamefont
  {Green}}]{liu2010mid}%
  \BibitemOpen
  \bibfield  {author} {\bibinfo {author} {\bibfnamefont {X.}~\bibnamefont
  {Liu}}, \bibinfo {author} {\bibfnamefont {R.~M.}\ \bibnamefont {Osgood}},
  \bibinfo {author} {\bibfnamefont {Y.~A.}\ \bibnamefont {Vlasov}}, \ and\
  \bibinfo {author} {\bibfnamefont {W.~M.}\ \bibnamefont {Green}},\ }\href@noop
  {} {\bibfield  {journal} {\bibinfo  {journal} {Nature Photonics}\ }\textbf
  {\bibinfo {volume} {4}},\ \bibinfo {pages} {557} (\bibinfo {year}
  {2010})}\BibitemShut {NoStop}%
\bibitem [{\citenamefont {Hacker}\ \emph {et~al.}(2016)\citenamefont {Hacker},
  \citenamefont {Welte}, \citenamefont {Rempe},\ and\ \citenamefont
  {Ritter}}]{hacker2016photon}%
  \BibitemOpen
  \bibfield  {author} {\bibinfo {author} {\bibfnamefont {B.}~\bibnamefont
  {Hacker}}, \bibinfo {author} {\bibfnamefont {S.}~\bibnamefont {Welte}},
  \bibinfo {author} {\bibfnamefont {G.}~\bibnamefont {Rempe}}, \ and\ \bibinfo
  {author} {\bibfnamefont {S.}~\bibnamefont {Ritter}},\ }\href@noop {}
  {\bibfield  {journal} {\bibinfo  {journal} {Nature}\ }\textbf {\bibinfo
  {volume} {536}},\ \bibinfo {pages} {193} (\bibinfo {year}
  {2016})}\BibitemShut {NoStop}%
\bibitem [{\citenamefont {Reimer}\ \emph {et~al.}(2016)\citenamefont {Reimer},
  \citenamefont {Kues}, \citenamefont {Roztocki}, \citenamefont {Wetzel},
  \citenamefont {Grazioso}, \citenamefont {Little}, \citenamefont {Chu},
  \citenamefont {Johnston}, \citenamefont {Bromberg}, \citenamefont {Caspani}
  \emph {et~al.}}]{reimer2016generation}%
  \BibitemOpen
  \bibfield  {author} {\bibinfo {author} {\bibfnamefont {C.}~\bibnamefont
  {Reimer}}, \bibinfo {author} {\bibfnamefont {M.}~\bibnamefont {Kues}},
  \bibinfo {author} {\bibfnamefont {P.}~\bibnamefont {Roztocki}}, \bibinfo
  {author} {\bibfnamefont {B.}~\bibnamefont {Wetzel}}, \bibinfo {author}
  {\bibfnamefont {F.}~\bibnamefont {Grazioso}}, \bibinfo {author}
  {\bibfnamefont {B.~E.}\ \bibnamefont {Little}}, \bibinfo {author}
  {\bibfnamefont {S.~T.}\ \bibnamefont {Chu}}, \bibinfo {author} {\bibfnamefont
  {T.}~\bibnamefont {Johnston}}, \bibinfo {author} {\bibfnamefont
  {Y.}~\bibnamefont {Bromberg}}, \bibinfo {author} {\bibfnamefont
  {L.}~\bibnamefont {Caspani}},  \emph {et~al.},\ }\href@noop {} {\bibfield
  {journal} {\bibinfo  {journal} {Science}\ }\textbf {\bibinfo {volume}
  {351}},\ \bibinfo {pages} {1176} (\bibinfo {year} {2016})}\BibitemShut
  {NoStop}%
\bibitem [{\citenamefont {Kockum}\ \emph {et~al.}(2017)\citenamefont {Kockum},
  \citenamefont {Miranowicz}, \citenamefont {Macr{\`\i}}, \citenamefont
  {Savasta},\ and\ \citenamefont {Nori}}]{kockum2017deterministic}%
  \BibitemOpen
  \bibfield  {author} {\bibinfo {author} {\bibfnamefont {A.~F.}\ \bibnamefont
  {Kockum}}, \bibinfo {author} {\bibfnamefont {A.}~\bibnamefont {Miranowicz}},
  \bibinfo {author} {\bibfnamefont {V.}~\bibnamefont {Macr{\`\i}}}, \bibinfo
  {author} {\bibfnamefont {S.}~\bibnamefont {Savasta}}, \ and\ \bibinfo
  {author} {\bibfnamefont {F.}~\bibnamefont {Nori}},\ }\href@noop {} {\bibfield
   {journal} {\bibinfo  {journal} {arXiv preprint arXiv:1701.05038}\ }
  (\bibinfo {year} {2017})}\BibitemShut {NoStop}%
\bibitem [{\citenamefont {Mabuchi}(2012)}]{mabuchi2012qubit}%
  \BibitemOpen
  \bibfield  {author} {\bibinfo {author} {\bibfnamefont {H.}~\bibnamefont
  {Mabuchi}},\ }\href@noop {} {\bibfield  {journal} {\bibinfo  {journal}
  {Physical Review A}\ }\textbf {\bibinfo {volume} {85}},\ \bibinfo {pages}
  {015806} (\bibinfo {year} {2012})}\BibitemShut {NoStop}%
\bibitem [{\citenamefont {Brasch}\ \emph {et~al.}(2016)\citenamefont {Brasch},
  \citenamefont {Geiselmann}, \citenamefont {Herr}, \citenamefont {Lihachev},
  \citenamefont {Pfeiffer}, \citenamefont {Gorodetsky},\ and\ \citenamefont
  {Kippenberg}}]{brasch2016photonic}%
  \BibitemOpen
  \bibfield  {author} {\bibinfo {author} {\bibfnamefont {V.}~\bibnamefont
  {Brasch}}, \bibinfo {author} {\bibfnamefont {M.}~\bibnamefont {Geiselmann}},
  \bibinfo {author} {\bibfnamefont {T.}~\bibnamefont {Herr}}, \bibinfo {author}
  {\bibfnamefont {G.}~\bibnamefont {Lihachev}}, \bibinfo {author}
  {\bibfnamefont {M.~H.}\ \bibnamefont {Pfeiffer}}, \bibinfo {author}
  {\bibfnamefont {M.~L.}\ \bibnamefont {Gorodetsky}}, \ and\ \bibinfo {author}
  {\bibfnamefont {T.~J.}\ \bibnamefont {Kippenberg}},\ }\href@noop {}
  {\bibfield  {journal} {\bibinfo  {journal} {Science}\ }\textbf {\bibinfo
  {volume} {351}},\ \bibinfo {pages} {357} (\bibinfo {year}
  {2016})}\BibitemShut {NoStop}%
\bibitem [{\citenamefont {Del~Haye}\ \emph {et~al.}(2007)\citenamefont
  {Del~Haye}, \citenamefont {Schliesser}, \citenamefont {Arcizet},
  \citenamefont {Wilkins}, \citenamefont {Holzwarth},\ and\ \citenamefont
  {Kippenberg}}]{del2007optical}%
  \BibitemOpen
  \bibfield  {author} {\bibinfo {author} {\bibfnamefont {P.}~\bibnamefont
  {Del~Haye}}, \bibinfo {author} {\bibfnamefont {A.}~\bibnamefont
  {Schliesser}}, \bibinfo {author} {\bibfnamefont {O.}~\bibnamefont {Arcizet}},
  \bibinfo {author} {\bibfnamefont {T.}~\bibnamefont {Wilkins}}, \bibinfo
  {author} {\bibfnamefont {R.}~\bibnamefont {Holzwarth}}, \ and\ \bibinfo
  {author} {\bibfnamefont {T.}~\bibnamefont {Kippenberg}},\ }\href@noop {}
  {\bibfield  {journal} {\bibinfo  {journal} {arXiv preprint arXiv:0708.0611}\
  } (\bibinfo {year} {2007})}\BibitemShut {NoStop}%
\bibitem [{\citenamefont {Guo}\ \emph {et~al.}(2016)\citenamefont {Guo},
  \citenamefont {Zou}, \citenamefont {Jung},\ and\ \citenamefont
  {Tang}}]{guo2016chip}%
  \BibitemOpen
  \bibfield  {author} {\bibinfo {author} {\bibfnamefont {X.}~\bibnamefont
  {Guo}}, \bibinfo {author} {\bibfnamefont {C.-L.}\ \bibnamefont {Zou}},
  \bibinfo {author} {\bibfnamefont {H.}~\bibnamefont {Jung}}, \ and\ \bibinfo
  {author} {\bibfnamefont {H.~X.}\ \bibnamefont {Tang}},\ }\href@noop {}
  {\bibfield  {journal} {\bibinfo  {journal} {Physical review letters}\
  }\textbf {\bibinfo {volume} {117}},\ \bibinfo {pages} {123902} (\bibinfo
  {year} {2016})}\BibitemShut {NoStop}%
\bibitem [{\citenamefont {Carlson}\ \emph {et~al.}(2017)\citenamefont
  {Carlson}, \citenamefont {Hickstein}, \citenamefont {Lind}, \citenamefont
  {Olson}, \citenamefont {Fox}, \citenamefont {Brown}, \citenamefont {Ludlow},
  \citenamefont {Li}, \citenamefont {Westly}, \citenamefont {Leopardi} \emph
  {et~al.}}]{carlson2017photonic}%
  \BibitemOpen
  \bibfield  {author} {\bibinfo {author} {\bibfnamefont {D.}~\bibnamefont
  {Carlson}}, \bibinfo {author} {\bibfnamefont {D.}~\bibnamefont {Hickstein}},
  \bibinfo {author} {\bibfnamefont {A.}~\bibnamefont {Lind}}, \bibinfo {author}
  {\bibfnamefont {J.}~\bibnamefont {Olson}}, \bibinfo {author} {\bibfnamefont
  {R.}~\bibnamefont {Fox}}, \bibinfo {author} {\bibfnamefont {R.}~\bibnamefont
  {Brown}}, \bibinfo {author} {\bibfnamefont {A.}~\bibnamefont {Ludlow}},
  \bibinfo {author} {\bibfnamefont {Q.}~\bibnamefont {Li}}, \bibinfo {author}
  {\bibfnamefont {D.}~\bibnamefont {Westly}}, \bibinfo {author} {\bibfnamefont
  {H.}~\bibnamefont {Leopardi}},  \emph {et~al.},\ }\href@noop {} {\bibfield
  {journal} {\bibinfo  {journal} {arXiv preprint arXiv:1702.03269}\ } (\bibinfo
  {year} {2017})}\BibitemShut {NoStop}%
\bibitem [{\citenamefont {Hsieh}\ \emph {et~al.}(2007)\citenamefont {Hsieh},
  \citenamefont {Chen}, \citenamefont {Liu}, \citenamefont {Dadap},
  \citenamefont {Panoiu}, \citenamefont {Chou}, \citenamefont {Xia},
  \citenamefont {Green}, \citenamefont {Vlasov},\ and\ \citenamefont
  {Osgood}}]{hsieh2007supercontinuum}%
  \BibitemOpen
  \bibfield  {author} {\bibinfo {author} {\bibfnamefont {I.-W.}\ \bibnamefont
  {Hsieh}}, \bibinfo {author} {\bibfnamefont {X.}~\bibnamefont {Chen}},
  \bibinfo {author} {\bibfnamefont {X.}~\bibnamefont {Liu}}, \bibinfo {author}
  {\bibfnamefont {J.~I.}\ \bibnamefont {Dadap}}, \bibinfo {author}
  {\bibfnamefont {N.~C.}\ \bibnamefont {Panoiu}}, \bibinfo {author}
  {\bibfnamefont {C.-Y.}\ \bibnamefont {Chou}}, \bibinfo {author}
  {\bibfnamefont {F.}~\bibnamefont {Xia}}, \bibinfo {author} {\bibfnamefont
  {W.~M.}\ \bibnamefont {Green}}, \bibinfo {author} {\bibfnamefont {Y.~A.}\
  \bibnamefont {Vlasov}}, \ and\ \bibinfo {author} {\bibfnamefont {R.~M.}\
  \bibnamefont {Osgood}},\ }\href@noop {} {\bibfield  {journal} {\bibinfo
  {journal} {Optics express}\ }\textbf {\bibinfo {volume} {15}},\ \bibinfo
  {pages} {15242} (\bibinfo {year} {2007})}\BibitemShut {NoStop}%
\bibitem [{\citenamefont {Ajayan}\ \emph {et~al.}(2016)\citenamefont {Ajayan},
  \citenamefont {Kim},\ and\ \citenamefont {Banerjee}}]{ajayan2016two}%
  \BibitemOpen
  \bibfield  {author} {\bibinfo {author} {\bibfnamefont {P.}~\bibnamefont
  {Ajayan}}, \bibinfo {author} {\bibfnamefont {P.}~\bibnamefont {Kim}}, \ and\
  \bibinfo {author} {\bibfnamefont {K.}~\bibnamefont {Banerjee}},\ }\href@noop
  {} {\bibfield  {journal} {\bibinfo  {journal} {Physics Today}\ }\textbf
  {\bibinfo {volume} {69}},\ \bibinfo {pages} {38} (\bibinfo {year}
  {2016})}\BibitemShut {NoStop}%
\bibitem [{\citenamefont {Mak}\ \emph {et~al.}(2010)\citenamefont {Mak},
  \citenamefont {Lee}, \citenamefont {Hone}, \citenamefont {Shan},\ and\
  \citenamefont {Heinz}}]{mak2010atomically}%
  \BibitemOpen
  \bibfield  {author} {\bibinfo {author} {\bibfnamefont {K.~F.}\ \bibnamefont
  {Mak}}, \bibinfo {author} {\bibfnamefont {C.}~\bibnamefont {Lee}}, \bibinfo
  {author} {\bibfnamefont {J.}~\bibnamefont {Hone}}, \bibinfo {author}
  {\bibfnamefont {J.}~\bibnamefont {Shan}}, \ and\ \bibinfo {author}
  {\bibfnamefont {T.~F.}\ \bibnamefont {Heinz}},\ }\href@noop {} {\bibfield
  {journal} {\bibinfo  {journal} {Physical review letters}\ }\textbf {\bibinfo
  {volume} {105}},\ \bibinfo {pages} {136805} (\bibinfo {year}
  {2010})}\BibitemShut {NoStop}%
\bibitem [{\citenamefont {Xiao}\ \emph {et~al.}(2012)\citenamefont {Xiao},
  \citenamefont {Liu}, \citenamefont {Feng}, \citenamefont {Xu},\ and\
  \citenamefont {Yao}}]{xiao2012coupled}%
  \BibitemOpen
  \bibfield  {author} {\bibinfo {author} {\bibfnamefont {D.}~\bibnamefont
  {Xiao}}, \bibinfo {author} {\bibfnamefont {G.-B.}\ \bibnamefont {Liu}},
  \bibinfo {author} {\bibfnamefont {W.}~\bibnamefont {Feng}}, \bibinfo {author}
  {\bibfnamefont {X.}~\bibnamefont {Xu}}, \ and\ \bibinfo {author}
  {\bibfnamefont {W.}~\bibnamefont {Yao}},\ }\href@noop {} {\bibfield
  {journal} {\bibinfo  {journal} {Physical Review Letters}\ }\textbf {\bibinfo
  {volume} {108}},\ \bibinfo {pages} {196802} (\bibinfo {year}
  {2012})}\BibitemShut {NoStop}%
\bibitem [{\citenamefont {Haug}\ and\ \citenamefont
  {Koch}(2009)}]{haug2009quantum}%
  \BibitemOpen
  \bibfield  {author} {\bibinfo {author} {\bibfnamefont {H.}~\bibnamefont
  {Haug}}\ and\ \bibinfo {author} {\bibfnamefont {S.~W.}\ \bibnamefont
  {Koch}},\ }\href@noop {} {\emph {\bibinfo {title} {Quantum theory of the
  optical and electronic properties of semiconductors}}}\ (\bibinfo
  {publisher} {World Scientific Publishing Co Inc},\ \bibinfo {year}
  {2009})\BibitemShut {NoStop}%
\bibitem [{\citenamefont {Wang}\ \emph {et~al.}(2017)\citenamefont {Wang},
  \citenamefont {Chernikov}, \citenamefont {Glazov}, \citenamefont {Heinz},
  \citenamefont {Marie}, \citenamefont {Amand},\ and\ \citenamefont
  {Urbaszek}}]{wang2017excitons}%
  \BibitemOpen
  \bibfield  {author} {\bibinfo {author} {\bibfnamefont {G.}~\bibnamefont
  {Wang}}, \bibinfo {author} {\bibfnamefont {A.}~\bibnamefont {Chernikov}},
  \bibinfo {author} {\bibfnamefont {M.~M.}\ \bibnamefont {Glazov}}, \bibinfo
  {author} {\bibfnamefont {T.~F.}\ \bibnamefont {Heinz}}, \bibinfo {author}
  {\bibfnamefont {X.}~\bibnamefont {Marie}}, \bibinfo {author} {\bibfnamefont
  {T.}~\bibnamefont {Amand}}, \ and\ \bibinfo {author} {\bibfnamefont
  {B.}~\bibnamefont {Urbaszek}},\ }\href@noop {} {\bibfield  {journal}
  {\bibinfo  {journal} {arXiv preprint arXiv:1707.05863}\ } (\bibinfo {year}
  {2017})}\BibitemShut {NoStop}%
\bibitem [{\citenamefont {Wang}\ \emph {et~al.}(2016)\citenamefont {Wang},
  \citenamefont {Zhang}, \citenamefont {Chan}, \citenamefont {Manolatou},
  \citenamefont {Tiwari},\ and\ \citenamefont {Rana}}]{wang2016radiative}%
  \BibitemOpen
  \bibfield  {author} {\bibinfo {author} {\bibfnamefont {H.}~\bibnamefont
  {Wang}}, \bibinfo {author} {\bibfnamefont {C.}~\bibnamefont {Zhang}},
  \bibinfo {author} {\bibfnamefont {W.}~\bibnamefont {Chan}}, \bibinfo {author}
  {\bibfnamefont {C.}~\bibnamefont {Manolatou}}, \bibinfo {author}
  {\bibfnamefont {S.}~\bibnamefont {Tiwari}}, \ and\ \bibinfo {author}
  {\bibfnamefont {F.}~\bibnamefont {Rana}},\ }\href@noop {} {\bibfield
  {journal} {\bibinfo  {journal} {Physical Review B}\ }\textbf {\bibinfo
  {volume} {93}},\ \bibinfo {pages} {045407} (\bibinfo {year}
  {2016})}\BibitemShut {NoStop}%
\bibitem [{\citenamefont {Wang}\ \emph
  {et~al.}(2015{\natexlab{a}})\citenamefont {Wang}, \citenamefont {Strait},
  \citenamefont {Zhang}, \citenamefont {Chan}, \citenamefont {Manolatou},
  \citenamefont {Tiwari},\ and\ \citenamefont {Rana}}]{wang2015fast}%
  \BibitemOpen
  \bibfield  {author} {\bibinfo {author} {\bibfnamefont {H.}~\bibnamefont
  {Wang}}, \bibinfo {author} {\bibfnamefont {J.~H.}\ \bibnamefont {Strait}},
  \bibinfo {author} {\bibfnamefont {C.}~\bibnamefont {Zhang}}, \bibinfo
  {author} {\bibfnamefont {W.}~\bibnamefont {Chan}}, \bibinfo {author}
  {\bibfnamefont {C.}~\bibnamefont {Manolatou}}, \bibinfo {author}
  {\bibfnamefont {S.}~\bibnamefont {Tiwari}}, \ and\ \bibinfo {author}
  {\bibfnamefont {F.}~\bibnamefont {Rana}},\ }\href@noop {} {\bibfield
  {journal} {\bibinfo  {journal} {Physical Review B}\ }\textbf {\bibinfo
  {volume} {91}},\ \bibinfo {pages} {165411} (\bibinfo {year}
  {2015}{\natexlab{a}})}\BibitemShut {NoStop}%
\bibitem [{\citenamefont {Wang}\ \emph {et~al.}(2012)\citenamefont {Wang},
  \citenamefont {Kalantar-Zadeh}, \citenamefont {Kis}, \citenamefont
  {Coleman},\ and\ \citenamefont {Strano}}]{wang2012electronics}%
  \BibitemOpen
  \bibfield  {author} {\bibinfo {author} {\bibfnamefont {Q.~H.}\ \bibnamefont
  {Wang}}, \bibinfo {author} {\bibfnamefont {K.}~\bibnamefont
  {Kalantar-Zadeh}}, \bibinfo {author} {\bibfnamefont {A.}~\bibnamefont {Kis}},
  \bibinfo {author} {\bibfnamefont {J.~N.}\ \bibnamefont {Coleman}}, \ and\
  \bibinfo {author} {\bibfnamefont {M.~S.}\ \bibnamefont {Strano}},\
  }\href@noop {} {\bibfield  {journal} {\bibinfo  {journal} {Nature
  nanotechnology}\ }\textbf {\bibinfo {volume} {7}},\ \bibinfo {pages} {699}
  (\bibinfo {year} {2012})}\BibitemShut {NoStop}%
\bibitem [{\citenamefont {Zhang}\ \emph
  {et~al.}(2014{\natexlab{a}})\citenamefont {Zhang}, \citenamefont {Wang},
  \citenamefont {Chan}, \citenamefont {Manolatou},\ and\ \citenamefont
  {Rana}}]{zhang2014absorption}%
  \BibitemOpen
  \bibfield  {author} {\bibinfo {author} {\bibfnamefont {C.}~\bibnamefont
  {Zhang}}, \bibinfo {author} {\bibfnamefont {H.}~\bibnamefont {Wang}},
  \bibinfo {author} {\bibfnamefont {W.}~\bibnamefont {Chan}}, \bibinfo {author}
  {\bibfnamefont {C.}~\bibnamefont {Manolatou}}, \ and\ \bibinfo {author}
  {\bibfnamefont {F.}~\bibnamefont {Rana}},\ }\href@noop {} {\bibfield
  {journal} {\bibinfo  {journal} {Physical Review B}\ }\textbf {\bibinfo
  {volume} {89}},\ \bibinfo {pages} {205436} (\bibinfo {year}
  {2014}{\natexlab{a}})}\BibitemShut {NoStop}%
\bibitem [{\citenamefont {Hill}\ \emph {et~al.}(2015)\citenamefont {Hill},
  \citenamefont {Rigosi}, \citenamefont {Roquelet}, \citenamefont {Chernikov},
  \citenamefont {Berkelbach}, \citenamefont {Reichman}, \citenamefont
  {Hybertsen}, \citenamefont {Brus},\ and\ \citenamefont
  {Heinz}}]{hill2015observation}%
  \BibitemOpen
  \bibfield  {author} {\bibinfo {author} {\bibfnamefont {H.~M.}\ \bibnamefont
  {Hill}}, \bibinfo {author} {\bibfnamefont {A.~F.}\ \bibnamefont {Rigosi}},
  \bibinfo {author} {\bibfnamefont {C.}~\bibnamefont {Roquelet}}, \bibinfo
  {author} {\bibfnamefont {A.}~\bibnamefont {Chernikov}}, \bibinfo {author}
  {\bibfnamefont {T.~C.}\ \bibnamefont {Berkelbach}}, \bibinfo {author}
  {\bibfnamefont {D.~R.}\ \bibnamefont {Reichman}}, \bibinfo {author}
  {\bibfnamefont {M.~S.}\ \bibnamefont {Hybertsen}}, \bibinfo {author}
  {\bibfnamefont {L.~E.}\ \bibnamefont {Brus}}, \ and\ \bibinfo {author}
  {\bibfnamefont {T.~F.}\ \bibnamefont {Heinz}},\ }\href@noop {} {\bibfield
  {journal} {\bibinfo  {journal} {Nano letters}\ }\textbf {\bibinfo {volume}
  {15}},\ \bibinfo {pages} {2992} (\bibinfo {year} {2015})}\BibitemShut
  {NoStop}%
\bibitem [{\citenamefont {Selig}\ \emph {et~al.}(2016)\citenamefont {Selig},
  \citenamefont {Bergh{\"a}user}, \citenamefont {Raja}, \citenamefont {Nagler},
  \citenamefont {Sch{\"u}ller}, \citenamefont {Heinz}, \citenamefont {Korn},
  \citenamefont {Chernikov}, \citenamefont {Malic},\ and\ \citenamefont
  {Knorr}}]{selig2016excitonic}%
  \BibitemOpen
  \bibfield  {author} {\bibinfo {author} {\bibfnamefont {M.}~\bibnamefont
  {Selig}}, \bibinfo {author} {\bibfnamefont {G.}~\bibnamefont
  {Bergh{\"a}user}}, \bibinfo {author} {\bibfnamefont {A.}~\bibnamefont
  {Raja}}, \bibinfo {author} {\bibfnamefont {P.}~\bibnamefont {Nagler}},
  \bibinfo {author} {\bibfnamefont {C.}~\bibnamefont {Sch{\"u}ller}}, \bibinfo
  {author} {\bibfnamefont {T.~F.}\ \bibnamefont {Heinz}}, \bibinfo {author}
  {\bibfnamefont {T.}~\bibnamefont {Korn}}, \bibinfo {author} {\bibfnamefont
  {A.}~\bibnamefont {Chernikov}}, \bibinfo {author} {\bibfnamefont
  {E.}~\bibnamefont {Malic}}, \ and\ \bibinfo {author} {\bibfnamefont
  {A.}~\bibnamefont {Knorr}},\ }\href@noop {} {\bibfield  {journal} {\bibinfo
  {journal} {Nature communications}\ }\textbf {\bibinfo {volume} {7}},\
  \bibinfo {pages} {13279} (\bibinfo {year} {2016})}\BibitemShut {NoStop}%
\bibitem [{\citenamefont {Kumar}\ \emph {et~al.}(2013)\citenamefont {Kumar},
  \citenamefont {Najmaei}, \citenamefont {Cui}, \citenamefont {Ceballos},
  \citenamefont {Ajayan}, \citenamefont {Lou},\ and\ \citenamefont
  {Zhao}}]{kumar2013second}%
  \BibitemOpen
  \bibfield  {author} {\bibinfo {author} {\bibfnamefont {N.}~\bibnamefont
  {Kumar}}, \bibinfo {author} {\bibfnamefont {S.}~\bibnamefont {Najmaei}},
  \bibinfo {author} {\bibfnamefont {Q.}~\bibnamefont {Cui}}, \bibinfo {author}
  {\bibfnamefont {F.}~\bibnamefont {Ceballos}}, \bibinfo {author}
  {\bibfnamefont {P.~M.}\ \bibnamefont {Ajayan}}, \bibinfo {author}
  {\bibfnamefont {J.}~\bibnamefont {Lou}}, \ and\ \bibinfo {author}
  {\bibfnamefont {H.}~\bibnamefont {Zhao}},\ }\href@noop {} {\bibfield
  {journal} {\bibinfo  {journal} {Physical Review B}\ }\textbf {\bibinfo
  {volume} {87}},\ \bibinfo {pages} {161403} (\bibinfo {year}
  {2013})}\BibitemShut {NoStop}%
\bibitem [{\citenamefont {Clark}\ \emph {et~al.}(2014)\citenamefont {Clark},
  \citenamefont {Senthilkumar}, \citenamefont {Le}, \citenamefont {Weerawarne},
  \citenamefont {Shim}, \citenamefont {Jang}, \citenamefont {Shim},
  \citenamefont {Cho}, \citenamefont {Sim}, \citenamefont {Seong} \emph
  {et~al.}}]{clark2014strong}%
  \BibitemOpen
  \bibfield  {author} {\bibinfo {author} {\bibfnamefont {D.}~\bibnamefont
  {Clark}}, \bibinfo {author} {\bibfnamefont {V.}~\bibnamefont {Senthilkumar}},
  \bibinfo {author} {\bibfnamefont {C.}~\bibnamefont {Le}}, \bibinfo {author}
  {\bibfnamefont {D.}~\bibnamefont {Weerawarne}}, \bibinfo {author}
  {\bibfnamefont {B.}~\bibnamefont {Shim}}, \bibinfo {author} {\bibfnamefont
  {J.}~\bibnamefont {Jang}}, \bibinfo {author} {\bibfnamefont {J.}~\bibnamefont
  {Shim}}, \bibinfo {author} {\bibfnamefont {J.}~\bibnamefont {Cho}}, \bibinfo
  {author} {\bibfnamefont {Y.}~\bibnamefont {Sim}}, \bibinfo {author}
  {\bibfnamefont {M.-J.}\ \bibnamefont {Seong}},  \emph {et~al.},\ }\href@noop
  {} {\bibfield  {journal} {\bibinfo  {journal} {Physical Review B}\ }\textbf
  {\bibinfo {volume} {90}},\ \bibinfo {pages} {121409} (\bibinfo {year}
  {2014})}\BibitemShut {NoStop}%
\bibitem [{\citenamefont {Merano}(2016)}]{merano2016nonlinear}%
  \BibitemOpen
  \bibfield  {author} {\bibinfo {author} {\bibfnamefont {M.}~\bibnamefont
  {Merano}},\ }\href@noop {} {\bibfield  {journal} {\bibinfo  {journal} {Optics
  letters}\ }\textbf {\bibinfo {volume} {41}},\ \bibinfo {pages} {187}
  (\bibinfo {year} {2016})}\BibitemShut {NoStop}%
\bibitem [{\citenamefont {Pedersen}(2015)}]{pedersen2015intraband}%
  \BibitemOpen
  \bibfield  {author} {\bibinfo {author} {\bibfnamefont {T.~G.}\ \bibnamefont
  {Pedersen}},\ }\href@noop {} {\bibfield  {journal} {\bibinfo  {journal}
  {Physical Review B}\ }\textbf {\bibinfo {volume} {92}},\ \bibinfo {pages}
  {235432} (\bibinfo {year} {2015})}\BibitemShut {NoStop}%
\bibitem [{\citenamefont {Trolle}\ \emph {et~al.}(2014)\citenamefont {Trolle},
  \citenamefont {Seifert},\ and\ \citenamefont {Pedersen}}]{trolle2014theory}%
  \BibitemOpen
  \bibfield  {author} {\bibinfo {author} {\bibfnamefont {M.~L.}\ \bibnamefont
  {Trolle}}, \bibinfo {author} {\bibfnamefont {G.}~\bibnamefont {Seifert}}, \
  and\ \bibinfo {author} {\bibfnamefont {T.~G.}\ \bibnamefont {Pedersen}},\
  }\href@noop {} {\bibfield  {journal} {\bibinfo  {journal} {Physical Review
  B}\ }\textbf {\bibinfo {volume} {89}},\ \bibinfo {pages} {235410} (\bibinfo
  {year} {2014})}\BibitemShut {NoStop}%
\bibitem [{\citenamefont {Gr{\"u}ning}\ and\ \citenamefont
  {Attaccalite}(2014)}]{gruning2014second}%
  \BibitemOpen
  \bibfield  {author} {\bibinfo {author} {\bibfnamefont {M.}~\bibnamefont
  {Gr{\"u}ning}}\ and\ \bibinfo {author} {\bibfnamefont {C.}~\bibnamefont
  {Attaccalite}},\ }\href@noop {} {\bibfield  {journal} {\bibinfo  {journal}
  {Physical Review B}\ }\textbf {\bibinfo {volume} {89}},\ \bibinfo {pages}
  {081102} (\bibinfo {year} {2014})}\BibitemShut {NoStop}%
\bibitem [{\citenamefont {Seyler}\ \emph {et~al.}(2015)\citenamefont {Seyler},
  \citenamefont {Schaibley}, \citenamefont {Gong}, \citenamefont {Rivera},
  \citenamefont {Jones}, \citenamefont {Wu}, \citenamefont {Yan}, \citenamefont
  {Mandrus}, \citenamefont {Yao},\ and\ \citenamefont
  {Xu}}]{seyler2015electrical}%
  \BibitemOpen
  \bibfield  {author} {\bibinfo {author} {\bibfnamefont {K.~L.}\ \bibnamefont
  {Seyler}}, \bibinfo {author} {\bibfnamefont {J.~R.}\ \bibnamefont
  {Schaibley}}, \bibinfo {author} {\bibfnamefont {P.}~\bibnamefont {Gong}},
  \bibinfo {author} {\bibfnamefont {P.}~\bibnamefont {Rivera}}, \bibinfo
  {author} {\bibfnamefont {A.~M.}\ \bibnamefont {Jones}}, \bibinfo {author}
  {\bibfnamefont {S.}~\bibnamefont {Wu}}, \bibinfo {author} {\bibfnamefont
  {J.}~\bibnamefont {Yan}}, \bibinfo {author} {\bibfnamefont {D.~G.}\
  \bibnamefont {Mandrus}}, \bibinfo {author} {\bibfnamefont {W.}~\bibnamefont
  {Yao}}, \ and\ \bibinfo {author} {\bibfnamefont {X.}~\bibnamefont {Xu}},\
  }\href@noop {} {\bibfield  {journal} {\bibinfo  {journal} {Nature
  nanotechnology}\ }\textbf {\bibinfo {volume} {10}},\ \bibinfo {pages} {407}
  (\bibinfo {year} {2015})}\BibitemShut {NoStop}%
\bibitem [{\citenamefont {Xiao}\ \emph {et~al.}(2015)\citenamefont {Xiao},
  \citenamefont {Ye}, \citenamefont {Wang}, \citenamefont {Zhu}, \citenamefont
  {Wang},\ and\ \citenamefont {Zhang}}]{xiao2015nonlinear}%
  \BibitemOpen
  \bibfield  {author} {\bibinfo {author} {\bibfnamefont {J.}~\bibnamefont
  {Xiao}}, \bibinfo {author} {\bibfnamefont {Z.}~\bibnamefont {Ye}}, \bibinfo
  {author} {\bibfnamefont {Y.}~\bibnamefont {Wang}}, \bibinfo {author}
  {\bibfnamefont {H.}~\bibnamefont {Zhu}}, \bibinfo {author} {\bibfnamefont
  {Y.}~\bibnamefont {Wang}}, \ and\ \bibinfo {author} {\bibfnamefont
  {X.}~\bibnamefont {Zhang}},\ }\href@noop {} {\bibfield  {journal} {\bibinfo
  {journal} {Light: Science \& Applications}\ }\textbf {\bibinfo {volume}
  {4}},\ \bibinfo {pages} {e366} (\bibinfo {year} {2015})}\BibitemShut
  {NoStop}%
\bibitem [{\citenamefont {Dresselhaus}(1956)}]{dresselhaus1956effective}%
  \BibitemOpen
  \bibfield  {author} {\bibinfo {author} {\bibfnamefont {G.}~\bibnamefont
  {Dresselhaus}},\ }\href@noop {} {\bibfield  {journal} {\bibinfo  {journal}
  {Journal of Physics and Chemistry of Solids}\ }\textbf {\bibinfo {volume}
  {1}},\ \bibinfo {pages} {14} (\bibinfo {year} {1956})}\BibitemShut {NoStop}%
\bibitem [{\citenamefont {Cudazzo}\ \emph {et~al.}(2011)\citenamefont
  {Cudazzo}, \citenamefont {Tokatly},\ and\ \citenamefont
  {Rubio}}]{cudazzo2011dielectric}%
  \BibitemOpen
  \bibfield  {author} {\bibinfo {author} {\bibfnamefont {P.}~\bibnamefont
  {Cudazzo}}, \bibinfo {author} {\bibfnamefont {I.~V.}\ \bibnamefont
  {Tokatly}}, \ and\ \bibinfo {author} {\bibfnamefont {A.}~\bibnamefont
  {Rubio}},\ }\href@noop {} {\bibfield  {journal} {\bibinfo  {journal}
  {Physical Review B}\ }\textbf {\bibinfo {volume} {84}},\ \bibinfo {pages}
  {085406} (\bibinfo {year} {2011})}\BibitemShut {NoStop}%
\bibitem [{\citenamefont {Wu}\ \emph {et~al.}(2015)\citenamefont {Wu},
  \citenamefont {Qu},\ and\ \citenamefont {MacDonald}}]{wu2015exciton}%
  \BibitemOpen
  \bibfield  {author} {\bibinfo {author} {\bibfnamefont {F.}~\bibnamefont
  {Wu}}, \bibinfo {author} {\bibfnamefont {F.}~\bibnamefont {Qu}}, \ and\
  \bibinfo {author} {\bibfnamefont {A.}~\bibnamefont {MacDonald}},\ }\href@noop
  {} {\bibfield  {journal} {\bibinfo  {journal} {Physical Review B}\ }\textbf
  {\bibinfo {volume} {91}},\ \bibinfo {pages} {075310} (\bibinfo {year}
  {2015})}\BibitemShut {NoStop}%
\bibitem [{\citenamefont {Robert}\ \emph {et~al.}(2017)\citenamefont {Robert},
  \citenamefont {Semina}, \citenamefont {Cadiz}, \citenamefont {Manca},
  \citenamefont {Courtade}, \citenamefont {Taniguchi}, \citenamefont
  {Watanabe}, \citenamefont {Cai}, \citenamefont {Tongay}, \citenamefont
  {Lassagne} \emph {et~al.}}]{robert2017optical}%
  \BibitemOpen
  \bibfield  {author} {\bibinfo {author} {\bibfnamefont {C.}~\bibnamefont
  {Robert}}, \bibinfo {author} {\bibfnamefont {M.}~\bibnamefont {Semina}},
  \bibinfo {author} {\bibfnamefont {F.}~\bibnamefont {Cadiz}}, \bibinfo
  {author} {\bibfnamefont {M.}~\bibnamefont {Manca}}, \bibinfo {author}
  {\bibfnamefont {E.}~\bibnamefont {Courtade}}, \bibinfo {author}
  {\bibfnamefont {T.}~\bibnamefont {Taniguchi}}, \bibinfo {author}
  {\bibfnamefont {K.}~\bibnamefont {Watanabe}}, \bibinfo {author}
  {\bibfnamefont {H.}~\bibnamefont {Cai}}, \bibinfo {author} {\bibfnamefont
  {S.}~\bibnamefont {Tongay}}, \bibinfo {author} {\bibfnamefont
  {B.}~\bibnamefont {Lassagne}},  \emph {et~al.},\ }\href@noop {} {\bibfield
  {journal} {\bibinfo  {journal} {arXiv preprint arXiv:1712.01548}\ } (\bibinfo
  {year} {2017})}\BibitemShut {NoStop}%
\bibitem [{\citenamefont {Kyl{\"a}np{\"a}{\"a}}\ and\ \citenamefont
  {Komsa}(2015)}]{kylanpaa2015binding}%
  \BibitemOpen
  \bibfield  {author} {\bibinfo {author} {\bibfnamefont {I.}~\bibnamefont
  {Kyl{\"a}np{\"a}{\"a}}}\ and\ \bibinfo {author} {\bibfnamefont {H.-P.}\
  \bibnamefont {Komsa}},\ }\href@noop {} {\bibfield  {journal} {\bibinfo
  {journal} {Physical Review B}\ }\textbf {\bibinfo {volume} {92}},\ \bibinfo
  {pages} {205418} (\bibinfo {year} {2015})}\BibitemShut {NoStop}%
\bibitem [{\citenamefont {Klots}\ \emph {et~al.}(2014)\citenamefont {Klots},
  \citenamefont {Newaz}, \citenamefont {Wang}, \citenamefont {Prasai},
  \citenamefont {Krzyzanowska}, \citenamefont {Lin}, \citenamefont {Caudel},
  \citenamefont {Ghimire}, \citenamefont {Yan}, \citenamefont {Ivanov} \emph
  {et~al.}}]{klots2014probing}%
  \BibitemOpen
  \bibfield  {author} {\bibinfo {author} {\bibfnamefont {A.}~\bibnamefont
  {Klots}}, \bibinfo {author} {\bibfnamefont {A.}~\bibnamefont {Newaz}},
  \bibinfo {author} {\bibfnamefont {B.}~\bibnamefont {Wang}}, \bibinfo {author}
  {\bibfnamefont {D.}~\bibnamefont {Prasai}}, \bibinfo {author} {\bibfnamefont
  {H.}~\bibnamefont {Krzyzanowska}}, \bibinfo {author} {\bibfnamefont
  {J.}~\bibnamefont {Lin}}, \bibinfo {author} {\bibfnamefont {D.}~\bibnamefont
  {Caudel}}, \bibinfo {author} {\bibfnamefont {N.}~\bibnamefont {Ghimire}},
  \bibinfo {author} {\bibfnamefont {J.}~\bibnamefont {Yan}}, \bibinfo {author}
  {\bibfnamefont {B.}~\bibnamefont {Ivanov}},  \emph {et~al.},\ }\href@noop {}
  {\bibfield  {journal} {\bibinfo  {journal} {Scientific reports}\ }\textbf
  {\bibinfo {volume} {4}},\ \bibinfo {pages} {6608} (\bibinfo {year}
  {2014})}\BibitemShut {NoStop}%
\bibitem [{\citenamefont {Chiu}\ \emph {et~al.}(2015)\citenamefont {Chiu},
  \citenamefont {Zhang}, \citenamefont {Shiu}, \citenamefont {Chuu},
  \citenamefont {Chen}, \citenamefont {Chang}, \citenamefont {Chen},
  \citenamefont {Chou}, \citenamefont {Shih},\ and\ \citenamefont
  {Li}}]{chiu2015determination}%
  \BibitemOpen
  \bibfield  {author} {\bibinfo {author} {\bibfnamefont {M.-H.}\ \bibnamefont
  {Chiu}}, \bibinfo {author} {\bibfnamefont {C.}~\bibnamefont {Zhang}},
  \bibinfo {author} {\bibfnamefont {H.-W.}\ \bibnamefont {Shiu}}, \bibinfo
  {author} {\bibfnamefont {C.-P.}\ \bibnamefont {Chuu}}, \bibinfo {author}
  {\bibfnamefont {C.-H.}\ \bibnamefont {Chen}}, \bibinfo {author}
  {\bibfnamefont {C.-Y.~S.}\ \bibnamefont {Chang}}, \bibinfo {author}
  {\bibfnamefont {C.-H.}\ \bibnamefont {Chen}}, \bibinfo {author}
  {\bibfnamefont {M.-Y.}\ \bibnamefont {Chou}}, \bibinfo {author}
  {\bibfnamefont {C.-K.}\ \bibnamefont {Shih}}, \ and\ \bibinfo {author}
  {\bibfnamefont {L.-J.}\ \bibnamefont {Li}},\ }\href@noop {} {\bibfield
  {journal} {\bibinfo  {journal} {Nature communications}\ }\textbf {\bibinfo
  {volume} {6}} (\bibinfo {year} {2015})}\BibitemShut {NoStop}%
\bibitem [{\citenamefont {Cao}\ \emph {et~al.}(2012)\citenamefont {Cao},
  \citenamefont {Wang}, \citenamefont {Han}, \citenamefont {Ye}, \citenamefont
  {Zhu}, \citenamefont {Shi}, \citenamefont {Niu}, \citenamefont {Tan},
  \citenamefont {Wang}, \citenamefont {Liu} \emph {et~al.}}]{cao2012valley}%
  \BibitemOpen
  \bibfield  {author} {\bibinfo {author} {\bibfnamefont {T.}~\bibnamefont
  {Cao}}, \bibinfo {author} {\bibfnamefont {G.}~\bibnamefont {Wang}}, \bibinfo
  {author} {\bibfnamefont {W.}~\bibnamefont {Han}}, \bibinfo {author}
  {\bibfnamefont {H.}~\bibnamefont {Ye}}, \bibinfo {author} {\bibfnamefont
  {C.}~\bibnamefont {Zhu}}, \bibinfo {author} {\bibfnamefont {J.}~\bibnamefont
  {Shi}}, \bibinfo {author} {\bibfnamefont {Q.}~\bibnamefont {Niu}}, \bibinfo
  {author} {\bibfnamefont {P.}~\bibnamefont {Tan}}, \bibinfo {author}
  {\bibfnamefont {E.}~\bibnamefont {Wang}}, \bibinfo {author} {\bibfnamefont
  {B.}~\bibnamefont {Liu}},  \emph {et~al.},\ }\href@noop {} {\bibfield
  {journal} {\bibinfo  {journal} {Nature communications}\ }\textbf {\bibinfo
  {volume} {3}},\ \bibinfo {pages} {887} (\bibinfo {year} {2012})}\BibitemShut
  {NoStop}%
\bibitem [{\citenamefont {Mak}\ \emph {et~al.}(2012)\citenamefont {Mak},
  \citenamefont {He}, \citenamefont {Shan},\ and\ \citenamefont
  {Heinz}}]{mak2012control}%
  \BibitemOpen
  \bibfield  {author} {\bibinfo {author} {\bibfnamefont {K.~F.}\ \bibnamefont
  {Mak}}, \bibinfo {author} {\bibfnamefont {K.}~\bibnamefont {He}}, \bibinfo
  {author} {\bibfnamefont {J.}~\bibnamefont {Shan}}, \ and\ \bibinfo {author}
  {\bibfnamefont {T.~F.}\ \bibnamefont {Heinz}},\ }\href@noop {} {\bibfield
  {journal} {\bibinfo  {journal} {Nature nanotechnology}\ }\textbf {\bibinfo
  {volume} {7}},\ \bibinfo {pages} {494} (\bibinfo {year} {2012})}\BibitemShut
  {NoStop}%
\bibitem [{\citenamefont {Sallen}\ \emph {et~al.}(2012)\citenamefont {Sallen},
  \citenamefont {Bouet}, \citenamefont {Marie}, \citenamefont {Wang},
  \citenamefont {Zhu}, \citenamefont {Han}, \citenamefont {Lu}, \citenamefont
  {Tan}, \citenamefont {Amand}, \citenamefont {Liu} \emph
  {et~al.}}]{sallen2012robust}%
  \BibitemOpen
  \bibfield  {author} {\bibinfo {author} {\bibfnamefont {G.}~\bibnamefont
  {Sallen}}, \bibinfo {author} {\bibfnamefont {L.}~\bibnamefont {Bouet}},
  \bibinfo {author} {\bibfnamefont {X.}~\bibnamefont {Marie}}, \bibinfo
  {author} {\bibfnamefont {G.}~\bibnamefont {Wang}}, \bibinfo {author}
  {\bibfnamefont {C.}~\bibnamefont {Zhu}}, \bibinfo {author} {\bibfnamefont
  {W.}~\bibnamefont {Han}}, \bibinfo {author} {\bibfnamefont {Y.}~\bibnamefont
  {Lu}}, \bibinfo {author} {\bibfnamefont {P.}~\bibnamefont {Tan}}, \bibinfo
  {author} {\bibfnamefont {T.}~\bibnamefont {Amand}}, \bibinfo {author}
  {\bibfnamefont {B.}~\bibnamefont {Liu}},  \emph {et~al.},\ }\href@noop {}
  {\bibfield  {journal} {\bibinfo  {journal} {Physical Review B}\ }\textbf
  {\bibinfo {volume} {86}},\ \bibinfo {pages} {081301} (\bibinfo {year}
  {2012})}\BibitemShut {NoStop}%
\bibitem [{\citenamefont {Sallen}\ \emph {et~al.}(2014)\citenamefont {Sallen},
  \citenamefont {Bouet}, \citenamefont {Marie}, \citenamefont {Wang},
  \citenamefont {Zhu}, \citenamefont {Han}, \citenamefont {Lu}, \citenamefont
  {Tan}, \citenamefont {Amand}, \citenamefont {Liu} \emph
  {et~al.}}]{sallen2014erratum}%
  \BibitemOpen
  \bibfield  {author} {\bibinfo {author} {\bibfnamefont {G.}~\bibnamefont
  {Sallen}}, \bibinfo {author} {\bibfnamefont {L.}~\bibnamefont {Bouet}},
  \bibinfo {author} {\bibfnamefont {X.}~\bibnamefont {Marie}}, \bibinfo
  {author} {\bibfnamefont {G.}~\bibnamefont {Wang}}, \bibinfo {author}
  {\bibfnamefont {C.}~\bibnamefont {Zhu}}, \bibinfo {author} {\bibfnamefont
  {W.}~\bibnamefont {Han}}, \bibinfo {author} {\bibfnamefont {Y.}~\bibnamefont
  {Lu}}, \bibinfo {author} {\bibfnamefont {P.}~\bibnamefont {Tan}}, \bibinfo
  {author} {\bibfnamefont {T.}~\bibnamefont {Amand}}, \bibinfo {author}
  {\bibfnamefont {B.}~\bibnamefont {Liu}},  \emph {et~al.},\ }\href@noop {}
  {\bibfield  {journal} {\bibinfo  {journal} {Physical Review B}\ }\textbf
  {\bibinfo {volume} {89}},\ \bibinfo {pages} {079903} (\bibinfo {year}
  {2014})}\BibitemShut {NoStop}%
\bibitem [{\citenamefont {Blount}(1962)}]{blount1962formalisms}%
  \BibitemOpen
  \bibfield  {author} {\bibinfo {author} {\bibfnamefont {E.}~\bibnamefont
  {Blount}},\ }\href@noop {} {\bibfield  {journal} {\bibinfo  {journal} {Solid
  state physics}\ }\textbf {\bibinfo {volume} {13}},\ \bibinfo {pages} {305}
  (\bibinfo {year} {1962})}\BibitemShut {NoStop}%
\bibitem [{\citenamefont {Simon}\ and\ \citenamefont
  {Bloembergen}(1968)}]{simon1968second}%
  \BibitemOpen
  \bibfield  {author} {\bibinfo {author} {\bibfnamefont {H.}~\bibnamefont
  {Simon}}\ and\ \bibinfo {author} {\bibfnamefont {N.}~\bibnamefont
  {Bloembergen}},\ }\href@noop {} {\bibfield  {journal} {\bibinfo  {journal}
  {Physical Review}\ }\textbf {\bibinfo {volume} {171}},\ \bibinfo {pages}
  {1104} (\bibinfo {year} {1968})}\BibitemShut {NoStop}%
\bibitem [{\citenamefont {Boyd}(2003)}]{boyd2003nonlinear}%
  \BibitemOpen
  \bibfield  {author} {\bibinfo {author} {\bibfnamefont {R.~W.}\ \bibnamefont
  {Boyd}},\ }\href@noop {} {\emph {\bibinfo {title} {Nonlinear optics}}}\
  (\bibinfo  {publisher} {Academic press},\ \bibinfo {year} {2003})\BibitemShut
  {NoStop}%
\bibitem [{\citenamefont {Radisavljevic}\ \emph {et~al.}(2011)\citenamefont
  {Radisavljevic}, \citenamefont {Radenovic}, \citenamefont {Brivio},
  \citenamefont {Giacometti},\ and\ \citenamefont
  {Kis}}]{radisavljevic2011single}%
  \BibitemOpen
  \bibfield  {author} {\bibinfo {author} {\bibfnamefont {B.}~\bibnamefont
  {Radisavljevic}}, \bibinfo {author} {\bibfnamefont {A.}~\bibnamefont
  {Radenovic}}, \bibinfo {author} {\bibfnamefont {J.}~\bibnamefont {Brivio}},
  \bibinfo {author} {\bibfnamefont {i.~V.}\ \bibnamefont {Giacometti}}, \ and\
  \bibinfo {author} {\bibfnamefont {A.}~\bibnamefont {Kis}},\ }\href@noop {}
  {\bibfield  {journal} {\bibinfo  {journal} {Nature nanotechnology}\ }\textbf
  {\bibinfo {volume} {6}},\ \bibinfo {pages} {147} (\bibinfo {year}
  {2011})}\BibitemShut {NoStop}%
\bibitem [{\citenamefont {Elliott}(1957)}]{elliott1957intensity}%
  \BibitemOpen
  \bibfield  {author} {\bibinfo {author} {\bibfnamefont {R.}~\bibnamefont
  {Elliott}},\ }\href@noop {} {\bibfield  {journal} {\bibinfo  {journal}
  {Physical Review}\ }\textbf {\bibinfo {volume} {108}},\ \bibinfo {pages}
  {1384} (\bibinfo {year} {1957})}\BibitemShut {NoStop}%
\bibitem [{\citenamefont {Klingshirn}(2012)}]{klingshirn2012semiconductor}%
  \BibitemOpen
  \bibfield  {author} {\bibinfo {author} {\bibfnamefont {C.~F.}\ \bibnamefont
  {Klingshirn}},\ }\href@noop {} {\emph {\bibinfo {title} {Semiconductor
  optics}}}\ (\bibinfo  {publisher} {Springer},\ \bibinfo {year}
  {2012})\BibitemShut {NoStop}%
\bibitem [{\citenamefont {Moody}\ \emph {et~al.}(2015)\citenamefont {Moody},
  \citenamefont {Dass}, \citenamefont {Hao}, \citenamefont {Chen},
  \citenamefont {Li}, \citenamefont {Singh}, \citenamefont {Tran},
  \citenamefont {Clark}, \citenamefont {Xu}, \citenamefont {Bergh{\"a}user}
  \emph {et~al.}}]{moody2015intrinsic}%
  \BibitemOpen
  \bibfield  {author} {\bibinfo {author} {\bibfnamefont {G.}~\bibnamefont
  {Moody}}, \bibinfo {author} {\bibfnamefont {C.~K.}\ \bibnamefont {Dass}},
  \bibinfo {author} {\bibfnamefont {K.}~\bibnamefont {Hao}}, \bibinfo {author}
  {\bibfnamefont {C.-H.}\ \bibnamefont {Chen}}, \bibinfo {author}
  {\bibfnamefont {L.-J.}\ \bibnamefont {Li}}, \bibinfo {author} {\bibfnamefont
  {A.}~\bibnamefont {Singh}}, \bibinfo {author} {\bibfnamefont
  {K.}~\bibnamefont {Tran}}, \bibinfo {author} {\bibfnamefont {G.}~\bibnamefont
  {Clark}}, \bibinfo {author} {\bibfnamefont {X.}~\bibnamefont {Xu}}, \bibinfo
  {author} {\bibfnamefont {G.}~\bibnamefont {Bergh{\"a}user}},  \emph
  {et~al.},\ }\href@noop {} {\bibfield  {journal} {\bibinfo  {journal} {Nature
  communications}\ }\textbf {\bibinfo {volume} {6}},\ \bibinfo {pages} {8315}
  (\bibinfo {year} {2015})}\BibitemShut {NoStop}%
\bibitem [{\citenamefont {Molina-S{\'a}nchez}\ \emph
  {et~al.}(2013)\citenamefont {Molina-S{\'a}nchez}, \citenamefont {Sangalli},
  \citenamefont {Hummer}, \citenamefont {Marini},\ and\ \citenamefont
  {Wirtz}}]{molina2013effect}%
  \BibitemOpen
  \bibfield  {author} {\bibinfo {author} {\bibfnamefont {A.}~\bibnamefont
  {Molina-S{\'a}nchez}}, \bibinfo {author} {\bibfnamefont {D.}~\bibnamefont
  {Sangalli}}, \bibinfo {author} {\bibfnamefont {K.}~\bibnamefont {Hummer}},
  \bibinfo {author} {\bibfnamefont {A.}~\bibnamefont {Marini}}, \ and\ \bibinfo
  {author} {\bibfnamefont {L.}~\bibnamefont {Wirtz}},\ }\href@noop {}
  {\bibfield  {journal} {\bibinfo  {journal} {Physical Review B}\ }\textbf
  {\bibinfo {volume} {88}},\ \bibinfo {pages} {045412} (\bibinfo {year}
  {2013})}\BibitemShut {NoStop}%
\bibitem [{\citenamefont {Rogers}\ \emph {et~al.}(shed)\citenamefont {Rogers},
  \citenamefont {Gray},\ and\ \citenamefont {Mabuchi}}]{rogers2017absorption}%
  \BibitemOpen
  \bibfield  {author} {\bibinfo {author} {\bibfnamefont {C.}~\bibnamefont
  {Rogers}}, \bibinfo {author} {\bibfnamefont {D.}~\bibnamefont {Gray}}, \ and\
  \bibinfo {author} {\bibfnamefont {H.}~\bibnamefont {Mabuchi}},\ }\href@noop
  {} {\  (\bibinfo {year} {To be published})}\BibitemShut {NoStop}%
\bibitem [{\citenamefont {He}\ \emph {et~al.}(2013)\citenamefont {He},
  \citenamefont {Poole}, \citenamefont {Mak},\ and\ \citenamefont
  {Shan}}]{he2013experimental}%
  \BibitemOpen
  \bibfield  {author} {\bibinfo {author} {\bibfnamefont {K.}~\bibnamefont
  {He}}, \bibinfo {author} {\bibfnamefont {C.}~\bibnamefont {Poole}}, \bibinfo
  {author} {\bibfnamefont {K.~F.}\ \bibnamefont {Mak}}, \ and\ \bibinfo
  {author} {\bibfnamefont {J.}~\bibnamefont {Shan}},\ }\href@noop {} {\bibfield
   {journal} {\bibinfo  {journal} {Nano letters}\ }\textbf {\bibinfo {volume}
  {13}},\ \bibinfo {pages} {2931} (\bibinfo {year} {2013})}\BibitemShut
  {NoStop}%
\bibitem [{\citenamefont {Qiu}\ \emph {et~al.}(2013)\citenamefont {Qiu},
  \citenamefont {Felipe},\ and\ \citenamefont {Louie}}]{qiu2013optical}%
  \BibitemOpen
  \bibfield  {author} {\bibinfo {author} {\bibfnamefont {D.~Y.}\ \bibnamefont
  {Qiu}}, \bibinfo {author} {\bibfnamefont {H.}~\bibnamefont {Felipe}}, \ and\
  \bibinfo {author} {\bibfnamefont {S.~G.}\ \bibnamefont {Louie}},\ }\href@noop
  {} {\bibfield  {journal} {\bibinfo  {journal} {Physical review letters}\
  }\textbf {\bibinfo {volume} {111}},\ \bibinfo {pages} {216805} (\bibinfo
  {year} {2013})}\BibitemShut {NoStop}%
\bibitem [{\citenamefont {Malard}\ \emph {et~al.}(2013)\citenamefont {Malard},
  \citenamefont {Alencar}, \citenamefont {Barboza}, \citenamefont {Mak},\ and\
  \citenamefont {de~Paula}}]{malard2013observation}%
  \BibitemOpen
  \bibfield  {author} {\bibinfo {author} {\bibfnamefont {L.~M.}\ \bibnamefont
  {Malard}}, \bibinfo {author} {\bibfnamefont {T.~V.}\ \bibnamefont {Alencar}},
  \bibinfo {author} {\bibfnamefont {A.~P.~M.}\ \bibnamefont {Barboza}},
  \bibinfo {author} {\bibfnamefont {K.~F.}\ \bibnamefont {Mak}}, \ and\
  \bibinfo {author} {\bibfnamefont {A.~M.}\ \bibnamefont {de~Paula}},\
  }\href@noop {} {\bibfield  {journal} {\bibinfo  {journal} {Physical Review
  B}\ }\textbf {\bibinfo {volume} {87}},\ \bibinfo {pages} {201401} (\bibinfo
  {year} {2013})}\BibitemShut {NoStop}%
\bibitem [{\citenamefont {Woodward}\ \emph {et~al.}(2016)\citenamefont
  {Woodward}, \citenamefont {Murray}, \citenamefont {Phelan}, \citenamefont
  {de~Oliveira}, \citenamefont {Runcorn}, \citenamefont {Kelleher},
  \citenamefont {Li}, \citenamefont {de~Oliveira}, \citenamefont {Fechine},
  \citenamefont {Eda} \emph {et~al.}}]{woodward2016characterization}%
  \BibitemOpen
  \bibfield  {author} {\bibinfo {author} {\bibfnamefont {R.}~\bibnamefont
  {Woodward}}, \bibinfo {author} {\bibfnamefont {R.}~\bibnamefont {Murray}},
  \bibinfo {author} {\bibfnamefont {C.}~\bibnamefont {Phelan}}, \bibinfo
  {author} {\bibfnamefont {R.}~\bibnamefont {de~Oliveira}}, \bibinfo {author}
  {\bibfnamefont {T.}~\bibnamefont {Runcorn}}, \bibinfo {author} {\bibfnamefont
  {E.}~\bibnamefont {Kelleher}}, \bibinfo {author} {\bibfnamefont
  {S.}~\bibnamefont {Li}}, \bibinfo {author} {\bibfnamefont {E.}~\bibnamefont
  {de~Oliveira}}, \bibinfo {author} {\bibfnamefont {G.}~\bibnamefont
  {Fechine}}, \bibinfo {author} {\bibfnamefont {G.}~\bibnamefont {Eda}},  \emph
  {et~al.},\ }\href@noop {} {\bibfield  {journal} {\bibinfo  {journal} {2D
  Materials}\ }\textbf {\bibinfo {volume} {4}},\ \bibinfo {pages} {011006}
  (\bibinfo {year} {2016})}\BibitemShut {NoStop}%
\bibitem [{\citenamefont {Soh}\ \emph {et~al.}(2016)\citenamefont {Soh},
  \citenamefont {Hamerly},\ and\ \citenamefont
  {Mabuchi}}]{soh2016comprehensive}%
  \BibitemOpen
  \bibfield  {author} {\bibinfo {author} {\bibfnamefont {D.~B.}\ \bibnamefont
  {Soh}}, \bibinfo {author} {\bibfnamefont {R.}~\bibnamefont {Hamerly}}, \ and\
  \bibinfo {author} {\bibfnamefont {H.}~\bibnamefont {Mabuchi}},\ }\href@noop
  {} {\bibfield  {journal} {\bibinfo  {journal} {Physical Review A}\ }\textbf
  {\bibinfo {volume} {94}},\ \bibinfo {pages} {023845} (\bibinfo {year}
  {2016})}\BibitemShut {NoStop}%
\bibitem [{\citenamefont {Korm{\'a}nyos}\ \emph {et~al.}(2013)\citenamefont
  {Korm{\'a}nyos}, \citenamefont {Z{\'o}lyomi}, \citenamefont {Drummond},
  \citenamefont {Rakyta}, \citenamefont {Burkard},\ and\ \citenamefont
  {Fal'ko}}]{kormanyos2013monolayer}%
  \BibitemOpen
  \bibfield  {author} {\bibinfo {author} {\bibfnamefont {A.}~\bibnamefont
  {Korm{\'a}nyos}}, \bibinfo {author} {\bibfnamefont {V.}~\bibnamefont
  {Z{\'o}lyomi}}, \bibinfo {author} {\bibfnamefont {N.~D.}\ \bibnamefont
  {Drummond}}, \bibinfo {author} {\bibfnamefont {P.}~\bibnamefont {Rakyta}},
  \bibinfo {author} {\bibfnamefont {G.}~\bibnamefont {Burkard}}, \ and\
  \bibinfo {author} {\bibfnamefont {V.~I.}\ \bibnamefont {Fal'ko}},\
  }\href@noop {} {\bibfield  {journal} {\bibinfo  {journal} {Physical review
  b}\ }\textbf {\bibinfo {volume} {88}},\ \bibinfo {pages} {045416} (\bibinfo
  {year} {2013})}\BibitemShut {NoStop}%
\bibitem [{\citenamefont {Ridolfi}\ \emph {et~al.}(2015)\citenamefont
  {Ridolfi}, \citenamefont {Le}, \citenamefont {Rahman}, \citenamefont
  {Mucciolo},\ and\ \citenamefont {Lewenkopf}}]{ridolfi2015tight}%
  \BibitemOpen
  \bibfield  {author} {\bibinfo {author} {\bibfnamefont {E.}~\bibnamefont
  {Ridolfi}}, \bibinfo {author} {\bibfnamefont {D.}~\bibnamefont {Le}},
  \bibinfo {author} {\bibfnamefont {T.}~\bibnamefont {Rahman}}, \bibinfo
  {author} {\bibfnamefont {E.}~\bibnamefont {Mucciolo}}, \ and\ \bibinfo
  {author} {\bibfnamefont {C.}~\bibnamefont {Lewenkopf}},\ }\href@noop {}
  {\bibfield  {journal} {\bibinfo  {journal} {Journal of Physics: Condensed
  Matter}\ }\textbf {\bibinfo {volume} {27}},\ \bibinfo {pages} {365501}
  (\bibinfo {year} {2015})}\BibitemShut {NoStop}%
\bibitem [{\citenamefont {Fang}\ \emph {et~al.}(2015)\citenamefont {Fang},
  \citenamefont {Defo}, \citenamefont {Shirodkar}, \citenamefont {Lieu},
  \citenamefont {Tritsaris},\ and\ \citenamefont {Kaxiras}}]{fang2015ab}%
  \BibitemOpen
  \bibfield  {author} {\bibinfo {author} {\bibfnamefont {S.}~\bibnamefont
  {Fang}}, \bibinfo {author} {\bibfnamefont {R.~K.}\ \bibnamefont {Defo}},
  \bibinfo {author} {\bibfnamefont {S.~N.}\ \bibnamefont {Shirodkar}}, \bibinfo
  {author} {\bibfnamefont {S.}~\bibnamefont {Lieu}}, \bibinfo {author}
  {\bibfnamefont {G.~A.}\ \bibnamefont {Tritsaris}}, \ and\ \bibinfo {author}
  {\bibfnamefont {E.}~\bibnamefont {Kaxiras}},\ }\href@noop {} {\bibfield
  {journal} {\bibinfo  {journal} {Physical Review B}\ }\textbf {\bibinfo
  {volume} {92}},\ \bibinfo {pages} {205108} (\bibinfo {year}
  {2015})}\BibitemShut {NoStop}%
\bibitem [{\citenamefont {Rasmussen}\ and\ \citenamefont
  {Thygesen}(2015)}]{rasmussen2015computational}%
  \BibitemOpen
  \bibfield  {author} {\bibinfo {author} {\bibfnamefont {F.~A.}\ \bibnamefont
  {Rasmussen}}\ and\ \bibinfo {author} {\bibfnamefont {K.~S.}\ \bibnamefont
  {Thygesen}},\ }\href@noop {} {\bibfield  {journal} {\bibinfo  {journal} {The
  Journal of Physical Chemistry C}\ }\textbf {\bibinfo {volume} {119}},\
  \bibinfo {pages} {13169} (\bibinfo {year} {2015})}\BibitemShut {NoStop}%
\bibitem [{\citenamefont {Korm{\'a}nyos}\ \emph {et~al.}(2015)\citenamefont
  {Korm{\'a}nyos}, \citenamefont {Burkard}, \citenamefont {Gmitra},
  \citenamefont {Fabian}, \citenamefont {Z{\'o}lyomi}, \citenamefont
  {Drummond},\ and\ \citenamefont {Fal'ko}}]{kormanyos2015k}%
  \BibitemOpen
  \bibfield  {author} {\bibinfo {author} {\bibfnamefont {A.}~\bibnamefont
  {Korm{\'a}nyos}}, \bibinfo {author} {\bibfnamefont {G.}~\bibnamefont
  {Burkard}}, \bibinfo {author} {\bibfnamefont {M.}~\bibnamefont {Gmitra}},
  \bibinfo {author} {\bibfnamefont {J.}~\bibnamefont {Fabian}}, \bibinfo
  {author} {\bibfnamefont {V.}~\bibnamefont {Z{\'o}lyomi}}, \bibinfo {author}
  {\bibfnamefont {N.~D.}\ \bibnamefont {Drummond}}, \ and\ \bibinfo {author}
  {\bibfnamefont {V.}~\bibnamefont {Fal'ko}},\ }\href@noop {} {\bibfield
  {journal} {\bibinfo  {journal} {2D Materials}\ }\textbf {\bibinfo {volume}
  {2}},\ \bibinfo {pages} {022001} (\bibinfo {year} {2015})}\BibitemShut
  {NoStop}%
\bibitem [{\citenamefont {Zhang}\ \emph
  {et~al.}(2014{\natexlab{b}})\citenamefont {Zhang}, \citenamefont {Johnson},
  \citenamefont {Hsu}, \citenamefont {Li},\ and\ \citenamefont
  {Shih}}]{zhang2014direct}%
  \BibitemOpen
  \bibfield  {author} {\bibinfo {author} {\bibfnamefont {C.}~\bibnamefont
  {Zhang}}, \bibinfo {author} {\bibfnamefont {A.}~\bibnamefont {Johnson}},
  \bibinfo {author} {\bibfnamefont {C.-L.}\ \bibnamefont {Hsu}}, \bibinfo
  {author} {\bibfnamefont {L.-J.}\ \bibnamefont {Li}}, \ and\ \bibinfo {author}
  {\bibfnamefont {C.-K.}\ \bibnamefont {Shih}},\ }\href@noop {} {\bibfield
  {journal} {\bibinfo  {journal} {Nano letters}\ }\textbf {\bibinfo {volume}
  {14}},\ \bibinfo {pages} {2443} (\bibinfo {year}
  {2014}{\natexlab{b}})}\BibitemShut {NoStop}%
\bibitem [{\citenamefont {Cheiwchanchamnangij}\ and\ \citenamefont
  {Lambrecht}(2012)}]{cheiwchanchamnangij2012quasiparticle}%
  \BibitemOpen
  \bibfield  {author} {\bibinfo {author} {\bibfnamefont {T.}~\bibnamefont
  {Cheiwchanchamnangij}}\ and\ \bibinfo {author} {\bibfnamefont {W.~R.}\
  \bibnamefont {Lambrecht}},\ }\href@noop {} {\bibfield  {journal} {\bibinfo
  {journal} {Physical Review B}\ }\textbf {\bibinfo {volume} {85}},\ \bibinfo
  {pages} {205302} (\bibinfo {year} {2012})}\BibitemShut {NoStop}%
\bibitem [{\citenamefont {Berkelbach}\ \emph {et~al.}(2013)\citenamefont
  {Berkelbach}, \citenamefont {Hybertsen},\ and\ \citenamefont
  {Reichman}}]{berkelbach2013theory}%
  \BibitemOpen
  \bibfield  {author} {\bibinfo {author} {\bibfnamefont {T.~C.}\ \bibnamefont
  {Berkelbach}}, \bibinfo {author} {\bibfnamefont {M.~S.}\ \bibnamefont
  {Hybertsen}}, \ and\ \bibinfo {author} {\bibfnamefont {D.~R.}\ \bibnamefont
  {Reichman}},\ }\href@noop {} {\bibfield  {journal} {\bibinfo  {journal}
  {Physical Review B}\ }\textbf {\bibinfo {volume} {88}},\ \bibinfo {pages}
  {045318} (\bibinfo {year} {2013})}\BibitemShut {NoStop}%
\bibitem [{\citenamefont {Ramasubramaniam}(2012)}]{ramasubramaniam2012large}%
  \BibitemOpen
  \bibfield  {author} {\bibinfo {author} {\bibfnamefont {A.}~\bibnamefont
  {Ramasubramaniam}},\ }\href@noop {} {\bibfield  {journal} {\bibinfo
  {journal} {Physical Review B}\ }\textbf {\bibinfo {volume} {86}},\ \bibinfo
  {pages} {115409} (\bibinfo {year} {2012})}\BibitemShut {NoStop}%
\bibitem [{\citenamefont {Novotny}\ and\ \citenamefont
  {Hecht}(2012)}]{novotny2012principles}%
  \BibitemOpen
  \bibfield  {author} {\bibinfo {author} {\bibfnamefont {L.}~\bibnamefont
  {Novotny}}\ and\ \bibinfo {author} {\bibfnamefont {B.}~\bibnamefont
  {Hecht}},\ }\href@noop {} {\emph {\bibinfo {title} {Principles of
  nano-optics}}}\ (\bibinfo  {publisher} {Cambridge university press},\
  \bibinfo {year} {2012})\BibitemShut {NoStop}%
\bibitem [{\citenamefont {Li}\ \emph {et~al.}(2014)\citenamefont {Li},
  \citenamefont {Chernikov}, \citenamefont {Zhang}, \citenamefont {Rigosi},
  \citenamefont {Hill}, \citenamefont {van~der Zande}, \citenamefont {Chenet},
  \citenamefont {Shih}, \citenamefont {Hone},\ and\ \citenamefont
  {Heinz}}]{li2014measurement}%
  \BibitemOpen
  \bibfield  {author} {\bibinfo {author} {\bibfnamefont {Y.}~\bibnamefont
  {Li}}, \bibinfo {author} {\bibfnamefont {A.}~\bibnamefont {Chernikov}},
  \bibinfo {author} {\bibfnamefont {X.}~\bibnamefont {Zhang}}, \bibinfo
  {author} {\bibfnamefont {A.}~\bibnamefont {Rigosi}}, \bibinfo {author}
  {\bibfnamefont {H.~M.}\ \bibnamefont {Hill}}, \bibinfo {author}
  {\bibfnamefont {A.~M.}\ \bibnamefont {van~der Zande}}, \bibinfo {author}
  {\bibfnamefont {D.~A.}\ \bibnamefont {Chenet}}, \bibinfo {author}
  {\bibfnamefont {E.-M.}\ \bibnamefont {Shih}}, \bibinfo {author}
  {\bibfnamefont {J.}~\bibnamefont {Hone}}, \ and\ \bibinfo {author}
  {\bibfnamefont {T.~F.}\ \bibnamefont {Heinz}},\ }\href@noop {} {\bibfield
  {journal} {\bibinfo  {journal} {Physical Review B}\ }\textbf {\bibinfo
  {volume} {90}},\ \bibinfo {pages} {205422} (\bibinfo {year}
  {2014})}\BibitemShut {NoStop}%
\bibitem [{\citenamefont {Trolle}\ \emph {et~al.}(2015)\citenamefont {Trolle},
  \citenamefont {Tsao}, \citenamefont {Pedersen},\ and\ \citenamefont
  {Pedersen}}]{trolle2015observation}%
  \BibitemOpen
  \bibfield  {author} {\bibinfo {author} {\bibfnamefont {M.~L.}\ \bibnamefont
  {Trolle}}, \bibinfo {author} {\bibfnamefont {Y.-C.}\ \bibnamefont {Tsao}},
  \bibinfo {author} {\bibfnamefont {K.}~\bibnamefont {Pedersen}}, \ and\
  \bibinfo {author} {\bibfnamefont {T.~G.}\ \bibnamefont {Pedersen}},\
  }\href@noop {} {\bibfield  {journal} {\bibinfo  {journal} {Physical Review
  B}\ }\textbf {\bibinfo {volume} {92}},\ \bibinfo {pages} {161409} (\bibinfo
  {year} {2015})}\BibitemShut {NoStop}%
\bibitem [{\citenamefont {Wang}\ \emph
  {et~al.}(2015{\natexlab{b}})\citenamefont {Wang}, \citenamefont {Marie},
  \citenamefont {Gerber}, \citenamefont {Amand}, \citenamefont {Lagarde},
  \citenamefont {Bouet}, \citenamefont {Vidal}, \citenamefont {Balocchi},\ and\
  \citenamefont {Urbaszek}}]{wang2015giant}%
  \BibitemOpen
  \bibfield  {author} {\bibinfo {author} {\bibfnamefont {G.}~\bibnamefont
  {Wang}}, \bibinfo {author} {\bibfnamefont {X.}~\bibnamefont {Marie}},
  \bibinfo {author} {\bibfnamefont {I.}~\bibnamefont {Gerber}}, \bibinfo
  {author} {\bibfnamefont {T.}~\bibnamefont {Amand}}, \bibinfo {author}
  {\bibfnamefont {D.}~\bibnamefont {Lagarde}}, \bibinfo {author} {\bibfnamefont
  {L.}~\bibnamefont {Bouet}}, \bibinfo {author} {\bibfnamefont
  {M.}~\bibnamefont {Vidal}}, \bibinfo {author} {\bibfnamefont
  {A.}~\bibnamefont {Balocchi}}, \ and\ \bibinfo {author} {\bibfnamefont
  {B.}~\bibnamefont {Urbaszek}},\ }\href@noop {} {\bibfield  {journal}
  {\bibinfo  {journal} {Physical review letters}\ }\textbf {\bibinfo {volume}
  {114}},\ \bibinfo {pages} {097403} (\bibinfo {year}
  {2015}{\natexlab{b}})}\BibitemShut {NoStop}%
\bibitem [{\citenamefont {Schweiner}\ \emph {et~al.}(2016)\citenamefont
  {Schweiner}, \citenamefont {Main},\ and\ \citenamefont
  {Wunner}}]{schweiner2016linewidths}%
  \BibitemOpen
  \bibfield  {author} {\bibinfo {author} {\bibfnamefont {F.}~\bibnamefont
  {Schweiner}}, \bibinfo {author} {\bibfnamefont {J.}~\bibnamefont {Main}}, \
  and\ \bibinfo {author} {\bibfnamefont {G.}~\bibnamefont {Wunner}},\
  }\href@noop {} {\bibfield  {journal} {\bibinfo  {journal} {Physical Review
  B}\ }\textbf {\bibinfo {volume} {93}},\ \bibinfo {pages} {085203} (\bibinfo
  {year} {2016})}\BibitemShut {NoStop}%
\bibitem [{\citenamefont {Liu}\ \emph {et~al.}(2013)\citenamefont {Liu},
  \citenamefont {Shan}, \citenamefont {Yao}, \citenamefont {Yao},\ and\
  \citenamefont {Xiao}}]{liu2013three}%
  \BibitemOpen
  \bibfield  {author} {\bibinfo {author} {\bibfnamefont {G.-B.}\ \bibnamefont
  {Liu}}, \bibinfo {author} {\bibfnamefont {W.-Y.}\ \bibnamefont {Shan}},
  \bibinfo {author} {\bibfnamefont {Y.}~\bibnamefont {Yao}}, \bibinfo {author}
  {\bibfnamefont {W.}~\bibnamefont {Yao}}, \ and\ \bibinfo {author}
  {\bibfnamefont {D.}~\bibnamefont {Xiao}},\ }\href@noop {} {\bibfield
  {journal} {\bibinfo  {journal} {Physical Review B}\ }\textbf {\bibinfo
  {volume} {88}},\ \bibinfo {pages} {085433} (\bibinfo {year}
  {2013})}\BibitemShut {NoStop}%
\bibitem [{\citenamefont {Rigosi}\ \emph {et~al.}(2016)\citenamefont {Rigosi},
  \citenamefont {Hill}, \citenamefont {Rim}, \citenamefont {Flynn},\ and\
  \citenamefont {Heinz}}]{rigosi2016electronic}%
  \BibitemOpen
  \bibfield  {author} {\bibinfo {author} {\bibfnamefont {A.~F.}\ \bibnamefont
  {Rigosi}}, \bibinfo {author} {\bibfnamefont {H.~M.}\ \bibnamefont {Hill}},
  \bibinfo {author} {\bibfnamefont {K.~T.}\ \bibnamefont {Rim}}, \bibinfo
  {author} {\bibfnamefont {G.~W.}\ \bibnamefont {Flynn}}, \ and\ \bibinfo
  {author} {\bibfnamefont {T.~F.}\ \bibnamefont {Heinz}},\ }\href@noop {}
  {\bibfield  {journal} {\bibinfo  {journal} {Physical Review B}\ }\textbf
  {\bibinfo {volume} {94}},\ \bibinfo {pages} {075440} (\bibinfo {year}
  {2016})}\BibitemShut {NoStop}%
\end{thebibliography}%

\end{document}